\documentclass[preprint,aps,prd,floatfix,nofootinbib,11pt]{revtex4-2}
\pdfoutput=1
\usepackage{graphicx}
\usepackage{amsfonts}
\usepackage{nicefrac}
\usepackage{xcolor}
\usepackage{natbib}

\usepackage{hyperref}

\usepackage{amsmath}
\usepackage{rotating}
\usepackage{amssymb,amsthm}
\usepackage{soul}

\usepackage{xcolor}
\usepackage[normalem]{ulem}

\usepackage{latexsym}
\usepackage{graphics}

\usepackage{amsfonts}
\usepackage{amsmath}
\usepackage{rotating}
\usepackage{amssymb}

\usepackage{amsmath}
\usepackage{rotating}
\usepackage{amssymb}
\usepackage{soul}

\usepackage{xcolor}

\newcommand\omegaLC{{\overset{\circ}{\omega}}{}}

\newcommand\GLC{{\overset{\ \circ}{G}}{}}

\newtheorem{thm}{Theorem}[section]

\newtheorem{prop}[thm]{Proposition}

\newcommand{\mbold}[1]{\mbox{\boldmath{\ensuremath{#1}}}}  
 
\usepackage{xcolor}
\def\beq{\begin{eqnarray}}  
\def\eeq{\end{eqnarray}}

\def \bh {\mbox{{\bf h}}}
\def \bomega {\mbox{{\mbold \omega}}}

\begin{document}
\title{Spherically symmetric teleparallel geometries}

\author{A. A. Coley}
\email{aac@mathstat.dal.ca}
\affiliation{Department of Mathematics and Statistics, Dalhousie University, Halifax, Nova Scotia, Canada, B3H 3J5}

\author{A. Landry}
\email{a.landry@dal.ca}
\affiliation{Department of Mathematics and Statistics, Dalhousie University, Halifax, Nova Scotia, Canada, B3H 3J5}

\author{R. J. van den Hoogen}
\email{rvandenh@stfx.ca}
\affiliation{Department of Mathematics and Statistics, St. Francis Xavier University, Antigonish, Nova Scotia, Canada, B2G 2W5}

\author{D. D. McNutt}
\email{mcnuttdd@gmail.com}
\affiliation{Centrum Fizyki Teoretycznej, Polska Akademia Nauk, Al. Lotnik\'ow 32/46, 02-668, Warszawa, Poland}

%
%
%
%

\begin{abstract}

We are interested in the development of spherically symmetric geometries in $F(T)$ teleparallel gravity which are of physical importance. We first express the general forms for the spherically  symmetric frame and the zero curvature, metric compatible, spin connection. We then analyse the antisymmetric field equations (the solutions of which split into two cases, which we subsequently consider separately), and derive and analyse the resulting symmetric field equations. In order to further study the applications of spherically symmetric teleparallel models, we study $3$ subcases in which there is an additional affine symmetry so that the resulting field equations reduce to a system of ordinary differential equations. First, we study static spherical symmetric geometries and solve the antisymmetric field equations and subsequently derive the full set of symmetric field equations. In particular, we investigate vacuum spacetimes and obtain a number of new solutions. Second, we consider an additional affine frame symmetry in order to expand the affine frame symmetry group to that of a spatially homogeneous Kantowski-Sachs geometry. Third, we study the special case of spherical symmetry with an additional fourth similarity affine vector.
%

%

\end{abstract}

\maketitle

%
\section{Introduction}

In Einstein's General theory of Relativity (GR), the spherically symmetric vacuum solution, the Schwarzschild solution,  is likely the most recognized solution in all of GR.  The Schwarzschild solution, one of the first solutions to the Einstein field equations (FEs), provides an excellent theoretical model for describing gravitational effects in vacuum around a non-rotating spherical mass. Within our solar system, the Schwarzschild solution provides a mechanism to explain most of the perihelion precession of Mercury that could not be ascribed to Newtonian mechanics alone.  This explanation provided one of the early successful tests of GR. The importance of the Schwarzschild solution in GR cannot be understated.  Indeed, Birkhoff's theorem states that the Schwarzschild solution is the only spherically symmetric and asymptotically flat solution to the Einstein FEs.  Furthermore, these Schwarzschild solutions can give rise to black hole solutions in which there is a horizon in which light ray trajectories which enter the horizon, or initially start inside the horizon, cannot escape and will approach a singularity.

Spherically symmetric solutions in any gravitational theory are clearly of the utmost importance.  Most astrophysical objects are spherically symmetric to some degree, and therefore any theory of gravity should have well described spherically symmetric solutions developed so that the theory can be tested against these astrophysical observations. In this paper, a general class of spherically symmetric geometries is developed in an effort to find spherically symmetric solutions in an alternative theory of gravity to GR.

Teleparallel gravity (see \cite{Bahamonde:2021gfp,cai2016f} and references within for comprehensive reviews) is an alternative to GR theory of gravity. In GR the geometry is characterized by the curvature of the Levi-Civita connection which is a function of the metric. In teleparallel gravity, the geometrical quantity of interest is the torsion which is computed from the frame (or co-frame) and a zero curvature, metric compatible, spin connection.  While the theories appear very different, interestingly, the Teleparallel Equivalent to General Relativity (TEGR) is a subclass of teleparallel gravitational theories that is dynamically equivalent to GR \cite{Aldrovandi_Pereira2013}.  These TEGR theories are based on a scalar, $T$, called the torsion scalar, constructed from a particular linear combination of various scalar invariants of the torsion tensor \cite{Aldrovandi_Pereira2013}. An important generalization of TEGR is $F(T)$ teleparallel gravity, where $F$ is an arbitrary twice differentiable function of the torsion scalar  \cite{Bahamonde:2021gfp,cai2016f,Ferraro:2006jd,Ferraro:2008ey,Linder:2010py,Krssak_Pereira2015,Krssak:2018ywd}.

When the geometrical framework describing $F(T)$ teleparallel gravity is defined in a gauge invariant manner as a geometry with a connection having zero curvature which is also compatible with the metric (that is, the non-metricity is also zero), the resulting FEs derived from such a theory are fully Lorentz covariant \cite{Krssak_Pereira2015,Krssak:2018ywd}.  Due to the Lorentz invariance, there will exist a frame/spin connection pair in which the spin connection identically vanishes.  In this very special class of frame/spin connection pairs, the frame is called a \emph{proper frame}. This is also known as the Weitzenbock gauge. Any teleparallel theory of gravity that restricts itself to only the proper frame is not locally Lorentz invariant. To have a locally Lorentz invariant teleparallel theory of gravity necessarily requires that we do not restrict ourselves to proper frames, and therefore the spin connection may be non-zero in these other frames \cite{Aldrovandi_Pereira2013,Krssak:2018ywd,Lucas_Obukhov_Pereira2009}. For example, if one was to use spherical coordinates to describe a vacuum Minkowski geometry and then assume a diagonal form for the frame and a trivial spin connection, one quickly concludes that something is amiss, as the torsion would be non-nero.  It is important to carefully choose a frame/spin connection pair that is in alignment with the symmetry conditions imposed.

Therefore to construct gravitational models in teleparallel gravity, one requires the development of suitable ansatzes for the frame (or co-frame) and the spin connection that respect the assumed symmetries of the situation. Since the geometrical framework is no longer that of a Riemannian geometry in which the tools to study symmetries have been well developed, one must adapt and generalize these tools to the arena of Riemann-Cartan geometries, and in our case restrict them to teleparallel geometries. The development of these tools to study symmetries in Riemann-Cartan geometries and/or teleparallel geometries is an active area of current research \cite{hohmann2019modified,Coley:2019zld,Hohmann:2021ast,pfeifer2022quick,BohmerJensko,bohmer1,MCH}. Indeed the development of a more robust understanding of symmetries in teleparallel geometries is the first step in the development physical gravitational models in teleparallel gravity. Refs. \cite{hohmann2019modified} and \cite{MCH} are equivalent. However, \cite{MCH} avoids the problems in representation theory that is necessary in \cite{hohmann2019modified} for their approach, which can be difficult to solve for larger groups.

In this paper we shall present the frame/spin connection pair that describes the most general spherically symmetric teleparallel geometries.  We then employ this frame/spin connection pair in the $F(T)$ teleparallel gravity FEs.  With the assumption of additional symmetries the FEs reduce to a system of ordinary differential equations (ODEs) suitable for further study. We study $3$ such cases in which there is an additional affine symmetry. First, we investigate the FEs assuming the geometry is static and determine solutions in vacuum. We then study spatially homogeneous Kankowski-Sachs geometries. Further, with the assumption of a non-trivial similarity condition, the FEs again reduce to a system of ODEs which we analyze.

We will follow the notation of \cite{MCH} and denote the coordinate indices by $\mu, \nu, \ldots$ and the tangent space indices by $a,b,\ldots$. Unless otherwise indicated the spacetime coordinates will be $x^\mu$. Round brackets surrounding indices represents symmetrization, while square brackets represents anti-symmetrization. Any underlined index is not included in the symmetrization. The orthonormal frame fields are denoted as $\bh_a$ and the dual coframe one-forms are $\bh^a$. The vierbein components are $h_a^{~\mu}$ or $h^a_{~\mu}$. The spacetime metric will be denoted as $g_{\mu \nu}$  and since we have an orthonormal frame the tangent space metric is $\eta_{ab}=Diag[-1,1,1,1]$. To denote a local Lorentz transformation leaving $\eta_{ab}$ unchanged, we write $\Lambda_a^{~b}(x^\mu)$ and express the inverse as $\Lambda_b^{~a}=(\Lambda^{-1})^a_{~b}$. The spin connection one-form $\bomega^a_{~b}$, is designated by $\bomega^a_{~b} = \omega^a_{~bc} \bh^c$. The spin connection one-form $\bomega^a_{~b}$ defines a covariant differentiation process on a tensor $T^a_{~~b}$ as $\nabla_c\,T^a_{~~b}=h_c\,\left(T^a_{~~b}\right)+\omega^a_{~ec}\,T^e_{~~b}-\omega^e_{~bc}\,T^a_{~~e}=T^a_{~~b;c}$. The Torsion tensor will be defined as $T^a_{~~bc}=2h_b^{~\mu}\,h_c^{~\nu}\left(h^a_{~[\nu,\mu]}-\omega^a_{~d[\mu} h^a_{~\nu]}\right)$.

\subsection{Review of the Literature}

Some simple spherically symmetric teleparallel solutions, and especially static and vacuum solutions, were studied in \cite{Bahamonde:2021gfp} (also see references within). For example, the Birkhoff theorem does not generally hold in $F(T)$ theories. In particular, a perturbed solution around Minkowski spacetime was first used to put solar system constraints on the parameters in $F(T)$ gravity \cite{Ruggiero,Ruggiero2}.

Cosmological models in flat ($k = 0$) teleparallel Robertson-Walker (TRW) models have been studied in \cite{Bahamonde:2021gfp} (also see references within). For example, simple power-law scale factor solutions in particular 
$F(T)$ models (using a priori ansatz such as a polynomial function) were investigated. In particular, reconstruction methods have been explored extensively in the literature  (in which the function $F(T)$ is  reconstructed from the underlying assumptions of the models). Dynamical systems methods \cite{coley03} (e.g., fixed point and stability analysis) in flat TRW models have been widely utilized \cite{Bahamonde:2021gfp} (also see \cite{BahamondeBohmer}, and \cite{Kofinas,BohmerJensko}), which includes the study of the stability of de Sitter fixed point. In future work we will be interested in further studying the curved $k \neq 0$ TRW models.

Almost all astrophysical solutions found within $F(T)$ theories have been restricted to the static case \cite{Bahamonde:2021gfp}. Static spherically symmetric solutions have been reviewed in \cite{Bahamonde:2021gfp}, and some more recent references are noted below. Here we are only interested in {\em{exact}} solutions in non-trivial $F(T)$ theories (i.e., $F(T) \neq T + \Lambda$ and $T \neq$ const.) that are physical (e.g., real valued tetrads). We note that many of the solutions presented in the literature are incorrect.

Finding exact vacuum static spherially symmetric $F(T)$ teleparallel gravity solutions is an open problem, and there are no known exact vacuum black hole solutions. Indeed, the only exact non-trivial vacuum solution known to the authors in $F(T)$ teleparallel gravity is the non-black hole power law solution presented in \cite{golov1}.

There are a number of special static spherically symmetric (anisotropic) non-vacuum solutions known \cite{Bahamonde:2021gfp}, including both black hole and non-black hole solutions. Some recent special non-trivial non-vacuum exact solutions include those presented in \cite{awad1} and \cite{bahagolov1}, and especially solutions in teleparallel Born-Infeld inspired theories \cite{baha6} (and \cite{bahagolov1}), solutions with electric and magnetic charges \cite{awad1,bahagolov1} and solutions with scalar fields including the scalarized black hole solutions of \cite{awad1,baha4} (also see \cite{golov1,baha1}).


\section{Symmetries and $F(T)$ teleparallel gravity} \label{sec:FSs}

\subsection{Affine frame symmetries with isotropy}

In Riemannian geometry in which the torsion and non-metricity are zero, the geometry is characterized by the metric $g_{ab}$ and its corresponding Levi-Civita connection, $\omegaLC^a_{~bc}$.  If ${\bf X}$ is the infinitesimal generator of a given symmetry, then the symmetry must satisfy Killing's equations (eqns.):
\begin{subequations}\label{Killing}
\begin{align}
\mathcal{L}_{{\bf X}} g_{ab} &= 0, \label{Killing1}\\
\mathcal{L}_{{\bf X}} \omegaLC^a_{~bc} &= 0, \label{Killing2}
\end{align}
\end{subequations}
where $\mathcal{L}$ is the Lie derivative.  However eqn. \eqref{Killing2} is identically satisfied by definition of the Levi-Civita connection as it is computed from the metric. We also note that eqns.  \eqref{Killing} are invariant under local Lorentz transformations $\Lambda^a_{~b}$ of an orthonormal frame, the remaining gauge freedom present in Riemannian geometries.

The class of geometries in which the non-metricity and curvature tensors are zero is a \emph{teleparallel geometry}. In such geometries, the geometry is characterized by the co-frame $\bh^a$ and a purely inertial Lorentz spin connection $\bomega^a_{~b}$  \cite{Aldrovandi_Pereira2013,Krssak:2018ywd}.  In Riemannian geometry, we see that the spin connection is computed from the metric, whereas in teleparallel geometry there exists some Lorentz transformation $\Lambda^a_{~b}$ such that
\begin{equation}
\omega^a_{~bc}=(\Lambda^{-1})^a_{~c}\bh_c\left(\Lambda^c_{~d}\right).
\end{equation}
Since we have assumed an orthonormal frame, the geometry is invariant under Lorentz transformations, and therefore the spin connection $\bomega^a_{~b}$ is not a dynamical quantity, and is only identically zero in a very special class of frames called \emph{Proper Frames} \cite{Aldrovandi_Pereira2013,Krssak:2018ywd}.  Clearly one cannot use Killings eqn. (\ref{Killing}) to define or fix the symmetry.  Any symmetry requirements needs to take into account the non-trivial nature of the spin connection resulting from the invariance of the geometry under Lorentz transformations.

Symmetries of a geometry where frames are the primary object describing the geometry have been previously explored in  \cite{chinea1988symmetries,estabrook1996moving,papadopoulos2012locally}.  A complication arises when the group of symmetries contains a non-trivial linear isotropy group. If there is non-trivial linear isotropy group, then the determination of the group of symmetries necessarily requires the solution to a system of inhomogeneous differential equations (DEs) \cite{olver1995equivalence}:
\begin{subequations}\label{Affine}
\begin{align}
\mathcal{L}_{{\bf X}} \bh^{a} &= \lambda^a_{~b} \bh^b, \label{Liederivative:frame}\\
\mathcal{L}_{{\bf X}} \omega^a_{~bc} &= 0, \label{Liederivative:Con}
\end{align}
\end{subequations}
with unknown local functions $\lambda^a_{~b}=\lambda^a_{~b}(x^\mu)$ which are elements of the Lie algebra of Lorentz transformations. We note that if the linear isotropy group is trivial \cite{Coley:2019zld}, then $\lambda^a_{~b} = 0$.

From eqns. \eqref{Liederivative:frame} and \eqref{Liederivative:Con} it follows that the Lie derivative of the torsion tensor and its covariant derivatives with respect to ${\bf X}$ are identically zero. We will call any vector field ${\bf X}$ satisfying eqns. \eqref{Liederivative:frame} and \eqref{Liederivative:Con} an {\it affine frame symmetry} of the geometry. We note that \eqref{Liederivative:frame} implies eqns. \eqref{Killing} while the second condition \eqref{Liederivative:Con} is the definition for an affine collineation \cite{aaman1998riemann}.

A novel approach to explore the symmetries of any Riemann-Cartan geometry characterized by the frame and a metric compatible spin connection was proposed in \cite{MCH}. In such Riemann-Cartan geometries, the connection is an independent object, and only if the curvature is zero does it reduce to teleparallel geometries.  The approach in \cite{MCH} relies on the existence and determination of a class of invariantly defined frames, which facilitates the solving the DEs \eqref{Affine} by fixing the functions $\lambda^a_{~b}$ in an invariant way.

The Cartan-Karlhede (CK) algorithm \cite{Coley:2019zld,fonseca1996algebraic} provides a mechanism to determine a class of invariantly defined co-frames, $\bh^a$, up to the linear isotropy $H_q$ present in the geometry. An initial frame is chosen and the parameters of the Lorentz frame transformations $\lambda^a_{~b}$ are fixed by normalizing the components of the  torsion tensor and its covariant derivatives in an invariant manner. The CK algorithm provides a finite number of invariants that can be employed to determine the dimension of the symmetry group and locally characterize a geometry uniquely. One sequence of discrete invariants are of particular importance; the dimension of the linear isotropy group, $dim~H_{p}$. The linear isotropy group, $H_p$ is a subgroup of the Lorentz frame transformations that leave the torsion tensor and its covariant derivatives up to $p^{th}$ order invariant.

We summarize the approach in \cite{MCH} that determines the class of invariantly define co-frames that can be employed as an ansatz for the characterization of the geometry satisfying an assumed group of symmetries. Consider a Riemann-Cartan geometry characterized by $({\bh}^a, \bomega^a_{~b})$ that admits an affine frame symmetry generated by the vector field ${\bf X}$.  The class of {\bf symmetry frames} are defined as the frames $\bh^a$ that satisfy
\begin{equation}
\mathcal{L}_{{\bf X}} \bh^a = f_{X}^{~\hat{i}} \lambda^{~a}_{\hat{i}~b} \bh^b, \label{TP:frm:sym}
\end{equation}
where $\lambda^{~a}_{\hat{i}~b}$ are basis elements of the Lie algebra of the isotropy group ($\hat{i}$ ranges from 1 to the dimension of the isotropy group). The components of $f_X^{~\hat{i}}$ are local functions that depend only on the coordinates affected by the symmetry. This is a general definition and is not (strictly speaking) an invariantly defined frame until the components $f_{ X}^{~\hat{i}}$ are fixed in some coordinate independent way. The linear isotropy group associated with fixing the frame using eqn. \eqref{TP:frm:sym} is given a new symbol, $\bar{H}_q$ in order to distinguish it from the linear isotropy group of the Cartan-Karlhede algorithm. This distinction is important as the elements of $\bar{H}_q$ are more restricted than their counterparts in $H_q$.

Since the components of $f_X^{~\hat{i}}$ are associated with the Lie derivative of a symmetry generating vector field, they are tensorial in nature that depend on the co-frame. Through the use of the CK algorithm we can judiciously choose the functions in $f_X^{~\hat{i}}$ to construct invariantly defined frames up to some minimal subgroup of the isotropy group. For example, if the geometry is spherically symmetric with the usual spherical coordinates $(t,r,\theta,\phi)$, a choice for $f^{~\hat{i}}_{{X}_I}$, $I=1,2,3$, can be made so that the linear isotropy subgroup, $\bar{H}_q$, consists only of rotations in the subspace $\theta=\theta_0$ and $\phi=\phi_0$. In order to preserve the form of $f^{~\hat{i}}_{{X}_I}$ the Lorentz parameters are dependent on $t$ and $r$ only. As this procedure develops invariantly defined frames up to the linear isotropy group in Riemann-Cartan geometries, all that remains to restrict to teleparallel geometries is to solve the zero curvature constraint. In \cite{MCH} we outlined a detailed procedure to fix the components of $f_X^{~\hat{i}}$ and determine $\overline{H}_q$ for the spherically symmetric geometries explicitly.

\subsubsection{Determining the geometry for a group of affine symmetries with isotropy}

The DEs that determine the most general teleparallel geometry which has a non-trivial linear isotropy group is presented in \cite{MCH}. Here we quote Theorem III.2 in \cite{MCH} for reference. The most general teleparallel geometry which admits a given group of affine frame symmetries, ${\bf X}_I,~ I,J,K \in \{1, \ldots N\}$ with a non-trivial isotropy group of dimension $n$ can be determined by solving for the unknowns $h^a_{~\mu}$, $f_I^{~\hat{i}}$ (with $\hat{i}, \hat{j}, \hat{k} \in \{1, \ldots n\}$) and $\omega^a_{~bc}$ from the eqns.
\begin{subequations}
\begin{align}\label{Sym:RC:Prop}
& X_I^{~\nu} \partial_{\nu} h^a_{~\mu} + \partial_{\mu} X_I^{~\nu} h^a_{~\nu} = f_I^{~\hat{i}} \lambda^a_{\hat{i}~b} h^b_{~\mu}, \\
& 2X_{[I} ( f_{J]}^{~\hat{k}}) - f_I^{~\hat{i}} f_J^{~\hat{j}} C^{\hat{k}}_{~\hat{i} \hat{j}} = C^K_{~IJ} f_K^{~\hat{k}}, \\
& X_I^{~d} \bh_d( \omega^a_{~bc}) + \omega^d_{~bc} f_I^{~\hat{i}} \lambda^a_{\hat{i}~d} - \omega^a_{~dc} f_I^{~\hat{i}} \lambda^d_{\hat{i}~b} - \omega^a_{~bd} f_I^{~\hat{i}} \lambda^d_{\hat{i}~c} - \bh_c( f_I^{~\hat{i}}) \lambda^a_{\hat{i}~b} = 0,
\end{align} 
\end{subequations}
where $\{ \lambda^a_{\hat{i}~b}\}_{\hat{i}=1}^n$ are a basis of the Lie algebra of the isotropy group, $[\lambda_{\hat{i}}, \lambda_{\hat{j}}] = C^{\hat{k}}_{~\hat{i}\hat{j}} \lambda_{\hat{k}}$, and $[{\bf X}_I, {\bf X}_J] = C^K_{~IJ} {\bf X}_K$.
Along with the additional condition that the  curvature tensor expressed in terms of the spin connection and its derivatives must vanish in teleparallel geometries
\begin{equation}
R^a_{~bcd}=2\omega^a_{\phantom{a}b[d,c]}+2\omega^a_{\phantom{a}e[c}\omega^e_{\phantom{a}bd]}
+\omega^a_{\phantom{a}be}c^e_{\phantom{a}cd} = 0.\label{Sym:Prop1}
\end{equation}

\noindent Note the equivalent of eqn. (\ref{Sym:Prop1}) in \cite{MCH} has been corrected here to include the contribution of the anholonimity $c^e_{\phantom{a}cd}= 2h^e_{~[\mu,\nu]}h_c^{~\nu}h_d^{~\mu}$ to the expression for the curvature which is missing.

\subsection{Brief Overview of $F(T)$ Teleparallel Gravity}

The complete Lagrangian for $F(T)$ teleparallel gravity is
\begin{equation}
L= \frac{h}{2\kappa}F(T)+L_{Matt} \label{lagrangian}
\end{equation}
where $\kappa$ is the gravitational coupling constant. The torsion scalar, $T$, can be expressed in terms of the torsion and the super-potential, $S^a_{~\mu \nu}$, as follows
\begin{equation}
T=\frac{1}{2}T^a_{\phantom{a}\mu\nu}S_a^{\phantom{a}\mu\nu},
\end{equation}
where
\begin{equation}
S_a^{\phantom{a}\mu\nu}=\frac{1}{2}\left(T_a^{\phantom{a}\mu\nu}+T^{\nu\mu}_{\phantom{\nu\mu}a}
    -T^{\mu\nu}_{\phantom{\mu\nu}a}\right)-h_a^{\phantom{a}\nu}T^{\phi\mu}_{~~\phi} + h_a^{\phantom{a}\mu}T^{\phi\nu}_{~~\phi}. \label{super}
\end{equation}

Assuming an orthonormal frame in which the tangent frame metric is $\eta_{ab}=Diag[-1,1,1,1]$, variations of the Lagrangian, which include a non-trivial spin connection \cite{Krssak:2018ywd,Krssak_Saridakis2015}, yield {\it Lorentz covariant} FEs. Using Lagrange multipliers to impose the conditions that the spin connection is both metric compatible and flat, and performing the necessary variations with respect to these two multipliers, yields the following
\begin{equation}
\omega_{(ab)\mu} = 0 \ \mbox{and}\  \omega^a_{\phantom{a}b\mu} = (\Lambda^{-1})^a_{~c}\partial_\mu\Lambda^c_{~b}. \label{solution_omega}
\end{equation}
Since we have chosen an orthonormal gauge, there is a $\Lambda^a_{\phantom{a}b}\in SO(1,3)$ for which this is true.

The canonical energy momentum is defined as
\begin{equation}
h\Theta_a^{\phantom{a}\mu}=-\frac{\delta L_{Matt}}{\delta h^{\vphantom{\mu}a}_{\phantom{a}\mu}}.
\end{equation}
The invariance of the FEs under Lorentz transformations implies that the canonical energy momentum is symmetric and it can be shown that the usual metrical energy momentum $T_{ab}$ is related to the symmetric part of the canonical energy momentum, that is,
\begin{equation}
\Theta_{[ab]}=0,\qquad \Theta_{(ab)}= T_{ab}=-\frac{1}{2}\frac{\delta L_{Matt}}{\delta g_{ab}}.
\end{equation}

Variations of the Lagrangian, eqn. (\ref{lagrangian}), with respect to the co-frame yield FEs that can be decomposed into symmetric and antisymmetric parts as
\begin{subequations}\label{FE}
\begin{align}
\kappa \Theta_{(ab)}&= F''(T)S_{(ab)}^{\phantom{(ab)}\nu} \partial_{\nu} T+F'(T)\GLC_{ab}  + \frac{1}{2}g_{ab}\left(F(T)-TF'(T)\right),\label{SYM_FE}\\
             0      &= F''(T)S_{[ab]}^{\phantom{[ab]}\nu} \partial_{\nu} T,\label{ASYM_FE}
\end{align}
\end{subequations}
where $\GLC_{ab}$ is the Einstein tensor computed from the Levi-Civita connection of the metric.

From eqn. \eqref{FE} we observe that if $T=T_0$, a constant, then the FEs for $F(T)$ teleparallel gravity are equivalent to a rescaled version of TEGR (which looks like GR with a cosmological constant and a re-scaled coupling constant) \cite{Krssak:2018ywd}. In the case of TEGR, where $F(T)=T$, eqn. \eqref{ASYM_FE} is identically satisfied, and again the theory reduces to a theory of gravity that is dynamically equivalent to GR.  In general, if $F(T)$ is non-linear, and $T$ is not a constant, then the anti-symmetric part of the FEs \eqref{ASYM_FE} impose constraints on the geometry \cite{Coley:2022aty,vandenHoogen:2023pjs,Coley:2023ibm,golovadd1}.

\section{Spherical symmetry}

Working in coordinates $ x^\mu = (t, r, \theta, \phi)$, the affine frame symmetry generators associated with the three-dimensional spherical symmetry  group are: 
\beq \begin{aligned} & X_z = \partial_{\phi},~X_y = - \cos \phi \partial_{\theta} + \frac{\sin \phi}{\tan \theta} \partial_{\phi}, X_x = \sin \phi \partial_{\theta} + \frac{\cos \phi}{\tan \theta} \partial_{\phi}. \end{aligned} \eeq
Writing $\{ X_I\}_{I=1}^3 = \{ X_x, X_y, X_z\}$ the non-zero commutator (structure) constants, $C^I_{~JK} = -\epsilon^I_{~JK}$, are
$C^{{3}}_{~{1} {2}} = -1,~ C^{{2}}_{~{1} {3}} = 1,~C^{{1}}_{~{2} {3}} = -1$. The basis for the isotropy group is of the form:
\beq \lambda_{\hat{1}} = \left[ \begin{array}{cccc} 0 & 0 & 0 & 0 \\
0 & 0 & 0 & 0 \\ 0 & 0 & 0 & 1 \\ 0 & 0 & -1 & 0 \end{array} \right],~ \lambda_{\hat{2}} = -\left[ \begin{array}{cccc} 0 & 0 & 0 & 0 \\
0 & 0 & 1 & 0 \\ 0 & -1 & 0 & 0 \\ 0 & 0 & 0 & 0 \end{array} \right] ,~
\lambda_{\hat{3}} =  -\left[ \begin{array}{cccc} 0 & 0 & 0 & 0 \\
0 & 0 & 0 & 1 \\ 0 & 0 & 0 & 0 \\ 0 & -1 & 0 & 0 \end{array} \right] ,  \label{SSrep} \eeq

\noindent where the group $\overline{Iso}$ consists of all spatial rotations. Since $X_3 = X_z$ is a generator of a spatial rotation, we can choose our frame so that $X_3$ acts as a rotation on the basis elements $\bh^3$ and $\bh^4$ \cite{MCH}; i.e., 

\beq f_{3}^{~\hat{i}} \lambda_{\hat{i}} = \left[ \begin{array}{cccc} 0 & 0 & 0 & 0 \\ 0 & 0 & 0 & 0 \\  0 & 0 & 0 & f_3^{~\hat{1}}  \\ 0 & 0 & -f_3^{~\hat{1}} & 0  \end{array} \right], \nonumber \eeq

\noindent then by applying a rotation about $\bh^3$ and $\bh^4$, the remaining component, $f_3^{~\hat{1}}$, can be set to zero.
Choosing $f_3^{~\hat{i}}$ in this way, and using eqn. \eqref{Sym:RC:Prop}, we find the solution:
\beq f_2^{~\hat{i}} = f_2^{~\hat{i}}(\theta) \cos \phi + g_2^{~\hat{i}}(\theta) \sin \phi; ~~
f_1^{~\hat{i}} = \partial_{\phi} f_2^{~\hat{i}}. \eeq

If we use Lorentz transformations that are dependent on $\theta$ alone we can, without loss of generality, apply a rotation to set the coefficients of $\cos \phi$ in $f_2^{~\hat{i}}$ to zero. Eqn. \eqref{Sym:RC:Prop} with ${\bf X}_2$ and ${\bf X}_1$ then gives:
\beq \partial_{\theta} g_2^{~\hat{i}} + \tan(\theta) g_2^{~\hat{i}} = 0; ~~
g_2^{~\hat{i}} = \frac{C_2^{~\hat{i}}}{\sin(\theta)}. \eeq
This fully determines $f_1^{~\hat{i}}$ and $f_2^{~\hat{i}}$, as the first is the $\phi$-derivative of the other.  Applying a rotation with constant parameters, we can set $f_2^{~\hat{2}}= f_2^{~\hat{3}} = 0$, without loss of generality we will choose the first component to be non-zero and set the arbitrary constant to one, giving the following matrix representation of the coefficients:
\beq f_I^{~\hat{i}} = \left [ \begin{array}{ccc}  \frac{\cos(\phi)}{\sin(\theta)} & 0 & 0 \\ \frac{\sin(\phi)}{\sin(\theta)} & 0 & 0 \\ 0 & 0 & 0 \end{array} \right] \label{SS:fmat}.\eeq

\noindent By making this choice, we have effectively fixed the frame up to $\overline{H}_q$ which consists of rotations in the $\bh^3-\bh^4$ plane with parameters dependent on coordinates $t$ and $r$.

Relative to the representation for the isotropy group in eqn. \eqref{SSrep}, we can solve eqn. \eqref{Sym:Prop1} to determine the most general frame admitting this symmetry group. We can choose a new coordinate system to ``diagonalize" the frame.  The resulting vierbein is:
\beq h^a_{~\mu} = \left[ \begin{array}{cccc} A_1(t,r) & 0 & 0 & 0 \\ 0 & A_2(t,r) & 0 & 0 \\ 0 & 0 & A_3(t,r) & 0 \\ 
0 & 0 & 0 & A_3(t,r) \sin(\theta) \end{array}\right]. \label{VB:SS} \eeq

The form of the $f_I^{~\hat{i}}$ in eqn. \eqref{SS:fmat} has been defined invariantly. With this choice the frame in eqn. \eqref{VB:SS} is an invariant symmetry frame, and we can now obtain the most general  metric compatible connection:
\beq \begin{aligned} & \omega_{341} = W_1(t,r),\quad\quad\quad\quad\quad   \omega_{342} = W_2(t,r), \quad\quad\quad\quad\quad \omega_{233} = \omega_{244} = W_3(t,r),
\\
& \omega_{234} = -\omega_{243} = W_4(t,r), \quad \omega_{121} = W_5(t,r),\quad\quad\quad\quad\quad \omega_{122} = W_6(t,r), 
\\
& \omega_{133} = \omega_{144} = W_7(t,r),\quad\quad \omega_{134} = -\omega_{143} = W_8(t,r), \quad \omega_{344} = - \frac{\cos(\theta)}{A_3 \sin(\theta)}. \end{aligned} \label{Con:SS} \eeq
Finally, to determine the most general connection for a teleparallel geometry we must impose the flatness condition in eqn. \eqref{Sym:Prop1}. The resulting eqns. can be solved so that any spherically symmetric teleparallel geometry is defined by (the three arbitrary functions in the vierbein \eqref{VB:SS} along with) the following spin connection components:
\beq \begin{aligned} 
W_1 &= -\frac{\partial_t \chi}{A_1}, W_2 = -\frac{\partial_r \chi}{A_2}, W_3 = \frac{\cosh(\psi)\cos(\chi)}{A_3}, W_4 = \frac{\cosh(\psi)\sin(\chi)}{A_3},\\
W_5 &= -\frac{\partial_t \psi}{A_1}, W_6 = -\frac{\partial_r \psi}{A_2}, W_7 = \frac{\sinh(\psi) \cos(\chi)}{A_3}, W_8 = \frac{\sinh(\psi) \sin(\chi)}{A_3},  
\end{aligned} \label{SS:TPcon} \eeq

\noindent where $\chi$ and $\psi$ are arbitrary functions of the coordinates $t$ and $r$.

Any choice of the arbitrary functions, $\psi$ and $\chi$, picks out a unique teleparallel geometry, as any change in the form of the spin connection which could affect the form of $\psi$ or $\chi$ leads to a change in the form of the vierbein. For a given pair of functions, the invariantly defined frame up to the linear isotropy group $\bar{H}_q$ arising from the CK algorithm could be computed to provide further sub-classification. We note that there are only five (the minimum number of) arbitrary functions required to specify a geometry: $A_1, A_2, A_3, \psi$ and $\chi$. We also note that there are four inequivalent branches of spherically symmetric teleparallel geometries, depending on whether $\chi$ or $\psi$ vanish when solving the antisymmetric FE, as the frame in eqn. \eqref{VB:SS} has been fully fixed by restricting the vierbein to a particular form. We note that special subclasses of these teleparallel geometries have been studied earlier in teleparallel gravity \cite{sharif2009teleparallel} using the Killing eqns. for an arbitrary spherically symmetric metric and using the non-invariant proper frame approach \cite{hohmann2019modified,pfeifer2022quick,pfeifer2021static}.

\subsection{Antisymmetric FEs}

The antisymmetric FEs for the frame given by eqn. \eqref{VB:SS} and the connection given by eqn. \eqref{Con:SS} with $W_i$ given by eqn. \eqref{SS:TPcon} are:
\begin{subequations}
\begin{eqnarray}
0 &=& \frac{F''\left(T\right)}{\kappa\,A_1\,A_2\,A_3}\,\left[\partial_r\,T\,\left(A_1\,\cos\,\chi\,\sinh\,\psi+\partial_t\,A_3\right)-\partial_t\,T\,\left(A_2\,\cos\,\chi\,\cosh\,\psi+\partial_r\,A_3\right)\right], \quad \label{1001a}
\\
0 &=& \frac{F''\left(T\right)\,\sin\,\chi}{\kappa\,A_1\,A_2\,A_3}\,\left[A_1\,\cosh\,\psi\,\partial_r\,T-A_2\,\sinh\,\psi\,\partial_t\,T\right], \label{1001b}
\end{eqnarray}
\end{subequations}
where $T$ is the torsion scalar. There are two possible solutions:
\begin{itemize}
\item First case: $\sin\,\chi=0$: $\chi = n\,\pi$ where $n \in\,\mathbb{Z}$ is an integer and $\cos\,\chi=\cos\left(n\,\pi\right)=\pm 1 = \delta$, and
\begin{align}\label{1002}
\left(\partial_t\,T\right) = \left[\frac{\delta\,A_1\,\sinh\,\psi+\partial_t\,A_3}{\delta\,A_2\,\cosh\,\psi+\partial_r\,A_3}\right]\left(\partial_r\,T\right).
\end{align}

\item Second case: $A_1\,\cosh\,\psi\,\partial_r\,T=A_2\,\sinh\,\psi\,\partial_t\,T$.
\end{itemize}

There is a third case in which $T=\text{const}$, whence the antisymmetric FEs  \eqref{1001a} and \eqref{1001b} are satisfied trivially since $\partial_t\,T=0$ and $\partial_r\,T=0$. Hence in this case the non-trivial symmetric FE reduce to those of GR, with a rescaled gravitational constant and a cosmological constant. We shall not be interested in such solutions here.

\subsection{Symmetric FEs for first case}

There are four non-trivial components to the symmetric FEs: the three independent diagonal components (1,1), (2,2) and (3,3), and (1,2). On the diagonal, we obtained three independent and non-trivial eqns. because the components (3,3) and (4,4) are identical. However, we can also transform these four eqns. to obtain simpler eqns. This boils down to the eqns. ((2,2)-(3,3)), ((1,1)+(2,2)), (1,1) and (1,2) by substituting in eqn. \eqref{1002}. If we set $F\left(T\right)=T$, $F'\left(T\right)=1$ and $F''\left(T\right)=0$ for the TEGR case, the FEs will simplify further.

Before we attempt to solve the symmetric FEs for $T\left(t,\,r\right)$ and the $A_j\left(t,\,r\right)$ (an alternative form of the full FE is presented in Appendix \ref{appena}), let us express them in a simple way as follows:
\begin{subequations}
\begin{eqnarray}
0 &=& -F''\left(T\right)\,\left(\partial_r\,T\right)\,k_1\left(t,r\right)+F'\left(T\right)\,g_1\left(t,r\right), \label{1003a}
\\
\kappa\,\left[\rho+P\right] &=& -2\,F''\left(T\right)\,\left(\partial_r\,T\right)\,k_2\left(t,r\right)+2\,F'\left(T\right)\,g_2\left(t,r\right), \label{1003b}
\\
\left[\kappa\,\rho+\frac{F\left(T\right)}{2}\right] &=& -2\,F''\left(T\right)\,\left(\partial_r\,T\right)\,k_3\left(t,r\right)+2\,F'\left(T\right)\,g_3\left(t,r\right) , \label{1003c}
\\
0 &=& -F''\left(T\right)\,\left(\partial_r\,T\right)\,k_4\left(t,r\right)+F'\left(T\right)\,g_4\left(t,r\right), \label{1003d}
\end{eqnarray}
\end{subequations}
where the expressions $k_i$ and $g_i$ ($i=1\,-\,4$) are given explicitly by:
\small
\begin{subequations}
\begin{align}
g_1 =& \frac{1}{A_1^3\,A_2^3\,A_3^2}\Bigg[-A_1^2\,A_2\,A_3^2\,\left(\partial_r^2\,A_1\right)+A_1\,A_2^2\,A_3^2\,\left(\partial_t^2\,A_2\right)-A_1^3\,A_2\,A_3\,\left(\partial_r^2\,A_3\right)-A_1\,A_2^3\,A_3\,\left(\partial_t^2\,A_3\right)
\nonumber\\
&+A_1^3\,A_2\,\left(\partial_r\,A_3\right)^2-A_1^3\,A_2^3+\partial_r\,\left(A_1\,A_2\right)\,A_1^2\,A_3\,\left(\partial_r\,A_3\right)-A_1\,A_2^3\,\left(\partial_t\,A_3\right)^2
\nonumber\\
&+A_2^2\,A_3\,\partial_t\,\left(A_1\,A_2\right)\left(\partial_t\,A_3\right)+A_1^2\,A_3^2\left(\partial_r\,A_1\right)\left(\partial_r\,A_2\right)-A_2^2\,A_3^2\left(\partial_t\,A_1\right)\left(\partial_t\,A_2\right)\Bigg]  , \label{1004a}
\\
g_2 =& \frac{1}{A_1^3\,A_2^3\,A_3}\Bigg[-A_1^3\,A_2\,\left(\partial_r^2\,A_3\right)-A_1\,A_2^3\,\left(\partial_t^2\,A_3\right)+\partial_r\,\left(A_1\,A_2\right)\,A_1^2\,\left(\partial_r\,A_3\right)+A_2^2\,\partial_t\,\left(A_1\,A_2\right)\left(\partial_t\,A_3\right)\Bigg]  , \label{1004b}
\\
g_3 =& \frac{1}{A_1^2\,A_2^3\,A_3^2}\Bigg[-A_1^2\,A_2\,A_3\,\left(\partial_r^2\,A_3\right)-A_1^2\,A_2\,\left(\partial_r\,A_3\right)^2-A_1\,A_2\,A_3\left(\partial_r\,A_3\right)\left(\partial_r\,A_1\right)
\nonumber\\
&+A_1\,A_2^2\,A_3\left(\partial_r\,A_3\right)\left(\partial_t\,\psi\right)-\delta\,A_1^2\,A_2^2\,\left(\partial_r\,A_3\right)\,\cosh\,\psi+A_1^2\,A_3\,\left(\partial_r\,A_2\right)\left(\partial_r\,A_3\right)
\nonumber\\
&+A_2^3\,\left(\partial_t\,A_3\right)^2+2A_2^2\,A_3\left(\partial_t\,A_2\right)\left(\partial_t\,A_3\right)+\delta\,A_1\,A_2^3\left(\partial_t\,A_3\right)\,\sinh\psi-A_1\,A_2^2\,A_3\,\left(\partial_t\,A_3\right)\left(\partial_r\,\psi\right)
\nonumber\\
&-A_1\,A_2^2\,A_3\left(-\delta\,A_2\cosh\psi\,\left(\partial_t\,\psi\right)+\delta\,A_1\,\sinh\psi \left(\partial_r\,\psi\right)+\delta\,\cosh\psi\,\left(\partial_r\,A_1\right)-\delta\,\sinh\psi\,\left(\partial_t\,A_2\right)\right)
\Bigg] , \label{1004c}
\\
g_4 =& \frac{1}{A_1^2\,A_2^2\,A_3}\Bigg[-A_1\,A_2\left(\partial_r\,\partial_t\,A_3\right)+A_1\,\left(\partial_r\,A_3\right)\left(\partial_t\,A_2\right)+A_2\,\left(\partial_r\,A_1\right)\left(\partial_t\,A_3\right)\Bigg] , \label{1004d}
\end{align}

\newpage

\begin{align}
k_1 =& \frac{1}{A_1^3\,A_2^3\,A_3^2}\Bigg[\left[\frac{\delta\,A_1\,\sinh\,\psi+\partial_t\,A_3}{\delta\,A_2\,\cosh\,\psi+\partial_r\,A_3}\right] \Bigg[A_1\,A_2^3\,A_3\,\left(\partial_t\,A_3\right)
\nonumber\\
& +A_1\,A_2^2\,A_3\,\Bigg(-A_3\,\left(\partial_t\,A_2\right) +A_1\,\Bigg(\delta\,A_2\,\sinh\psi+A_3\,\left(\partial_r\,\psi\right)\Bigg)\Bigg)\Bigg]
\nonumber\\
& +\Bigg[A_1^3\,A_2\,A_3\,\left(\partial_r\,A_3\right)-A_1^2\,A_2\,A_3\left(-A_3\,\left(\partial_r\,A_1\right)
 +A_2\,\left(-\delta\,A_1\,\cosh \psi +A_3\,\left(\partial_t\,\psi\right)\right)\right)\Bigg]\Bigg] , \label{1005a}
\\
k_2 =& \frac{1}{A_1^2\,A_2^2\,A_3}\left[\frac{A_2^2\,\left[\delta\,A_1\,\sinh\,\psi+\partial_t\,A_3\right]^2+A_1^2\left[\partial_r\,A_3+\delta\,A_2\,\cosh\psi\right]^2}{\delta\,A_2\,\cosh\,\psi+\partial_r\,A_3}\right],  \label{1005b}
\\
k_3 =& \frac{1}{A_2^2\,A_3}\left[\delta\,A_2\,\cosh\psi+\partial_r\,A_3\right], \label{1005c}
\\
k_4 =& \frac{1}{A_1\,A_2\,A_3}\left[\delta\,A_1\,\sinh\,\psi+\partial_t\,A_3\right]. \label{1005d}
\end{align}
\end{subequations}
\normalsize

Eqns. \eqref{1003a} and \eqref{1003d} yield the following relation without $F\left(T\right)$ and its derivatives:
\begin{eqnarray}\label{1006}
g_1\left(t,r\right)\,k_4\left(t,r\right) = g_4\left(t,r\right)\,k_1\left(t,r\right) ,
\end{eqnarray}
which can be written explicitly as:
\small
\begin{multline}\label{1007}
\Bigg[-A_1^2\,A_2\,A_3^2\,\left(\partial_r^2\,A_1\right)+A_1\,A_2^2\,A_3^2\,\left(\partial_t^2\,A_2\right)-A_1^3\,A_2\,A_3\,\left(\partial_r^2\,A_3\right)-A_1\,A_2^3\,A_3\,\left(\partial_t^2\,A_3\right)+A_1^3\,A_2\,\left(\partial_r\,A_3\right)^2\\
-A_1^3\,A_2^3+\partial_r\,\left(A_1\,A_2\right)\,A_1^2\,A_3\,\left(\partial_r\,A_3\right)-A_1\,A_2^3\,\left(\partial_t\,A_3\right)^2+A_2^2\,A_3\,\partial_t\,\left(A_1\,A_2\right)\left(\partial_t\,A_3\right)\\
+A_1^2\,A_3^2\left(\partial_r\,A_1\right)\left(\partial_r\,A_2\right)-A_2^2\,A_3^2\left(\partial_t\,A_1\right)\left(\partial_t\,A_2\right)\Bigg]
\times\Bigg(A_1\,A_2\,\left[\delta\,A_1\,\sinh\,\psi+\partial_t\,A_3\right]\Bigg)\\
=\Bigg[\left[\frac{\delta\,A_1\,\sinh\,\psi+\partial_t\,A_3}{\delta\,A_2\,\cosh\,\psi+\partial_r\,A_3}\right] \Bigg[A_1\,A_2^3\,A_3\,\left(\partial_t\,A_3\right)+A_1\,A_2^2\,A_3\,\Bigg(-A_3\,\left(\partial_t\,A_2\right)
+A_1\,\Bigg(\delta\,A_2\,\sinh\psi+A_3\,\left(\partial_r\,\psi\right)\Bigg)\Bigg)\Bigg]\\+\Bigg[A_1^3\,A_2\,A_3\,\left(\partial_r\,A_3\right)-A_1^2\,A_2\,A_3\,\left(-A_3\,\left(\partial_r\,A_1\right)+A_2\,\left(-\delta\,A_1\,\cosh \psi +A_3\,\left(\partial_t\,\psi\right)\right)\right)\Bigg]\Bigg]\\
\times\Bigg(-A_1\,A_2\left(\partial_r\,\partial_t\,A_3\right)+A_1\,\left(\partial_r\,A_3\right)\left(\partial_t\,A_2\right)+A_2\,\left(\partial_r\,A_1\right)\left(\partial_t\,A_3\right)\Bigg).
\end{multline}
\normalsize
This is a partial differential equation (PDE) involving only the $A_j\left(t,r\right)$ ($j = 1-3$) and $\psi \left(t,r\right)$.

Eqns. \eqref{1003a} and \eqref{1003d} also yield
\begin{equation}\label{1008}
\partial_r\,\left[\ln\,F'\left(T\right)\right]=\frac{g_1\left(t,r\right)}{k_1\left(t,r\right)},
\end{equation}
which can be formally integrated to find
\begin{equation}\label{1009}
F'\left(T\right)=C\left(t\right)\,\exp\left[\int\,dr'\,\frac{g_1\left(t,r'\right)}{k_1\left(t,r'\right)}\right],
\end{equation}
where $C\left(t\right)$ is an arbitrary function of $t$ only. In principle, we can eventually solve this eqn. (\ref{1009}) for $F\left(T(t,r)\right)$. By substituting the eqns.  (\ref{1008}) and (\ref{1009}) in the eqns. \eqref{1003b} and \eqref{1003c}, we obtain the following:
\small
\begin{subequations}
\begin{eqnarray}
\kappa\,\left[\rho+P\right] &=& 2\,C\left(t\right)\,\exp\left[\int\,dr'\,\frac{g_1\left(t,r'\right)}{k_1\left(t,r'\right)}\right]\left[g_2\left(t,r\right)-k_2\left(t,r\right)\,\left(\frac{g_1\left(t,r\right)}{k_1\left(t,r\right)}\right)\right],  \label{1010a}
\\
\left[\kappa\,\rho+\frac{F\left(T\right)}{2}\right] &=& 2\,C\left(t\right)\,\exp\left[\int\,dr'\,\frac{g_1\left(t,r'\right)}{k_1\left(t,r'\right)}\right]\left[g_3\left(t,r\right)-k_3\left(t,r\right)\,\left(\frac{g_1\left(t,r\right)}{k_1\left(t,r\right)}\right)\right].  \label{1010b}
\end{eqnarray}
\end{subequations}
\normalsize
We may also replace $\frac{g_1\left(t,r'\right)}{k_1\left(t,r'\right)}$ terms by $\frac{g_4\left(t,r'\right)}{k_4\left(t,r'\right)}$ term inside eqns. \eqref{1010a} and \eqref{1010b} for more simplicity.

The form of the torsion scalar becomes (using $\sin\,\chi=0$ and  $\cos\,\chi= \delta$):
\small 
\begin{align} \label{1011}
T=& \left( -4\,\delta\,{\frac{\sinh \left(\psi( t,r)\right)}{A_2(t,r) A_3(t,r)}}
-4\,{\frac{\partial_t\,A_3(t,r)}{A_2(t,r) A_3(t,r) A_{1} \left( t,r \right)}} \right) \partial_r\,\psi( t,r)
+4\,\delta\,{\frac{ \left(\partial_t\psi( t,r)\right) \cosh \left(\psi( t,r)\right)}{A_1(t,r) A_3(t,r)}}
\nonumber\\
&
-4\,\delta\,{\frac{\left( {\partial_r}A_{1} \left( t,r \right)  \right) \cosh \left(\psi( t,r)\right)}{A_2(t,r) A_3(t,r) A_{1} \left( t,r \right) }}+4\,\delta\,{\frac{ \left( {\partial_t}A_2(t,r)\right) \sinh \left( \psi( t,r)\right) }{A_2(t,r) A_3(t,r) A_1(t,r)}}
-4\,\delta\,\frac{\cosh \left( \psi( t,r)\right){\partial_r}A_3(t,r)}{A_2(t,r)\left( A_3(t,r)\right)^{2}}
\nonumber\\
&+4\,\delta\,{\frac{\sinh \left( \psi( t,r)  \right) {\partial_t}A_3(t,r) }{A_1(t,r) \left( A_3(t,r)\right)^{2}}}
+4\,{\frac{\left({\partial_t}\psi( t,r)\right) {\partial_r}A_3(t,r) }{A_2(t,r) A_3(t,r) A_{1} \left( t,r\right)}}
-4\,{\frac{\left({\partial_r}A_1(t,r)\right) {\partial_r}A_3(t,r) }{A_3(t,r) \left( A_2(t,r)\right)^{2}A_1(t,r) }}
\nonumber\\
&+4\,{\frac{\left({\partial_t}A_2(t,r)\right) {\partial_t}A_{3} \left( t,r \right) }{A_3(t,r)\left( A_1(t,r)\right)^{2}A_2(t,r)}}
-2\,{\frac{\left({\partial_r}A_{3}\left( t,r \right)\right)^{2}}{\left( A_2(t,r)\right)^{2} \left( A_3(t,r) \right)^{2}}}
+2\,{\frac{\left({\partial_t}A_{3}\left( t,r \right)\right)^{2}}{\left( A_1(t,r)\right)^{2} \left( A_3(t,r) \right)^{2}}}
\nonumber\\
&-2\,\left( A_3(t,r)\right)^{-2} .
\end{align}
\normalsize

\subsection{Symmetric FEs for second case}

For the general solution in this case, we substitute the eqn. \eqref{1001b} into the eqn. \eqref{1001a} for the antisymmetric FEs. We recall for $\partial_i\,T \neq 0$ that $\partial_t\,T=\frac{A_1\,\cosh\,\psi}{A_2\,\sinh\,\psi}\,\partial_r\,T$, so that:
\begin{eqnarray}\label{1012}
\sinh\,\psi\,\left(A_1^{-1} \partial_t\,A_3\right) = \cos\,\chi+\cosh\,\psi\,\left(A_2^{-1} \partial_r\,A_3\right).
\end{eqnarray}
We can isolate $\chi\left(t,\,r\right)$ in the following way:
\begin{eqnarray}\label{1013}
\cos(\chi\left(t,\,r\right)) =\sinh\,\psi\,\left( A_1^{-1}\,\partial_t\,A_3\right)-\cosh\,\psi\,\left(A_2^{-1}\,\partial_r\,A_3\right)
\end{eqnarray}
We have in eqn. \eqref{1013} a relation giving $\chi$ as a function of $\psi$ and the other $A_j$ as a function of $t$ and $r$. For the eqn. giving $T$, this will reduce to the following PDE:
\begin{eqnarray}\label{1014}
\partial_t\,T = \left[\frac{A_1\,\coth \,\psi}{A_2}\right]\,\left(\partial_r\,T\right).
\end{eqnarray}
There are also four non-trivial components to the symmetric FEs: three independent diagonals plus the component (1,2) because components (3,3) and (4,4) are identical. However, we have also transformed these four eqns. to obtain much simpler eqns. Before solving the FEs for $T\left(t,\,r\right)$ and the $A_j\left( t,\,r\right)$ in general, we express the symmetric FEs in the same manner as in eqns. \eqref{1003a} -- \eqref{1003d}. The corresponding expressions for the functions $k_i$ and $g_i$ ($i=1\,-\,4$) are given in this case by:
\newpage
\small
\begin{subequations}
\begin{align}
g_1 =& \frac{1}{A_1^3\,A_2^3\,A_3^2}\Bigg[-A_1^2\,A_2\,A_3^2\,\left(\partial_r^2\,A_1\right)+A_1\,A_2^2\,A_3^2\,\left(\partial_t^2\,A_2\right)-A_1^3\,A_2\,A_3\,\left(\partial_r^2\,A_3\right)-A_1\,A_2^3\,A_3\,\left(\partial_t^2\,A_3\right)
\nonumber\\
&+A_1^3\,A_2\,\left(\partial_r\,A_3\right)^2+\partial_r\,\left(A_1\,A_2\right)\,A_1^2\,A_3\,\left(\partial_r\,A_3\right)-A_1\,A_2^3\,\left(\partial_t\,A_3\right)^2+A_2^2\,A_3\,\partial_t\,\left(A_1\,A_2\right)\left(\partial_t\,A_3\right)
\nonumber\\
&+A_1^2\,A_3^2\left(\partial_r\,A_1\right)\left(\partial_r\,A_2\right)-A_2^2\,A_3^2\left(\partial_t\,A_1\right)\left(\partial_t\,A_2\right)-A_1^3\,A_2^3\Bigg] ,  \label{1015a}
\\
g_2 =& \frac{1}{A_1^3\,A_2^3\,A_3} \Bigg[-A_1^3\,A_2\,\left(\partial_r^2\,A_3\right)-A_1\,A_2^3\,\left(\partial_t^2\,A_3\right)+\partial_r\,\left(A_1\,A_2\right)\,A_1^2\,\left(\partial_r\,A_3\right)+A_2^2\,\partial_t\,\left(A_1\,A_2\right)\left(\partial_t\,A_3\right)\Bigg] , \label{1015b}
\\
g_3 =& \frac{1}{A_1^3\,A_2^3\,A_3^2}\Bigg[A_1^3\,A_2\,A_3\,\left(\partial_r^2\,A_3\right)\,\left(\delta\,\cosh^2\psi-1\right)+A_1\,A_2^3\,A_3\,\delta\sinh^2\psi\left(\partial_t^2\,A_3\right)
\nonumber\\
&-2A_1^2\,A_2^2\,A_3\sinh\psi\,\cosh\psi\,\delta\,\left(\partial_r\,\partial_t\,A_3\right)+A_1^3\,A_2\,\sinh^2\psi\left(\partial_r\,A_3\right)^2
\nonumber\\
&-2A_1^2\,A_2^2\,\left(\partial_t\,A_3\right)\left(\partial_r\,A_3\right)\,\cosh\psi\,\sinh\psi+A_1^2\,A_3\,\left(\partial_r\,A_3\right)\Bigg(-A_2^2\sinh^2\psi\left(1+\delta\right)\left(\partial_t\psi\right)
\nonumber\\
&+A_2\,\left(\partial_r\,A_1\right)\sinh^2\psi+A_2\,\left(\partial_t\,A_2\right)\left(\delta-1\right)\sinh\psi\,\cosh\psi+A_1\,A_2\left(1+\delta\right)\cosh\psi\,\sinh\psi\,\left(\partial_r\,\psi\right)
\nonumber\\
&-A_1\,\left(\partial_r\,A_2\right)\left(\delta\cosh^2\psi-1\right)\Bigg)+A_1\,A_2^3\cosh^2\psi\left(\partial_t\,A_3\right)^2
\nonumber\\
&-A_2^2\,A_3\left(\partial_t\,A_3\right)\Bigg(-A_1\,A_2\cosh\psi\,\sinh\psi\left(\delta+1\right)\left(\partial_t\psi\right)-A_1\left(\partial_r\,A_1\right)\cosh\psi\,\sinh\psi\left(\delta-1\right)
\nonumber\\
&-A_1\left(\partial_t\,A_2\right)\left(\cosh^2\psi+1\right)+A_1^2\,\left(\delta+1\right)\cosh^2\psi\left(\partial_r\,\psi\right)+A_2\,\left(\partial_t\,A_1\right)\,\delta\,\sinh^2\psi\Bigg)\Bigg] , \label{1015c}
\\
g_4 =& \frac{1}{A_1^2\,A_2^2\,A_3}\Bigg[-A_1\,A_2\left(\partial_r\,\partial_t\,A_3\right)+A_1\,\left(\partial_r\,A_3\right)\left(\partial_t\,A_2\right)+A_2\,\left(\partial_r\,A_1\right)\left(\partial_t\,A_3\right)\Bigg] , \label{1015d}
\end{align}
\begin{align}
k_1 =& \frac{1}{A_1\,A_2^2\,A_3}\Bigg[\coth\,\psi\Bigg[\cosh\psi\left[A_1\left(\partial_r\,A_3\right)\,\sinh\psi-A_2\,\left(\partial_t\,A_3\right)\cosh\psi\right]-A_3\left(A_1\,\left(\partial_r\psi\right)-\left(\partial_t\,A_2\right)\right)\Bigg]
\nonumber\\
&+\Bigg[\sinh\psi\,\left[A_1\,\left(\partial_r\,A_3\right)\,\sinh\psi-A_2\,\left(\partial_t\,A_3\right)\cosh\psi\right]+A_3\,\left[A_2\,\left(\partial_t\psi\right)-\left(\partial_r\,A_1\right)\right]\Bigg]\Bigg] , \label{1016a}
\\
k_2 =& -\frac{1}{A_1\,A_2^2\,A_3}\left(\frac{1+2\,\sinh^2\psi}{\sinh\psi}\right)\left[A_1\,\sinh\psi\left(\partial_r\,A_3\right)-A_2\,\cosh\psi\,\left(\partial_t\,A_3\right)\right] , \label{1016b}
\\
k_3 =& -\frac{\sinh\psi}{A_1\,A_2^2\,A_3} \left[A_1\,\left(\partial_r\,A_3\right)\sinh\psi-A_2\,\cosh\psi\,\left(\partial_t\,A_3\right)\right], \label{1016c}
\\
k_4 =& -\frac{1}{A_1\,A_2^2\,A_3}\,\cosh\psi\,\left[A_1\,\left(\partial_r\,A_3\right)\sinh\psi-A_2\,\left(\partial_t\,A_3\right)\,\cosh\psi\right] . \label{1016d}
\end{align}
\end{subequations}
\normalsize

\noindent We will obtain the same types of relations as eqns. \eqref{1010a} and \eqref{1010b} for the above functions $k_i\left(t,r\right)$, and $g_i\left(t,r\right)$. The eqn. \eqref{1006} represents a relation involving only the $A_j\left(t,r\right)$ and $\psi \left(t,r\right)$. 

\newpage

The torsion scalar will be expressed as follows:
\small
\begin{align} \label{1018}
T=& -4\,{\frac{\sinh \left( \psi( t,r)  \right)  \left( {\partial_t}\arccos\left[\sinh\,\psi\,\left( A_1^{-1}\,\partial_t\,A_3\right)-\cosh\,\psi\,\left(A_2^{-1}\,\partial_r\,A_3\right)\right]  \right)}{A_{1} \left( t,r \right) A_3(t,r) }}
\nonumber\\
& \quad\;\;\times \sin \left( \arccos\left[\sinh\,\psi\,\left( A_1^{-1}\,\partial_t\,A_3\right)-\cosh\,\psi\,\left(A_2^{-1}\,\partial_r\,A_3\right)\right] \right)
\nonumber\\
&+4\,{\frac {\cosh \left( \psi( t,r)  \right)  \left( {\partial_r}\arccos\left[\sinh\,\psi\,\left( A_1^{-1}\,\partial_t\,A_3\right)-\cosh\,\psi\,\left(A_2^{-1}\,\partial_r\,A_3\right)\right]  \right) }{A_2(t,r) A_3(t,r) }}
\nonumber\\
&\quad\;\;\times \sin \left( \arccos\left[\sinh\,\psi\,\left( A_1^{-1}\,\partial_t\,A_3\right)-\cosh\,\psi\,\left(A_2^{-1}\,\partial_r\,A_3\right)\right]  \right)
\nonumber\\
&+ \left( -4\,{\frac{\sinh \left( \psi( t,r)  \right) \left[\sinh\,\psi\,\left( A_1^{-1}\,\partial_t\,A_3\right)-\cosh\,\psi\,\left(A_2^{-1}\,\partial_r\,A_3\right)\right]}{A_2(t,r) A_3(t,r) }}
-4\,{\frac{{\partial_t}A_3(t,r) }{A_2(t,r) A_3(t,r) A_{1} \left( t,r \right) }} \right)
\nonumber\\
& \times{\partial_r}\psi( t,r)+4\,{\frac{ \left( {\partial_t}\psi( t,r)  \right) \cosh \left( \psi( t,r)  \right) \left[\sinh\,\psi\,\left( A_1^{-1}\,\partial_t\,A_3\right)-\cosh\,\psi\,\left(A_2^{-1}\,\partial_r\,A_3\right)\right]}{A_1(t,r) A_3(t,r)}}
\nonumber\\
&-4\,{\frac{ \left( {\partial_r}A_{1} \left( t,r \right)  \right) \left[\sinh\,\psi\,\left( A_1^{-1}\,\partial_t\,A_3\right)-\cosh\,\psi\,\left(A_2^{-1}\,\partial_r\,A_3\right)\right]  \cosh \left( \psi( t,r)\right) }{A_2(t,r) A_3(t,r) A_{1} \left( t,r \right) }}
\nonumber\\
&+4\,{\frac{ \left( {\partial_t}A_2(t,r)  \right) \left[\sinh\,\psi\,\left( A_1^{-1}\,\partial_t\,A_3\right)-\cosh\,\psi\,\left(A_2^{-1}\,\partial_r\,A_3\right)\right]  \sinh \left( \psi( t,r)  \right) }{A_2(t,r) A_3(t,r) A_1(t,r) }}
\nonumber\\
&-4\,{\frac{\cosh \left( \psi( t,r)  \right) \left[\sinh\,\psi\,\left( A_1^{-1}\,\partial_t\,A_3\right)-\cosh\,\psi\,\left(A_2^{-1}\,\partial_r\,A_3\right)\right]  {\partial_r}A_3(t,r) }{A_2(t,r)  \left( A_3(t,r)  \right) ^{2}}}
\nonumber\\
&+4\,{\frac{\sinh \left( \psi( t,r)  \right) \left[\sinh\,\psi\,\left( A_1^{-1}\,\partial_t\,A_3\right)-\cosh\,\psi\,\left(A_2^{-1}\,\partial_r\,A_3\right)\right] {\partial_t}A_3(t,r) }{A_1(t,r)  \left( A_3(t,r)  \right) ^{2}}}
\nonumber\\
&+4\,{\frac{ \left( {\partial_t}\psi( t,r)  \right) {\partial_r}A_3(t,r) }{A_2(t,r) A_3(t,r) A_{1} \left( t,r \right) }}
-4\,{\frac{ \left( {\partial_r}A_1(t,r)  \right) {\partial_r}A_3(t,r) }{A_3(t,r)  \left( A_2(t,r)  \right) ^{2}A_1(t,r) }}
+4\,{\frac{ \left( {\partial_t}A_2(t,r)  \right) {\partial_t}A_{3} \left( t,r \right) }{A_3(t,r)  \left( A_1(t,r)  \right) ^{2}A_2(t,r) }}
\nonumber\\
& -2\,{\frac{ \left( {\partial_r}A_{3} \left( t,r \right)  \right) ^{2}}{ \left( A_2(t,r)  \right) ^{2} \left( A_3(t,r)  \right) ^{2}}}
+2\,{\frac{ \left( {\partial_t}A_{3} \left( t,r \right)  \right) ^{2}}{ \left( A_1(t,r)  \right) ^{2} \left( A_3(t,r)  \right) ^{2}}}
-2\, \left( A_3(t,r)\right)^{-2}.
\end{align}
\normalsize

Having found the general form of the spin connection, we can now in principle apply a Lorentz transformation to set the spin connection to zero and explicitly present the general spherically symmetric geometry in the proper frame.

\subsubsection{Coordinate choice}

There is a class of transformations that maintain the ``diagonal" form of the frame in \eqref{VB:SS}. But, in general, this will also lead to a change in the form of the connection and the FEs. Therefore, and in order to maintain the comoving nature of the perfect fluid source, we restrict ourselve to transformations of the form $t \rightarrow f(t)$ and $r \rightarrow g(r)$ (i.e., redefining the $r$ and $t$ coordinates). This may simplify the eqns. further. However, even then the explicit forms for, e.g., the symmetry vectors (and the similarity variable, $z$ -- see later) will change.

We shall look for solutions for particular functions $F(T)$, with an eqn. of state $P=\alpha\,\rho$, and with additional symmetries. First, let us review some special geometries.

\subsection{Robertson-Walker geometries}

In \cite{MCH} we discussed the teleparallel Robertson-Walker (TRW) geometries with a 6D group of affine symmetries. Working in the coordinate system  $(t, r, \theta, \phi)$ (and in a diagonal co-frame with $A_1=1,\,A_2=a(t)\,{\sqrt{1-kr^2}},\, A_3=r\,a(t)$), the additional three affine frame symmetry generators (in addition to the three associated with spherical symmetry) of the RW metric were displayed and the most general metric-compatible connection was obtained. In the flat ($k=0$) case the non-trivial components are:
\begin{equation} 
\omega_{233} = \omega_{244} = - \frac{1}{a(t)r}, \quad\quad \omega_{344} = - \frac{\cos(\theta)}{a(t) r \sin(\theta)}.  \label{Con:FLRW} 
\end{equation}
The TRW class was determined using the proper frame approach in \cite{hohmann2021complete}, which is not an invariant approach \cite{hohmann2021complete}. Many of the applications in the literature of these geometries have concentrated on the analysis of the flat ($k = 0$) TRW cosmological models \cite{Bahamonde:2021gfp,cai2016f}. But TRW models with non-zero $k$ have been studied \cite{hohmann2021teleparallel,hohmann2021general}. In addition, particular forms for $F(T)$ have been considered \cite{paliathanasis2022f}.

\subsection{de Sitter geometry}

We recall that in GR, the de Sitter solution is a special case of the flat FLRW solutions. Hence we consider the special subclass of flat  $k = 0$ TRW geometries which admit a seven-dimensional symmetry group. This is given by the TRW vierbein  and the connection  where (the functions $\chi = \psi =0$ and) $a(t)$ is defined by $a(t) = C_3 e^{C_4 t}$. 
In this case the (seventh) affine symmetry is given by
\beq X = \frac{1}{C_0} \partial_t + r \partial_r.\label{extra}
\eeq
where $C_0 \equiv -H_0$. The teleparallel de Sitter geometry (TdS) is defined to be the  teleparallel geometry with the $G_7$ Lie group of affine symmetries, $\mathcal{R} \rtimes E(3)$ which is a subgroup of $O(1,4)$.  We note in this geometry the covariant derivative of the torsion tensor is zero.

\subsection{Equivalence of FLRW subclasses}

Let us next show that the $k = +1$ and $k = -1$ cases are inequivalent and that the subcases with $\delta = \pm 1$ are equivalent using Cartan invariants \cite{Coley:2019zld}. For both signs of $k$ the Cartan invariants appear as the components of the irreducible parts of the torsion tensor and their first covariant derivative. This is due to the fact that the linear isotropy group is $SO(3)$ at all iterations of the CK algorithm, and that the number of functionally independent invariants is at most one. 
\begin{enumerate}
\item $k = +1$: 

At zeroth order, the invariants are 
\begin{eqnarray}
V_1 = - \frac{3 \dot{a}}{a},~~ A_1 = - \frac{2  \delta \sqrt{k}}{a}, \label{kp1:zero} \end{eqnarray}

\noindent while at first order, we find the following two invariants

\begin{eqnarray}
(\nabla {\bf V})_{11} = \dot{V_1}, ~~ (\nabla {\bf A})_{11} = - \frac{2 \delta \sqrt{k} \dot{a} }{a^2}. \label{kp1:first} \end{eqnarray}

\item $k = -1$: 

At zeroth order, the sole invariant is

\begin{eqnarray}
V_1 = 3 \left( \frac{\dot{a}}{a} + \frac{\delta \sqrt{-k}}{a}\right), \label{km1:zero} \end{eqnarray}

\noindent while at first order we find two invariants

\begin{eqnarray}
(\nabla {\bf V})_{11} = \dot{V_1},~~ (\nabla {\bf V})_{22} = (\nabla {\bf V})_{33} = (\nabla {\bf V})_{44} = \frac{\delta \sqrt{-k} V_1}{a}. \label{km1:first} \end{eqnarray} 
\end{enumerate}

It is clear that the cases $k = \pm 1$ must be inequivalent due to the differing invariant structure of the Cartan invariants. It is less obvious, if the $\delta = \pm 1$ are equivalent. 

Alternatively, the connection coefficients for this case can be determined using the connection coefficients of the spherically symmetric case in \cite{McNutt:2022} where the vierbein is given by eqn. \eqref{VB:SS}, while the spin-connection is given by eqns. \eqref{Con:SS} with the ansatz described by eqns. \eqref{SS:TPcon}. In the case of TRW symmetry, the affine frame symmetry conditions lead to further conditions on the $W_i$'s:
\begin{subequations}
\begin{eqnarray}
k<0 &&: \chi = 0, \pi,~\psi = \cosh^{-1} (-\cos(\chi) \sqrt{1-kr^2}), \\
k \geq 0 &&: \chi =  \cos^{-1} (- \sqrt{1-kr^2}), \psi = 0. \end{eqnarray}
\end{subequations}

\noindent This choice of representation of the arbitrary functions in the connection coefficients shows that the sign difference (i.e., $\delta=\pm 1$) in the case of $k >0$ is purely a choice of representation and hence these subcases are equivalent. 

We note that in the $k<0$ case, $\delta = \cos(\chi)$ and the two subcases $\delta = \pm 1$  are not related by a proper orthochronous Lorentz transformation, since transformations of the form  $r\,\rightarrow\,-r$ and $t \,\rightarrow\,-t$ are required to transform from one to the other. This implies that the two subcases are not equivalent under $SO^*\left(1,\,3\right)$.

\section{Static Spherical Symmetric spacetimes}

Within teleparallel gravity, stationary or static spherically symmetric geometries have been discussed in applications \cite{sharif2009teleparallel,bahamonde2021exploring,pfeifer2021static}. In these papers, the stationarity or static condition is not rigorously derived but instead arises from restricting the arbitrary functions to be dependent on the radial coordinate alone. Although such a choice intuitively makes sense, it does not necessarily constitute the most general class of stationary or static spherically symmetric geometries since it does not consider the inequivalent subclasses of spherically symmetric geometries.

Stationary spherically symmetric geometries can be produced by including an additional affine frame symmetry, ${\bf Y}$ with $Y^a Y_a < 0$ locally. This vector field must commute with the original frame symmetries, $[{\bf X}_I, {\bf Y}] = 0$ for $I \in \{1, \ldots, 3\}$. Therefore, a new coordinate system can be chosen so that ${\bf Y} = \partial_{t}$, while preserving the vierbein in {\em non-diagonal} form instead of the more specialized vierbein in eqn. \eqref{VB:SS}; that is, in general we cannot use a coordinate transformation for stationary solutions because our vierbein ansatz is set to a diagonal metric and in the coordinate system where ${\bf X} = \partial_t$ the metric will, in general, have a non-diagonal (i.e., $dtdr$) cross-term.
The timelike condition on ${\bf Y}$ imposes an additional condition on the vierbein functions $A_i$, $i\in \{1, \ldots, 3\}$.

Static spherically symmetric geometries are produced by requiring that ${\bf Y}$ is hypersurface orthogonal:  
\beq {\bf Y} \wedge d {\bf Y} = 0. \eeq
\noindent In the case of static solutions, the diagonal vierbein form \eqref{VB:SS} is applicable (unlike in the stationary case). The expansion of the symmetry group introduces three additional functions and this can be determined using \eqref{Sym:RC:Prop}. As the commutator constants are zero, the analysis of these eqns. is straightforward and imply that the vierbein and the components of the connection one-form must be dependent on the coordinate $r$ alone. As no other conditions are imposed, this proves that the static spherically symmetric solutions can be found by requiring that the functions $A_i,~i=1,2,3$, $\psi$ and $\chi$ are functions of the $r$ coordinate only.

\subsubsection{Static spherically symmetric affine symmetry}

We have here ${\bf X}=\frac{1}{C_0}\partial_t$ as the affine static spherically symmetric generator. The coframe components are now:
\begin{eqnarray}\label{2001}
A_1=A_1\left(r\right),\quad\quad A_2=A_2\left(r\right),\quad\quad A_3=A_3\left(r\right),
\end{eqnarray}
and $\chi=\chi(r), \psi=\psi(r)$. We still have the coordinate freedom to choose $A_3(r)=r$, which we will exploit later.

For the antisymmetric static FEs, we will use the following version of eqn. (\ref{VB:SS}) as follows:
\begin{eqnarray}\label{2002}
h^a_{\;\;\mu}=Diag\left[A_1\left(r\right), A_2\left(r\right), A_3\left(r\right), A_3\left(r\right)\,\sin \theta\right].
\end{eqnarray}
The functions parameterizing the spin-connection described by eqn. (\ref{SS:TPcon}) become:
\small
\begin{align}\label{2003}
W_1 \left(r\right) &= 0 , & W_2 \left(r\right) &= -\partial_r\,\chi\left(r\right),& W_3 \left(r\right) &= \cosh\left(\psi\left(r\right)\right)\cos\left(\chi\left(r\right)\right), 
\nonumber\\
W_4 \left(r\right) &= \cosh\left(\psi\left(r\right)\right)\sin\left(\chi\left(r\right)\right), & W_5 \left(r\right) &= 0 , & W_6 \left(r\right) &= -\partial_r\,\psi\left(r\right),
\nonumber\\
W_7 \left(r\right) &= \sinh\left(\psi\left(r\right)\right)\cos\left(\chi\left(r\right)\right), & W_8 \left(r\right) &= \sinh\left(\psi\left(r\right)\right)\sin\left(\chi\left(r\right)\right).
\end{align}
\normalsize

\subsection{Antisymmetric FEs}

Further, we assume that the source is a perfect fluid having energy density $\rho=\rho(r)$ and pressure $P=P(r)$ with comoving fluid velocity $${\bf u}  = \frac{1}{A_1}\partial_t.$$

For the antisymmetric FEs, the equivalent of eqns. \eqref{1001a} and \eqref{1001b} are described by the following relations:
\begin{eqnarray}\label{2004}
0 &=& \frac{F''\left(T\right)\,\partial_r\,T}{\kappa\,A_2\,A_3}\,\left[\cos\,\chi\,\sinh\,\psi\right] \quad \text{and}\quad 0 = \frac{F''\left(T\right)\,\partial_r\,T}{\kappa\,A_2\,A_3}\,\left[\sin\,\chi\,\cosh\,\psi\right] .
\end{eqnarray}
The eqn.  \eqref{2004} admits the solution (for $T \neq \text{const}$):
\begin{itemize}
\item $\sin\,\chi=0$ and $\sinh\psi=0$: $\chi = n\,\pi$ and $\psi=0$ where $n \in \mathbb{Z}$ is an integer and $\cos\,\chi=\cos\left(n\,\pi\right)=\pm 1 = \delta$ and $\cosh\psi=1$.
\end{itemize}

It is curious to note that in this case ($T\not = 0$), most of the spin connection coefficients become zero. The only non-trivial spin connection components are
\begin{equation}\label{2005}
\omega_{233}=\omega_{244}=\frac{\delta}{A_3}, \text{\ and\ } \omega_{344}=-\frac{\cos(\theta)}{A_3\sin(\theta)}.
\end{equation}
This, of course, is in agreement with what is already known in the literature \cite{Krssak:2018ywd,Krssak_Saridakis2015}. The torsion scalar becomes
\begin{equation}
T=-2\left(\frac{\delta}{A_3}+\frac{\partial_r A_3}{A_2 A_3}\right)\left(\frac{\delta}{A_3}+\frac{\partial_r A_3}{A_2A_3}+2\frac{\partial_r A_1}{A_2A_1}\right), \label{T_sim_form}
\end{equation}  
and we note that the ``axial'' part of the torsion scalar is identically zero \cite{Krssak_Saridakis2015}.

\subsection{Symmetric FEs}

The full set of  eqns. \eqref{1003a} -- \eqref{1003c} above (using the restrictions obtained from the antisymmetric FE) then becomes (removing all time dependencies):
\begin{subequations}
\begin{align} 
&\kappa \rho=-\frac{1}{2}\left(F(T)-TF'(T)\vphantom{\frac{1}{A_3^2}}\right)
-2F''(T)\left(\frac{\partial_r T}{A_2}\right)\left(\frac{\delta}{A_3}+\frac{\partial_r A_3}{A_2A_3}\right)
\nonumber\\
&\qquad\quad +F'(T)\left[-2\frac{\partial_r^2 A_3}{A_2^2A_3}+2\left(\frac{\partial_r A_2}{A_2^2}\right)\left(\frac{\partial_r A_3}{A_2 A_3}\right)-\left(\frac{\partial_r A_3}{A_2A_3}\right)^2+\frac{1}{A_3^2}\right] ,\label{2006a} 
\\
&\kappa P=\frac{1}{2}\left(F(T)-TF'(T)\vphantom{\frac{1}{A_3^2}}\right)
\nonumber\\
&\qquad\qquad\qquad\quad\, -F'(T)\left[-2\left(\frac{\partial_r A_1}{A_2A_1}\right)\left(\frac{\partial_r A_3}{A_2 A_3}\right)-\left(\frac{\partial_r A_3}{A_2A_3}\right)^2+\frac{1}{A_3^2}\right] ,\label{2006b}
\\
&\frac{\kappa}{2} (\rho+3P)=\frac{1}{2}\left(F(T)-TF'(T)\vphantom{\frac{1}{A_3^2}}\right)
+F''(T)\left(\frac{\partial_r T}{A_2}\right)\left(\frac{\partial_r A_1}{A_2A_1}\right)
\nonumber\\
&\qquad\quad +F'(T)\left[\frac{\partial_r^2 A_1}{A_2^2A_1}-\left(\frac{\partial_r A_1}{A_2A_1}\right)\left(\frac{\partial_r A_2}{A_2A_2}\right)+2\left(\frac{\partial_r A_1}{A_2A_1}\right)\left(\frac{\partial_r A_3}{A_2 A_3}\right) \right] .\label{2006c}
\end{align}
\end{subequations}
The eqns. above can be supplemented with the energy-momentum conservation eqn.
\begin{equation}  \label{con}
(\rho+P)\partial_rA_1+A_1\partial_r P = 0 ,
\end{equation}
which follows from eqns. \eqref{2006a}--\eqref{2006c}. In the case of TEGR, where $F(T)=T$, the right-hand side of the first line in each of eqns. \eqref{2006a} -- \eqref{2006c} above is trivial and only the second line in each eqn. remains.

We note that eqns. \eqref{2006a} -- \eqref{2006c} constitute three FEs (instead of four for the symmetric parts, because the component (1,2) becomes trivial without the time dependence) for  $A_j\left(r\right)$, $\rho(r), P(r)$ and $F\left(T\left(r\right)\right)$.  We are now able to solve these FEs for a given  $F\left(T\left(r\right)\right)$, using an eqn. of state between $\rho$ and $P$ and using the final coordinate condition.

We could also express the symmetric FEs as (see eqns. \eqref{1003a} -- \eqref{1003c}):
\begin{subequations}
\begin{eqnarray}
0 &=& -F''\left(T\right)\,\left(\partial_r\,T\right)\,k_1\left(r\right)+F'\left(T\right)\,g_1\left(r\right), \label{2007a}
\\
\kappa\,\left[\rho+P\right] &=& -2\,F''\left(T\right)\,\left(\partial_r\,T\right)\,k_2\left(r\right)+2\,F'\left(T\right)\,g_2\left(r\right) , \label{2007b}
\\
\left[\kappa\,\rho+\frac{F\left(T\right)}{2}\right] &=& -2\,F''\left(T\right)\,\left(\partial_r\,T\right)\,k_3\left(r\right)+2\,F'\left(T\right)\,g_3\left(r\right) , \label{2007c}
\end{eqnarray}
\end{subequations}
and the correspoding conservation eqn., where the functions $k_i$ and $g_i$ ($i=1\,-\,3$) are given by: 
\begin{subequations}
\begin{align}
g_1 =& \frac{1}{A_1\,A_2^3\,A_3^2}\Bigg[-A_2\,A_3^2\,\left(\partial_r^2\,A_1\right)-A_1\,A_2\,A_3\,\left(\partial_r^2\,A_3\right)+A_1\,A_2\,\left(\partial_r\,A_3\right)^2+\partial_r\,\left(A_1\,A_2\right)\,A_3\,\left(\partial_r\,A_3\right)
\nonumber\\
&+A_3^2\left(\partial_r\,A_1\right)\left(\partial_r\,A_2\right)-A_1\,A_2^3\Bigg] , \label{2007d}
\\
g_2 =& \frac{1}{A_1\,A_2^3\,A_3}\Bigg[-A_1\,A_2\,\left(\partial_r^2\,A_3\right)+\partial_r\,\left(A_1\,A_2\right)\,\left(\partial_r\,A_3\right)\Bigg] , \label{2007e}
\\
g_3 =& \frac{1}{A_1\,A_2^3\,A_3^2}\Bigg[-A_1\,A_2\,A_3\,\left(\partial_r^2\,A_3\right)-A_1\,A_2\,\left(\partial_r\,A_3\right)^2-A_2\,A_3\left(\partial_r\,A_3\right)\left(\partial_r\,A_1\right)
\nonumber\\
&-\delta\,A_1\,A_2^2\,\left(\partial_r\,A_3\right)+A_1\,A_3\,\left(\partial_r\,A_2\right)\left(\partial_r\,A_3\right)-\delta\,A_2^2\,A_3\left(\partial_r\,A_1\right)\Bigg] , \label{2007f}
\\
k_1 =& \frac{1}{A_1\,A_2^3\,A_3^2}\left[A_1\,A_2\,A_3\,\left(\partial_r\,A_3\right)+A_2\,A_3^2\left(\partial_r\,A_1\right)+\delta\,A_1\,A_2^2\,A_3\right] , \label{2007g}
\\
k_2=&k_3 = \frac{1}{A_2^2\,A_3}\left[\left(\partial_r\,A_3\right)+\delta\,A_2\right] . \label{2007h}
\end{align}
\end{subequations}
It is possible to rearrange the symmetric part of the FEs from eqn. \eqref{2007a} to get:
\begin{equation}\label{sim_form}
 F''(T(r))\left(\partial_r T(r)\right)\,k_1(r)=F'(T(r))\,g_1(r),
\end{equation}
where $k_1$ and $g_1$ are given by eqns. \eqref{2007d} and \eqref{2007g}.
We can write the eqn. \eqref{sim_form} as:
\begin{equation}\label{2009}
\partial_r\,\left[\ln\,F'\left(T(r)\right)\right]=\frac{g_1\left(r\right)}{k_1\left(r\right)} ,
\end{equation}
and obtain integral representations for $F'(T(r))$. 

We could look for solutions for a particular $F(T)$ or for a specific eqn. of state; for example, if we assume that $P=\alpha \rho$, after a redefinition of the spatial derivative an explicit integration is possible.

\subsection{Vacuum solutions}

Let us consider eqns. \eqref{2007a} -- \eqref{2007c} in the special case $P\left(r\right)=0$ and $\rho\left(r\right)=0$. The antisymmetric FEs are expressed by eqns. \eqref{2004} and we hence set $\sin \chi = 0$, $\cos \chi = \delta$ and $\psi=0$.  Eqns. \eqref{2007a} -- \eqref{2007c} can be written:
\begin{subequations}
\begin{eqnarray}
F''\left(T\right)\,\left(\partial_r\,T\right)\,k_1\left(r\right)&=& F'\left(T\right)\,g_1\left(r\right), \label{2010a}
\\
F''\left(T\right)\,\left(\partial_r\,T\right)\,k_2\left(r\right)&=& F'\left(T\right)\,g_2\left(r\right),  \label{2010b}
\\
\frac{F\left(T\right)}{4} &=& -F''\left(T\right)\,\left(\partial_r\,T\right)\,k_3\left(r\right)+F'\left(T\right)\,g_3\left(r\right) . \label{2010c}
\end{eqnarray}
\end{subequations}
where
\newpage

\small
\begin{subequations}
\begin{align}
& g_1 =\frac{1}{A_1\,A_2^3\,A_3^2}\Bigg[-A_2\,A_3^2\,\left(\partial_r^2\,A_1\right) -A_1\,A_2\,A_3\,\left(\partial_r^2\,A_3\right)+A_1\,A_2\,\left(\partial_r\,A_3\right)^2+\partial_r\,\left(A_1\,A_2\right)\,A_3\,\left(\partial_r\,A_3\right)
\nonumber\\
&+A_3^2\left(\partial_r\,A_1\right)\left(\partial_r\,A_2\right)-A_1\,A_2^3\Bigg], \label{2011a}
\\
& g_2 = \frac{1}{A_1\,A_2^3\,A_3}\Bigg[-A_1\,A_2\,\left(\partial_r^2\,A_3\right)+\partial_r\,\left(A_1\,A_2\right)\,\left(\partial_r\,A_3\right)\Bigg], \label{2011b}
\\
& g_3 = \frac{1}{A_1\,A_2^3\,A_3^2}\Bigg[-A_1\,A_2\,A_3\,\left(\partial_r^2\,A_3\right)-A_1\,A_2\,\left(\partial_r\,A_3\right)^2-A_2\,A_3\left(\partial_r\,A_3\right)\left(\partial_r\,A_1\right)-\delta\,A_1\,A_2^2\,\left(\partial_r\,A_3\right)  
\nonumber\\
& +A_1\,A_3\,\left(\partial_r\,A_2\right)\left(\partial_r\,A_3\right)-\delta\,A_2^2\,A_3\left(\partial_r\,A_1\right)\Bigg], \label{2011c}
\\
& k_1 = \frac{1}{A_1\,A_2^3\,A_3^2}\Bigg[A_1\,A_2\,A_3\,\left[\delta\,A_2+\left(\partial_r\,A_3\right)\right]+A_2\,A_3^2\left(\partial_r\,A_1\right)\Bigg], \label{2011d}
\\
& k_2 =k_3 = \frac{1}{A_2^2\,A_3}\left[\delta\,A_2+\left(\partial_r\,A_3\right)\right].  \label{2011f}
\end{align}
\end{subequations}
\normalsize

From the eqns. \eqref{2010a} and \eqref{2010b} we obtain the relation $g_1\,k_2=g_2\,k_1$:
\small
\begin{align}\label{2012}
& \Bigg[-A_2\,A_3^2\,\left(\partial_r^2\,A_1\right)-A_1\,A_2\,A_3\,\left(\partial_r^2\,A_3\right)+A_1\,A_2\,\left(\partial_r\,A_3\right)^2+\partial_r\,\left(A_1\,A_2\right)\,A_3\,\left(\partial_r\,A_3\right)
\nonumber\\
&+A_3^2\left(\partial_r\,A_1\right)\left(\partial_r\,A_2\right)-A_1\,A_2^3\Bigg]\,\left[A_1\,A_2\,\left(\partial_r\,A_3\right)+\delta\,A_1\,A_2^2\right]
\nonumber\\
&= \Bigg[-A_1\,A_2\,\left(\partial_r^2\,A_3\right)+\partial_r\,\left(A_1\,A_2\right)\,\left(\partial_r\,A_3\right)\Bigg]\,\left[A_1\,A_2\,A_3\,\left(\partial_r\,A_3\right)+A_2\,A_3^2\left(\partial_r\,A_1\right)+\delta\,A_1\,A_2^2\,A_3\right]
\end{align}
\normalsize
and from \eqref{2010c}, using \eqref{2010b}, the expression
\begin{align}\label{2013}
& \frac{{k_1\left(r\right)}}{4\left[{k_1\left(r\right)} g_3\left(r\right)- {g_1\left(r\right)\,k_3\left(r\right)}\right]} = \frac{d}{dT}\left(\ln F(T)\right) ,
\end{align}
which leads to an integral representations of the solutions, where $T\left(r\right)$ is given by eqn. \eqref{T_sim_form}, which further simplifies using the eqn. \eqref{2010c}.

It is of interest to consider the teleparallel version of {\em Minkowski} vacuum, in which additional conditions are imposed, and there are constraints on the form of the arbitary function $F(T)$ when $T$ vanishes. Perturbations of teleparallel {\em Minkowski} vacuum was studied in \cite{landryvandenhoogen1}.

\subsubsection{Specific coordinate choice}

Redefining the $r$  coordinate may simplify the governing eqns. further. We shall redefine $r$ to set $A_3\left(r\right)=r$ below. In this case,  eqns. \eqref{2007a} -- \eqref{2007c} have the components: 
\begin{subequations}
\begin{align}
g_1 =& \frac{1}{A_1\,A_2^3\,r^2}\Bigg[-A_2\,r^2\,\left(\partial_r^2\,A_1\right)+A_1\,A_2+\partial_r\,\left(A_1\,A_2\right)\,r+r^2\left(\partial_r\,A_1\right)\left(\partial_r\,A_2\right)-A_1\,A_2^3\Bigg] , \label{2014a}
\\
g_2 =& \frac{1}{A_1\,A_2^3\,r}\Bigg[\partial_r\,\left(A_1\,A_2\right)\Bigg] ,\label{2014b}
\\
g_3 =& \frac{1}{A_1\,A_2^3\,r^2}\Bigg[-A_1\,A_2-A_2\,r\left(\partial_r\,A_1\right)-\delta\,A_1\,A_2^2+A_1\,r\,\left(\partial_r\,A_2\right)-\delta\,A_2^2\,r\left(\partial_r\,A_1\right)\Bigg], \label{2014c}
\\
k_1 =& \frac{1}{A_1\,A_2^3\,r^2}\left[A_1\,A_2\,r+A_2\,r^2\left(\partial_r\,A_1\right)+\delta\,A_1\,A_2^2\,r\right], \label{2014d}
\\
k_2=& k_3 =  \frac{1}{A_2^2\,r}\left[1+\delta\,A_2\right]  .  \label{2014e}
\end{align}
\end{subequations}
The torsion scalar is expressed as:
\begin{equation}
T(r)=-2\left(\frac{\delta}{r}+\frac{1}{A_2\,r}\right)\left(\frac{\delta}{r}+\frac{1}{A_2\,r}+2\frac{\partial_r A_1}{A_2\,A_1}\right). \label{2015}
\end{equation}  
This can only be done in the general case $A_3 \neq $ constant. There are restrictions on constraining $A_1$, since the conservation eqn. relates its radial derivative with $\partial_r P = 0$. An alternative option might be to set $A_2 = 1/A_1$.

\subsubsection{Schwarzschild-like solutions}

Let us consider the coframe eqn. \eqref{2002} in which  $A_2\left(r\right)=\frac{1}{A_1\left(r\right)}$ and $A_3\left(r\right)=r$, corresponding to {\em{Schwarzschild}}-like solutions. Eqns. \eqref{2004} yield $\sin\,\chi=0$ and $\sinh\psi=0$: $\cos\,\chi=\cos\left(n\,\pi\right)=\pm 1 = \delta$ and $\cosh\psi=1$. The three non-trivial symmetric FEs are expressed as in eqns. \eqref{2007a} -- \eqref{2007c} with the components:
 
\begin{subequations}
\begin{align}
g_1 =& \frac{A_1^2}{r^2}\Bigg[-A_1^{-1}\,r^2\,\left(\partial_r^2\,A_1\right)+1-r^2\,A_1^{-2}\left(\partial_r\,A_1\right)^2-A_1^{-2}\Bigg] , \label{2222a}
\\
g_2 =& 0 , \label{2222b}
\\
g_3 =& \frac{A_1^2}{r^2}\Bigg[-1-2\,A_1^{-1}\,r\left(\partial_r\,A_1\right)-\delta\,A_1^{-1}-\delta\,A_1^{-2}\,r\left(\partial_r\,A_1\right)\Bigg] , \label{2222c}
\\
k_1 =& \frac{A_1^2}{r^2}\left[r+A_1^{-1}\,r^2\left(\partial_r\,A_1\right)+\delta\,A_1^{-1}\,r\right] , \label{2222d}
\\
k_2=k_3 =& \frac{A_1^2}{r}\left[1+\delta\,A_1^{-1}\right] . \label{2222e}
\end{align}
\end{subequations}
Finally, from eqn. \eqref{T_sim_form} we obtain the torsion scalar:
\begin{eqnarray}\label{2223}
T(r)=-4\delta\left[\frac{\frac{dA_1}{dr}}{r}+\frac{A_1}{r^2}\right]-\frac{2}{r^2}-\frac{4\,A_1\,\frac{dA_1}{dr}}{r}-\frac{2\,A_1^2}{r^2} .
\end{eqnarray}
The solutions of the eqns. \eqref{2007a} -- \eqref{2007c} for eqns. \eqref{2222a} -- \eqref{2222e} include the Schwarzschild spacetime ($A_1(r)=\sqrt{1-\frac{r_S}{r}}$) as well as the Reissner-Nordstrom spacetime ($A_1(r)=\sqrt{1-\frac{r_S}{r} +\frac{r_Q^2}{r^2}}$), as examples. We note that for $\rho=P=0$ we have that $F''=0$ or $T=$ constant, which leads to the GR case with a cosmological constant.

Note that the Birkhoff theorem does not generally hold in $F(T)$ theories. In addition, a perturbed solution around Minkowski spacetime and solar system  constraints on the parameters in $F(T)$ gravity were first discussed in \cite{Ruggiero,Ruggiero2}.

\subsection{Analysis and Power-law Solutions}

We summarize the governing eqns. in the vacuum case with $A_3=r$ (we also could consider the quadratic function $F$). The FEs \eqref{2010a} -- \eqref{2010c} with components \eqref{2014a} -- \eqref{2014e} are given by:
\begin{subequations}
\begin{eqnarray}
0 &=& g_1k_2-g_2k_1 , \label{2100a} 
\\
g_2 &=& k_2 \partial_r\,\left[\ln\,F'\left(T(r)\right)\right],  \label{2100b} 
\\
\frac{F(T)}{4} &=&F'(T)\left(g_3-g_2\right).  \label{2100c} 
\end{eqnarray}
\end{subequations}
or in terms of the original variables we have  explicitly
\small
\begin{subequations}
\begin{align}
0 =& \left(\delta\,A_2+1\right)\left(\frac{\partial_{r}^2 A_1}{A_1}\right) +\left(\frac{\partial_r A_1}{A_1}\right)^2-\delta\left( \frac{\partial_r A_1}{A_1}\right)\left(\partial_r A_2\right)+\frac{1}{r^2}\left(\delta\,A_2+1\right)\left(A_2^2-1\right), \label{2100d}
\\
0 =& \left(\delta\,A_2+1\right)\partial_r\,\left[\ln\,F'\left(T(r)\right)\right]-\partial_r\left[\ln\,(A_1 A_2)\right], \label{2100e}
\\
\frac{F(T)}{4}  =& \frac{F'(T)}{r^2 A_2^2}\left[-\left(\delta\,A_2+2\right)\left(\frac{r\,\partial_r A_1}{A_1}\right)-\left(\delta\,A_2+1\right) \right]. \label{2100f}
\end{align}
\end{subequations}
\normalsize
It is noteworthy that the torsion scalar, given explicitly earlier, can be written as
\begin{equation}
T(r)=2A_2^2\,(k_2^2-2k_1 k_2).
\end{equation}

We already know that solutions with $A_1\,A_2=$ constant lead to GR-like solutions. We can seek power-law solutions in general. The only case in which we cannot set $A_3=r$ is when $A_3=$ constant, which we must treat separately. Finally, we consider the case $A_2=$ constant, in which a number of new solutions can be obtained.

\subsubsection{Power-law solutions}

In general, for $A_1(r)=a_0\,r^a$ and $A_2(r)=b_0\,r^b$, after setting $A_3=r$ as a coordinate choice, we get from eqns. \eqref{2015} and \eqref{2100d} -- \eqref{2100f}:
\begin{subequations}
\begin{align}
0 =& \frac{1}{b_0^2\,r^{2b}}\left[\delta\,\left(a^2-a(1+b)-1\right)+\frac{\left(2a^2-a-1\right)}{b_0\,r^{b}}\right]+\left[\frac{1}{b_0\,r^{b}}+\delta\right] , \label{2114a}
\\
F'(T(r)) =& F_1\,\exp\left[(a+b)\int\,\frac{dr}{r\,\left(\delta\,b_0\,r^b+1\right)} \right] ,  \label{2114b}
\\
F'(T(r)) =&-\frac{b_0\,r^{b+2}\,F(T(r))}{4\left[\left(\delta+\frac{2}{b_0\,r^{b}}\right)\,a+\left(\delta+\frac{1}{b_0\,r^{b}}\right)\right]} , \label{2114c}
\\
T(r)=&-\frac{2}{r^2}\left[\left(\delta+\frac{1}{b_0\,r^{b}}\right)^2+\frac{2\,a}{b_0\,r^{b}}\,\left(\delta+\frac{1}{b_0\,r^{b}}\right)\right] . \label{2114d}
\end{align} 
\end{subequations}

\noindent The only possible solutions for eqn. \eqref{2114a} is obtained by setting $b=0$, which means we reduce to a special case of the general case in eqn. \eqref{2122a} with $y_1=0$ below.

\subsubsection{Solution for $A_3(r)=$ constant}

With $A_3=c_0$ and the coordinate choice $A_2=b_0=1$, using eqns. \eqref{T_sim_form} and \eqref{2007d}--\eqref{2007h} we obtain:
\begin{subequations}
\begin{align}
A_1''+\frac{A_1}{c_0^2}=& 0 , \label{2104a}
\\
\partial_r\left[\ln F'(T)\right] =& 0 , \label{2104b}
\\
\partial_r\left[\ln F(T)\right]=& -\frac{\delta\,c_0\,A_1\,T'(r)}{4\,A_1'}  , \label{2104c}
\\
T(r)=&-\frac{2}{c_0^2}-\frac{4\delta\,A_1'}{c_0\,A_1} . \label{2103}
\end{align}
\end{subequations}
From eqn. \eqref{2104b}, we find that $F'(T)=F_{1}=$ constant, leading to $F(T)=F_{1} \,T+ F_{0}$. From eqn. \eqref{2103}, we find that $\left( T(r)+\frac{2}{c_0^2}\right)=-\frac{4A_1'}{\delta\,c_0\,A_1}$ and then eqn. \eqref{2104c} becomes:
\begin{equation}
F(T)=F_1\left[T+\frac{2}{c_0^2}\right] .
\end{equation}
Since eqn. \eqref{2104a} leads to an oscillating $A_1(r)$ of frequency $\omega_0=\frac{1}{c_0}$, it is not physically relevant.

\subsection{Solutions for $A_3(r)=r$ and $A_2=$ constant}

We consider solutions for $F(T(r))$ by using eqns. \eqref{2014a} - \eqref{2014e} as $g_i$ and $k_i$ components in eqns. \eqref{2100a}--\eqref{2100c}, leading to the FEs described by eqns. \eqref{2007d}--\eqref{2007h}. By setting $A_2(r)=b_0=$ constant in eqns. \eqref{2007d}--\eqref{2007h}, this assumption leads to the simplified FEs: 
\begin{subequations}
\begin{align}
0=& \left[\frac{A_1''}{A_1}+\frac{1}{r^2}\,\left(b_0^2-1\right)\right]\left(\delta\,b_0+1\right)+\left(\frac{A_1'}{A_1}\right)^2=0 , \label{2120a}
\\
F'(T(r)) =& F_1\,A_1^{\frac{1}{\left(\delta\,b_0+1\right)}} , \label{2120b}
\\
F'(T(r)) =&-\frac{b_0^2\,r^2\,F(T(r))}{4\left[\left(\delta\,b_0+2\right)\,\left(\frac{r\,A_1'}{A_1}\right)+\left(\delta\,b_0+1\right)\right]}, \label{2120c}
\end{align}
\end{subequations}
where $F_1$ is an integration constant and $\delta b_0 \neq -1$. By setting $y(r)=\ln\,\left(A_1(r)\right)$, eqn. \eqref{2120a} becomes a non-linear ODE:
\begin{align}\label{2121}
0 =& \left(\delta\,b_0+1\right)\,y''+\left(\delta\,b_0+2\right)\,y'^2+\,\frac{\left(\delta\,b_0-1\right)\left(\delta\,b_0+1\right)^2}{r^2} .
\end{align}
We will solve eqn. \eqref{2121} in the general case and then we will consider the special values of $\delta\,b_0=+1$ and $\delta\,b_0=-2$. Note, if $\delta\,b_0=-1$, then the torsion scalar is identically zero (see eqn. \eqref{T_sim_form}) and is not of interest.

\subsubsection{General solution}\label{sub1}

The general solution of eqn. \eqref{2121} is ($\delta\,b_0\neq -2$ and $\delta\,b_0\neq \pm 1$):
\begin{align}\label{2122a}
y(r)=& y_2+\frac{(\delta\,b_0+1)}{2(\delta\,b_0+2)}\left(1-S\right)\,\ln\,r+\frac{(\delta\,b_0+1)}{(\delta\,b_0+2)}\ln\left[r^{S}+y_1\right],
\end{align}
where $y_1$ and $y_2$ are arbitrary constants and
\begin{align}\label{2122s}
S=\pm \sqrt{1-4\left(\delta\,b_0-1\right)\left(\delta\,b_0+2\right)}.
\end{align}
If $y_1=0$, then all solutions to eqn. \eqref{2122a} yield power-law solutions for $A_1(r)$. By taking the exponential of eqn. \eqref{2122a}, we obtain:
\begin{align}
A_1(r)= a_0\,r^{\frac{\left(\delta\,b_0+1\right)}{2\,\left(\delta\,b_0+2\right)}\left(1+S\right)}\,\left(1+y_1\,r^{-S}\right)^{\frac{\left(\delta\,b_0+1\right)}{\left(\delta\,b_0+2\right)}} ,\label{2122}
\end{align}
where $a_{0}$ is a constant. Eqns. \eqref{2120b} and \eqref{2120c} become:
\begin{subequations}
\begin{align}
F'(T(r)) =& F_2\,r^{\frac{\left(1+S\right)}{2\,\left(\delta\,b_0+2\right)}}\,\left(1+y_1\,r^{-S}\right)^{\frac{1}{\left(\delta\,b_0+2\right)}} , \label{2123b}
\\
F'(T(r)) =&-\frac{b_0^2\,r^2\,F(T)}{2\left(\delta\,b_0+1\right)\left( 3-S\right)\left[1+\frac{2S}{\left( 3-S\right)\left[1+y_1\,r^{-S}\right]}\right]} , \label{2123c}
\end{align}
\end{subequations}
where $S\neq 3$ and $F_2=F_1\,a_0^{\frac{1}{\left(\delta\,b_0+1\right)}}$. The eqn. \eqref{T_sim_form} for the torsion scalar becomes:
\begin{align}\label{2123d}
T(r)=& -\frac{2\left(\delta b_0+1\right)^2\,\left(3+\delta\,b_0-S\right)}{b_0^2\,(\delta\,b_0+2)\,r^2}\left[1+\frac{2S}{\left(3+\delta\,b_0-S\right)\left[1+y_1\,r^{-S}\right]}\right] ,
\nonumber\\
=& \frac{T_0}{r^2}\,\left[1+\frac{T_1}{\left[1+y_1\,r^{-S}\right]}\right] ,
\end{align} 
where $T_0$ and $T_1$ are constants depending on $b_0$ and $S$. By combining eqns. \eqref{2123b} and \eqref{2123c}, we obtain $F(T)$:
\begin{align}
F(T(r))=-\frac{2\left(\delta\,b_0+1\right)\left( 3-S\right)}{b_0^2}\,F_2\,\left[1 +F_3\,\left(1+y_1\,r^{-S}\right)^{-1}\right]\,\left(1+y_1\,r^{-S}\right)^{\frac{1}{\left(\delta\,b_0+2\right)}}\,r^{\frac{\left(1+S\right)}{2\,\left(\delta\,b_0+2\right)}-2},  \label{2124}
\end{align}
where $F_3=\frac{2S}{\left( 3-S\right)}$ (for $S \neq 3$). Eqn. \eqref{2124} is the general expression for $F(T(r))$ and is a new result. We consider the following cases:
\begin{enumerate}
\item ${\bf y_1=0}$ (Power-law solutions for $F(T)$): From eqns. \eqref{2123d} and \eqref{2124} for all $S$ values, we find:
\begin{subequations}
\begin{align}
T(r)=& \frac{T_0\left(1+T_1\right)}{r^2} , \label{2125a}
\\
F(T)=&\,{-\frac{2\left(\delta\,b_0+1\right)\left(3-S\right)}{b_0^2}\,F_2\,\left(1+F_3\right)\,\left(\frac{T}{T_0\left(1+T_1\right)}\right)^{1-\frac{\left(1+S\right)}{4\,\left(\delta\,b_0+2\right)}}.} \label{2125b}
\end{align}
\end{subequations}
We obtain from eqns. \eqref{2125a} and \eqref{2125b} the power-law solution of Golovnev-Guzman \cite{golov1}. 

\item ${\bf y_1 \neq 0}$: Eqn. \eqref{2123d} becomes:
\begin{align}\label{2126}
\left(\frac{T}{T_0}\right)\left(r^2+y_1\,r^{2-S}\right)-y_1\,r^{-S}-\left(1+T_1\right)=0.
\end{align} 
To find $F(T)$ explicitly, we need to find $r$ as a function of $T$ in eqn. \eqref{2126}. There are some solvable cases for specific values of $S$:
\begin{enumerate}
\item ${\bf S=0}$: We obtain that $\delta\,b_0=\frac{\pm \sqrt{10}-1}{2}$ from eqn. \eqref{2122s}, $T_0 =-\frac{2}{27}\left(25\pm 34\sqrt{10}\right)$, $T_1=0$, $F_2=F_1\,a_0^{\frac{2}{9}\left(-1\pm\sqrt{10}\right)}$ and $F_3=0$.  Eqn. \eqref{2126} leads to:
\begin{align}\label{2126a}
r(T)=\left(\frac{T}{T_0}\right)^{-\frac{1}{2}}.
\end{align}
Then eqn. \eqref{2124} will be:
\begin{align}\label{2127a}
F(T)= -\frac{4}{27}\,\left(31\pm13\sqrt{10}\right) F_2\,\left(1+y_1\right)^{\frac{2}{\left(3\pm\sqrt{10}\right)}}\,\left(\frac{T}{T_0}\right)^{1-\frac{1}{2\left(3\pm \sqrt{10}\right)}},
\end{align}
which is again a power-law solution for $F(T)$ \cite{golov1}.

\item ${\bf S=\pm 1}$: We obtain $\delta\,b_0=1$ or $-2$ from eqn. \eqref{2122s}, both special cases. We consider these specific cases below.

\item ${\bf S=2}$: We obtain $\delta\,b_0=\frac{\pm \sqrt{6}-1}{2}$ from eqn. \eqref{2122s}. We find that $T_0=-\frac{2}{75} \left(117\pm 62\sqrt{6}\right)$, $T_1=-\frac{8}{5}\left(1\mp\sqrt{6}\right)$, $F_2=F_1\,a_0^{\frac{2}{5}\left(-1\pm\sqrt{6}\right)}$ and $F_3=4$. Eqn. \eqref{2126} leads to:
\begin{align}\label{2126b}
r^2(T)=\frac{1}{2}\left(\frac{T_0 (1+T_1)}{T}-y_1\right)\left[1\pm\sqrt{1+\frac{4\,y_1\,\frac{T_0}{T}}{\left(\frac{T_0 (1+T_1)}{T}-y_1\right)^2}} \right] .
\end{align} 
Eqn. \eqref{2124} becomes:
\begin{align}\label{2127b}
F(T)=& -\frac{4}{25}\,\left(19\pm 9\sqrt{6}\right)\,F_2\,\frac{\left(5\,r^2(T)+y_1\right)}{\left(r^2(T)+y_1\right)^{1-\frac{2}{3\pm\sqrt{6}}}\,\left(r(T)\right)^{2+\frac{1}{3\pm\sqrt{6}}}},
\end{align}
where $r(T)$ is the eqn. \eqref{2126b}. This is a new result.

\item ${\bf S=-2}$: For $\delta\,b_0=\frac{\pm \sqrt{6}-1}{2}$ and by using eqn. \eqref{2122s}, we get that $T_0=-2\left(7\pm 2\sqrt{6}\right)$, $T_1=-\frac{8}{75}\left(9\mp\sqrt{6}\right)$, $F_2=F_1\,a_0^{\frac{2}{5}\left(-1\pm\sqrt{6}\right)}$ and $F_3=-\frac{4}{5}$.  Eqn. \eqref{2126} leads to:
\begin{align}\label{2126c}
r^2(T) =\frac{1}{2}\left[\left(\frac{T}{T_0}\right)^{-1}-\frac{1}{y_1}\right]\left[1\pm\sqrt{1+\frac{4(1+T_1)\left(\frac{T}{T_0}\right)}{y_1\left[1-\frac{1}{y_1}\left(\frac{T}{T_0}\right)\right]^2}} \right] .
\end{align} 
Eqn. \eqref{2124} becomes:
\begin{align}\label{2127c}
F(T)=& -\frac{4}{5}\,\left(19\pm 9\sqrt{6}\right)\,F_2\,\frac{\left(1+5y_1\,r^2(T)\right)}{\left(1+y_1\,r^2(T)\right)^{1-\frac{2}{3\pm\sqrt{6}}}\,\left(r(T)\right)^{2+\frac{1}{3\pm\sqrt{6}}}} ,
\end{align}
where $r(T)$ is the eqn. \eqref{2126c}. This is also a new result.

\item \textbf{Limits on ${\bf S}$}: By using eqn. \eqref{2122s}, we find that real values for $\delta b_0$ are only possible when ${-\sqrt{10}\, \leq\, S\, \leq\, +\sqrt{10}}$. By the extremum test applied to eqn. \eqref{2122s}, we find that these occur when $\delta b_0=-\frac{1}{2}$. However we are unable to find $r(T)$ explicitly from eqn. \eqref{2123d} in this case.

\end{enumerate}

For the $S=\pm 2$ cases, we obtain new results. For other values of $S$, we obtain a more complex form for the relationship between $T$ and $r$ in eqn. \eqref{2126} which might not yield an explicit expression for $r=r(T)$.

\end{enumerate}

\subsubsection{Special solution: $b_0=-2\delta$}\label{sub2}

\noindent By setting $b_0=-2\delta$, eqn. \eqref{2121} becomes:
\begin{align}\label{2128}
0 =& y''+\frac{3}{r^2} .
\end{align}
The solution is $y=\, 3\,\ln\,r+y_1\,r+y_2$, which yields
\begin{align}\label{2129b}
A_1(r)= a_0\,r^3\,e^{y_1\,r}.
\end{align}
Eqns. \eqref{2120b} and \eqref{2120c} then yield:
\begin{align}
F(T(r)) = F_2\,r^{-5}\,e^{-y_1\,r} , \label{2129c}
\end{align}
where $F_2=\frac{F_1}{a_0}$. Eqn. \eqref{T_sim_form} then gives $T(r)= \,\frac{5}{2}\,r^{-2}+y_1\,r^{-1}$, which yields
\begin{align}\label{2129e}
r^{-1}(T)=\,\frac{1}{5}\left[-y_1 \pm \sqrt{y_1^2+10\,T}\right],
\end{align}
and then eqn. \eqref{2129c} yields the new solution:
\begin{align}
F(T) = \frac{F_2}{3125}\,\left[-y_1 \pm \sqrt{y_1^2+10\,T}\right]^{5}\,\exp\left[\frac{5y_1}{\left[y_1 \mp \sqrt{y_1^2+10\,T}\right]}\right] . \label{2129f}
\end{align}

\noindent For $y_1=0$, eqns. \eqref{2129b} and \eqref{2129c} become, respectively, $A_1(r)= a_0\,r^3$ and $F(T(r)) = \frac{F_{1}}{a_0}\,r^{-5}$. Then eqn. \eqref{2129e} becomes $r^{-1}(T)=\sqrt{\frac{2}{5}}\,T^{\frac{1}{2}}$ and we find a new power-law solution:
\begin{align}\label{2129g}
F(T)=& \left(\frac{2}{5}\right)^{\frac{5}{2}}\,F_2\,T^{\frac{5}{2}}.
\end{align}
\noindent Note that the case $\delta\,b_0=-2$ in Golovnev-Guzman \cite{golov1} was treated via a complex ansatz in $A_1(r)$ and using $A_2(r)=\frac{\delta\left(1-2c_1\,r^2\right)}{c_1\,r^2+1}\,\neq$  constant. Taking the asymptotic limit $r\,\rightarrow\,\infty$ of $A_2(r)$, we obtain that $A_2(r)\,\rightarrow\,-2\delta$. However, eqn. \eqref{2129g} is not in \cite{golov1} and hence this as a new result.

\subsubsection{Special solution: $b_0=+\delta$}\label{sub3}

\noindent Eqn. \eqref{2121} for $\delta\,b_0=1$ reduces to:
\begin{align}\label{2151}
0 =& y''+\frac{3}{2}\,y'^2 .
\end{align}
By integrating, we obtain $y(r)=\frac{2}{3}\,\ln\left(r+y_1\right)+y_2$, which yields
\begin{align}\label{2152}
A_1(r)=a_0\,\left(r+y_1\right)^{\frac{2}{3}}.
\end{align}
Eqn. \eqref{T_sim_form} then yields:
\begin{align}
T(r) =& -\frac{8}{r^2}\left(1+\frac{2r}{3\left(r+y_1\right)}\right) . \label{2153d}
\end{align}
From eqn. \eqref{2153d}, it is possible to find an explicit expression for $r(T)$:
\begin{align}\label{2156}
r(T) =& -\frac{y_1}{3}+\frac{1}{3T}\Bigg[-\left(y_1\,T\right)^3+2\sqrt{2}\,T\,\sqrt{27\,y_1^4\,T^3-312\,y_1^2\,T^2+8000\,T}-48\,y_1\,T^2\Bigg]^{\frac{1}{3}}
\nonumber\\
&\quad-\frac{\left(40-y_1^2\,T\right)}{3}\Bigg[-\left(y_1\,T\right)^3+2\sqrt{2}\,T\,\sqrt{27\,y_1^4\,T^3-312\,y_1^2\,T^2+8000\,T}-48\,y_1\,T^2\Bigg]^{-\frac{1}{3}}.
\end{align} 
\normalsize
By combining eqns. \eqref{2120b} and \eqref{2120c}, we obtain the new result:
\begin{align}
F(T)=-8\,F_2\,\frac{\left(2r(T)+y_1 \right)}{\left(r(T)+y_1\right)^{\frac{2}{3}}\,r^{2}(T)} ,  \label{2157}
\end{align}
where $F_2=F_1\,\sqrt{a_0}$.

For $y_1=0$, eqn. \eqref{2152} becomes $A_1(r)=a_0\,r^{\frac{2}{3}}$, then eqn. \eqref{2156} leads to $r(T)=\sqrt{\frac{40}{3}}\,\left(-T\right)^{-\frac{1}{2}}$. Finally, eqn. \eqref{2157} becomes
\begin{align}\label{2157a}
F(T)=&-16\,\left(\frac{3}{40}\right)^{\frac{5}{6}}\,F_2\,\left(-T\right)^{\frac{5}{6}}, 
\end{align}
where $T \leq 0$. Eqn. \eqref{2157} is again the power-law solution \cite{golov1}.

In this subsection, we have presented a number of new solutions. We have also recovered the known Golovnev-Guzman power-law solution in a number of special subcases \cite{golov1}.

\section{Kantowski-Sachs Geometry}

In order to expand the affine frame symmetry group of the spherically symmetric case to that of a Kantowski-Sachs geometry, we need to consider the following four affine frame symmetry generators: 

\beq \begin{aligned} & {\bf X}_3 = \partial_{\phi},~{\bf X}_2 = - \cos \phi \partial_{\theta} + \frac{\sin \phi}{\tan \theta} \partial_{\phi}, {\bf X}_1 = \sin \phi \partial_{\theta} + \frac{\cos \phi}{\tan \theta} \partial_{\phi}, \
{\bf X}_4 = \partial_r. \end{aligned} \eeq

The resulting Lie algebra is then

\beq [{\bf X}_1, {\bf X}_2] = -{\bf X}_3, [{\bf X}_1, {\bf X}_3] = {\bf X}_2, [{\bf X}_2, {\bf X}_3] = -{\bf X}_1, [{\bf X}_{I'}, {\bf X_4}] = 0, \eeq

\noindent where in the last bracket, $I' = 1,2,3$. Using the approach in \cite{MCH}, we find that 

\beq \begin{aligned} \mathcal{L}_{{\bf X}_{I'}} \bh^a &= f_{I'}^{~\hat{i}} \lambda^a_{\hat{i}~b} \bh^b, \\
\mathcal{L}_{{\bf X}_4} \bh^a &= 0, \\
\mathcal{L}_{{\bf X}_I} \omega^a_{~bc} &= 0, I=1,\ldots 4.\end{aligned} \label{KSeqns} \eeq

\noindent where
\beq f_I^{\tilde{i}} = \left[ \begin{array}{cccc} \frac{\cos(\phi)}{\sin(\theta)} & 0 & 0 & 0 \\
\frac{\sin(\phi)}{\sin(\theta)} & 0 & 0 & 0 \\
0 & 0 & 0 & 0  \end{array}\right] \eeq

Solving these eqns. leads again to the spherically symmetric ansatz for a spherically symmetric Riemann-Cartan geometry, where now there is only $t$-dependence in the functions of the vierbein and spin-connection described by eqns. \eqref{VB:SS} and \eqref{Con:SS} with $A_i=A_i(t)$ ($i=1,\,...,\,3$) and $W_j=W_j(t)$ ($j=1,\,...\,8$). Requiring that the curvature vanishes then implies that the arbitrary functions described by eqns. \eqref{SS:TPcon} must take the form: 

\beq \begin{aligned} 
W_1 &= -\frac{\partial_t \chi}{A_1},\quad  W_2 = 0,\quad W_3 = \frac{\cosh(\psi)\cos(\chi)}{A_3}, \quad W_4 = \frac{\cosh(\psi)\sin(\chi)}{A_3},\\
W_5 &= -\frac{\partial_t \psi}{A_1}, \quad W_6 = 0, \quad W_7 = \frac{\sinh(\psi) \cos(\chi)}{A_3}, W_8 = \frac{\sinh(\psi) \sin(\chi)}{A_3},  
\end{aligned} \label{SS:TPcon3} \eeq

\noindent where $\chi$ and $\psi$ are arbitrary functions of the coordinate $t$.

Without loss of generality, the $t$-coordinate can be rescaled to set $A_1 = 1$, giving the vierbein:
\begin{eqnarray}
h^a_{~\mu} = \left[ \begin{array}{cccc} 1 & 0 & 0 & 0 \\ 0 & A_2(t) & 0 & 0 \\ 0 & 0 & A_3(t) & 0 \\ 
0 & 0 & 0 & A_3(t) \sin(\theta) \end{array}\right] . \label{VB:SS3} 
\end{eqnarray}
Denoting derivatives with respect to $t$ with a prime, $f_{,t} = f'$, the vector-part and axial-part of the torsion are

\beq \begin{aligned} 
{\bf V} &= \frac{2A_2 \sinh(\psi) \cos(\chi)+A_{2}'A_3 + 2A_{3}'A_2)}{A_2 A_3} \bh^1 - \frac{2\cosh(\psi) \cos(\chi) - \psi' A_3}{A_3} \bh^2, \\
{\bf A} &= \frac{4 \cosh(\psi)\sin(\chi)}{3A_3} \bh^1 - \frac{2(\chi' A_3 - 2\sinh(\psi) \sin(\chi))}{3A_3} \bh^2, \end{aligned} \eeq

\noindent while the tensor-part of the torsion, $$t_{(ab)c} =  \hat{T}_{abc} + \hat{T}_{bac},$$ where the $\hat{T}_{a[bc]}$ have components:

\beq \begin{aligned}
 \hat{T}_{112} &= \hat{T}_{332} = \hat{T}_{442} = \frac{-2(\cosh(\psi)\cos(\chi)+\psi' A_3)}{3 A_3}, \\
\hat{T}_{134} &= \hat{T}_{314} = -\hat{T}_{413} = \frac{-2(\sinh(\psi)\sin(\chi)+\chi' A_3)}{3 A_3}, \\ 
\hat{T}_{212} &= \hat{T}_{331} = \hat{T}_{441} = -\frac{2(\sinh(\psi)\cos(\chi)A_2 + A_{3}' A_2 - A_{2}' A_3)}{3A_2 A_3}, \\
\hat{T}_{234} &= \hat{T}_{324} = \hat{T}_{432} = -\frac{\sinh(\psi)\sin(\chi)}{3A_3}. \end{aligned} \eeq

We remark that the Kantowski-Sach solutions admitting a $G_6$ in GR, with functions $A_2 = e^{2\sqrt{\Lambda} t}$ and $A_3 = \frac{1}{\Lambda}$ for $\Lambda \in \mathbb{R}$, can at most admit a $G_5$, which arises when $\chi = \pi/2$ and $\psi = 0$. This is due to the fact that the boost and spin isotropy in this solution necessarily requires that all irreducible parts of the torsion tensor vanish. 

The antisymmetric part of the $F(T)$ FEs give
\begin{subequations}
\begin{align}
 \frac{F_{TT}(T)\, T' \cosh(\psi) \cos(\chi)}{A_3} &= 0, \\
 \frac{F_{TT}(T)\,T' \sinh(\psi) \sin(\chi)}{A_3} &=0. 
\end{align}
\end{subequations}
These eqns. imply that $\psi = 0$ and $\chi = \pi/2$ (or $T$ is constant).

\subsection{Constant Torsion Scalar}

\beq \begin{aligned} T &= -\frac{4(2 (\sinh(\psi) \cos(\chi)A_3 A_2)' -2 A_{2}' A_{3}'A_3 - {A_{3}'}^2 A_2 + A_2 )}{A_3^2 A_2}. \end{aligned} \eeq

\noindent The algebraically independent components of the symmetric part of the FEs are then 
\begin{subequations}
\begin{align}
\kappa \rho + \frac{F(T)}{2} &= \left( \frac{4 (\sinh(\psi) \cos(\chi)A_3 A_2)' + 4 {A_{3}'}^2 A_2 + 8 A_{3}' A_{2}' A_3}{2 A_3^2 A_2}  \right) F_T(T),\\
-\kappa(\rho + P) & = -\left( \frac{2(A_{3}'' A_2 - A_{3}' A_{2}')}{A_3 A_2} \right) F_T(T), \\
0 &= \left( \frac{{A_{3}'}^2 A_2 - A_{3}' A_{2}' A_3 + A_2 A_3 A_{3}'' - A_{2}'' A_3^2 + A_2}{A_3^2 A_2} \right) F_T(T).
\end{align}
\end{subequations}
As the constant $T$ case reduces to TEGR, $F(T)=T$, with some scaling of physical quantities, we will ignore this case. However, we will state the following result for Kantowski-Sachs geometries admitting a timelike affine frame symmetry.

\begin{prop}
Any teleparallel Kantowski-Sachs geometry, which admits the affine frame symmetry, ${\bf X}_5 = -\frac{1}{C_0} \partial_t + r \partial_r$ with the additional Lie bracket relations
\beq [{\bf X}_I, {\bf X}_5] = 0,~~ [{\bf X}_4, {\bf X}_5] = {\bf X}_4, \nonumber \eeq  

\noindent is given by the following:
\beq \psi = 0, \chi = \frac{\pi}{2}, A_3 = \frac{1}{C_0},~ A_2 = e^{C_0 t}, \nonumber \eeq
\noindent and is a solution of the constant torsion scalar eqn. with $\rho = p = 0$. 
\end{prop}

\subsection{Non-constant Torsion scalar}

The torsion scalar is
\beq T = 2\left(\frac{A_{3}'}{A_3} \right)^2 + \frac{4 A_{3}' A_{2}'}{A_3 A_2} - \frac{2}{A_3^2}. \eeq 

\noindent Writing $F_T(T)$ as $F'(T)$ in this subsection, the algebraically independent components of the symmetric part of the FEs are then:
\begin{subequations}\label{SymFEs}
\begin{align}
 \kappa \rho + \frac{F(T)}{2} &=  \left( T+\frac{2}{A_3^2} \right) F'(T), \\ 
-\kappa ( \rho + P) &= \left( \frac{ A_{3}'}{A_3}\right) T'\, F''(T) + \left( \frac{A_{3}''}{A_3} - \frac{A_{3}' A_{2}'}{A_3 A_2} \right) F'(T), \\
\left( \left(\ln(A_2)\right)' - \left(\ln(A_3)\right)'  \right) T'\,F''(T) &= - \left( \frac{A_{3}' A_{2}'}{A_2 A_3} - \left(\frac{A_{3}'}{A_3} \right)^2 - \frac{A_{3}''}{A_3} + \frac{A_{2}''}{A_2} - \frac{1}{A_3^2} \right) F'(T). 
\end{align}
\end{subequations}
\noindent Comparing with previous discussions of the Kantowski-Sachs $F(T)$ FEs in the literature \cite{Rodrigues2015,Amir2015}, we point out that the governing eqns. differ significantly. In particular, the anti-symmetric part of the FEs do not require $F(T) =T$ and the symmetric eqns. have subtle but important differences, such as the inclusion of $2 A_3^{-2}$ in the coefficient of $F'(T)$ in the first eqn.

To avoid $F'(T) = 0$ or $F''(T)=0$, the coefficients in the last eqn. in \eqref{SymFEs} cannot vanish (otherwise it would imply either $A_2$ or $A_3$ vanishes, which is not permitted). We may partially integrate $F'(T)$ from this eqn.:

\beq F'(T) = C_0  \exp  \left[ \int \frac{\left( -\frac{A_{3}' A_{2}'}{A_2 A_3} + \left(\frac{A_{3}'}{A_3}\right)^2 + \frac{A_{3}''}{A_3} - \frac{A_{2}''}{A_2} + \frac{1}{A_3^2} \right)}{\left(\ln(A_2)\right)' - \left(\ln(A_3)\right)'} dt \right] = C_0 \exp( B(t)). \eeq

\noindent Substituting this into \eqref{SymFEs}, the FEs become:
\begin{subequations}
\begin{align}
\kappa \rho + \frac{F(T)}{2} &= \left( T + \frac{2}{A_3^2} \right) F'(T), 
\\
-\kappa(\rho + P) &= \left( \frac{A_3'}{A_3} B' + \frac{A_3''}{A_3} - \frac{A_3' A_2'}{A_3 A_2} \right) F'(T).
\end{align}
\end{subequations}

\subsubsection{The vacuum $F(T)$ equations}

In the case of vacuum, where $\rho = P = 0$, we find the following conditions:
\begin{subequations}
\begin{align}
\frac{F(T)}{2} &= \left(T + \frac{2}{A_3^2}\right) F'(T), \\
0 &= \left( \frac{A_3'}{A_3} B' + \left( \frac{A_3'}{A_3} \right)' + \left(\frac{A_3'}{A_3} \right)^2 - \frac{A_3' A_2'}{A_3 A_2} \right) F'(T) .
\end{align}
\end{subequations}
To avoid $F'(T) = 0$, the second eqn. must vanish. The general solution to this eqn. is possible. However special solutions will be sufficient for our purposes. One solution to this eqn. follows from requiring $A_2=A_3^n$, where $n \in \mathbb{R}$. If $A_3$ is not constant, then the resulting functions and eqns. are: 
\begin{subequations}
\begin{align}
T =& 2(2n+1)\,\left[\left(\ln\,(A_3)\right)'\right]^2-2\,A_3^{-2},
\\
B'=& \frac{\left(1-n^2\right)\left[\left(\ln\,(A_3)\right)'\right]^2-(n-1)\,\frac{A_3''}{A_3}+A_3^{-2}}{(n-1)\,\left(\ln\,(A_3)\right)'},
\\
0 =& \left(\ln\,(A_3)\right)'\,B'+\frac{A_3''}{A_3}-n\,\left[\left(\ln\,(A_3)\right)'\right]^2,
\\
\frac{F(T)}{2} =& \left(T+\frac{2}{A_3^2}\right)\,F'(T).
\end{align}
\end{subequations}
Eliminating $B'$, we find the following eqn. for $A_3$:
\begin{align}
0=&(2n+1)(n-1)\left(\frac{A_3'}{A_3}\right)^2-\frac{1}{A_3^2},
\end{align}
with the corresponding solution:
\begin{align}
A_3 = \pm \frac{t}{\sqrt{(2n+1)(n-1)}},\quad\quad n>1.
\end{align}
Finally, writing $T$ in terms of these functions, we obtain:
\begin{align}
T = \frac{2(2-n)(2n+1)}{t^2}.
\end{align}
In the case where the constant of integration is zero, we find the following eqn. for $F(T)$: 
\begin{align}
\frac{F(T)}{2}=& \left(\frac{T}{2-n}\right)F'(T),
\end{align}
which integrates to 
\begin{align}
F(T) =F_1\,T^{1-\frac{n}{2}}.
\end{align}

\section{Special Similarity case}

We wish to study the special case of spherical symmetry with the 3 affine spherical symmetry vectors and the additional fourth affine vector  $X$ (or $X_4$) (\ref{extra}) which commutes with $\{X_x, X_y, X_z\}$.

To compute the conditions on the frame and the connection, we suppose that the affine vectors, $\{X_I\}_{I=1}^4 = \{ X_x, X_y, X_z, X_4\}$, act on the frame in the following way:

\beq \mathcal{L}_{{\bf X_I}} \bh^a = f_I^{~\hat{i}} \lambda_{\hat{i}~b}^a \bh^b , \nonumber \eeq

\noindent where $\lambda_{\hat{i}}$ is defined as before in eqn. \eqref{SSrep} and $f_{I}^{~\hat{i}}$ are functions of $re^{-C_0 t}, \theta$ and $\phi$ (where $C_0 \equiv C_3 C_4$; and $C_0\equiv -H_0$ to compare with the TdS solution). 

In order to determine $f_I^{\hat{i}}$, we must expand the Lie algebra constants $C^I_{~JK}$ and then solve the resulting DEs from \eqref{Sym:RC:Prop}. Using the transformation properties of $\overline{Iso}$ a frame can be chosen where 

\beq f_I^{\hat{i}} = \left[ \begin{array}{ccc} \frac{\cos \phi}{\sin \theta} & 0 & 0 \\ \frac{\sin \phi }{\sin \theta} & 0 & 0 \\
0 & 0 & 0 \\
0 & 0 & 0 \\ \end{array} \right]. \eeq	

\noindent Thus, to determine the conditions of the inclusion of the new symmetry generator, we need only examine the conditions 

\beq \mathcal{L}_{{\bf X}_4} \bh^a  = 0 . \eeq
\normalsize
Solving the resulting eqns., we find that 

\beq A_1 (t,r) = f_1(z),\quad A_2(t,r) =  \frac{f_2(z)}{r} , \quad A_3(t,r) = f_3(z), \eeq

\noindent where $z \equiv r e^{-C_0 t}$. Due to the simple form of $f_I^{~\hat{i}}$, the Lie derivative of the connection gives a simple condition, namely that 

\beq \mathcal{L}_{{\bf X_4}} \omega^a_{~bc} = {\bf X_4 } ( \omega^a_{~bc})  = 0 . \eeq

\noindent Therefore, the arbitrary functions in the spin-connection's components must be functions of the form:

\begin{equation}
\chi(t,r) = \chi(z), \quad \psi(t,r) = \psi(z).    
\end{equation}

\noindent We now have a resulting diagonal metric $g$ by substituting the coframe components, where $\mathcal{L}_{{\bf X}}\,g = 0$ as required.

\subsection{Similarity}

We can modify the frame ansatz on the vierbein functions: 

\beq A_1 = r^\lambda f_1(z), \quad A_2 = r^{\lambda-1} f_2(z), \quad A_3 = r^\lambda f_3(z), \eeq

\noindent where if $\lambda \neq 0$, 
${\bf X} = \frac{1}{C_0} \partial_t + r \partial_r$
corresponds to  a homothetic vector with $\mathcal{L}_{{\bf X}}\,g = 2\lambda\,g$, while if $\lambda = 0$ this is an affine frame symmetry and hence a Killing vector.

$\mathcal{L}_{{\bf X}}\,\omega$ can be calculated by considering the components for $\omega_{ab}$ defined as follows:
\begin{align}\label{3001}
\omega_{12}=& -W_5\,dt-W_6\,dr ,& \omega_{13} =& -W_7 d\theta -W_8\,\sin\theta\,d\phi, & \omega_{14} =& W_8\,d\theta -W_7\,\sin\theta\,d\phi,
\nonumber\\
\omega_{23}=& W_3\,d\theta + W_4\,\sin\theta d\phi, & \omega_{24} =& -W_4\,d\theta +W_3\,\sin\theta\,d\phi, & \omega_{34} =& W_1\,dt+W_2\,dr-\cos\theta\,d\phi,
\nonumber\\
\end{align}
from which we find that $\mathcal{L}_{{\bf X}}\,\omega=0$ for all values of $\lambda$ \cite{MCH}. For $\lambda=0$, we still get $\mathcal{L}_{{\bf X}}\,g=0$ and $\mathcal{L}_{{\bf X}}\,\omega^a_{~bc}=0$ as expected.

Note that
\begin{eqnarray}\label{3002}
r\,\partial_r\,T &=& r^{-2\lambda}\,\left(-2\,\lambda\,T_0(z)+z\,\partial_z\,T_0(z)\right) ,
\nonumber\\
z\,\partial_z\,T &=& r^{-2\lambda}\,z\,\partial_z\,T_0(z) .
\end{eqnarray}
For the general situation ($\lambda\neq 0$), we have that the torsion scalar $T=T(z,r)$ will also contain terms in $r^{\lambda}$; $T(z,r)=\frac{T_0(z)}{r^{2\lambda}}$.

The general torsion scalar (consistent with the form of these expressions) is given explicitly by:
\small
\begin{align}\label{3003}
& T(z, r)\equiv\frac{T_0(z)}{r^{2\lambda}} = \frac{1}{r^{2\lambda}f_1^2\,f_2^2\,f_3^2}\Bigg[-8\,f_1\,f_2\,\cos\chi\,\Bigg(\left[f_1\,\left(\lambda\,f_3+\frac{z\,f_3'}{2}\right)+\frac{z\,f_3\left(\psi'\,C_0\,f_2+f_1'\right)}{2}\right]\cosh\psi
\nonumber\\
&+\frac{z\,\sinh\psi}{2}\left[f_1\,f_3\,\psi'+C_0\,\left(f_2\,f_3\right)'\right]\Bigg)+4\,z\,\sin\chi\,\cosh\psi\,\chi'\,f_1^2\,f_2\,f_3+4\,z\,\sinh\psi\,\sin\chi\,\chi'\,C_0\,f_1\,f_2^2\,f_3
\nonumber\\
& +f_1^2\,\left[-6\lambda^2\,f_3^2-8\lambda\,z\,f_3\,f_3'-2\,z^2\,f_3'^2-2\,f_2^2\right]-4\,f_1\,f_3\left[\lambda\,C_0\,z\,f_2\,f_3\,\psi'+f_1'\left(\lambda\,z\,f_3+z^2\,f_3'\right)\right]
\nonumber\\
&+2\,C_0^2\,z^2\,f_2\,f_3'\left(f_2\,f_3'+2\,f_3\,f_2'\right)\Bigg] .
\end{align}
\normalsize

\subsection{The antisymmetric FEs}

Therefore, the non-trivial antisymmetric FEs will be as follows ($\lambda \neq 0$):
\small
\begin{subequations}
\begin{align}
& 0 = z\,\left(\ln\,T_0(z)\right)'\left[\cos\chi\,\left(C_0\,f_2\,\cosh\psi+f_1\,\sinh\psi\right)+\lambda\,C_0\,f_3\right]-2\,\lambda\,\left[\cos\chi\,\left(f_1\,\sinh\psi\right)-C_0\,z\,f_3'\right] , \label{3004a}
\\
& 0 =\sin\chi\,\Bigg[z\,\left(\ln\,T_0(z)\right)'\left[C_0\,f_2\,\sinh\psi+f_1\,\cosh\psi\right]-2\,\lambda\,\left[f_1\,\cosh\psi\right]\Bigg] ,  \label{3004b}
\end{align}
\end{subequations}
\normalsize
where $\chi=\chi(z)$, $\psi=\psi(z)$ and the $f_i=f_i(z)$ are defined earlier.

There are two subsolutions of eqns. \eqref{3004a} and \eqref{3004b} for $T_0(z)\neq 0$ as follows:
\begin{enumerate}
\item $\sin\,\chi(z)=0$: $\cos\,\chi(z)=\delta=\pm 1$, $\chi = k\pi$ and the eq. (\ref{3004a}) is:
\begin{align}\label{3005}
\left(\ln\,T_0(z)\right)'=2\,\lambda\,\frac{\left[\delta\,\left(f_1\,\sinh\psi\right)-C_0\,z\,f_3'\right]}{z\,\left[\delta\,\left(C_0\,f_2\,\cosh\psi+f_1\,\sinh\psi\right)+\lambda\,C_0\,f_3\right]} .
\end{align}
The general solution of eqn. \eqref{3005} can be obtained by integration:
\begin{align}\label{3006}
T_0(z)=T_0(0)\,\exp\left[2\,\lambda\,\int_0^{z}\,dz\,\frac{\left[\delta\,\left(f_1\,\sinh\psi\right)-C_0\,z\,f_3'\right]}{z\,\left[\delta\,\left(C_0\,f_2\,\cosh\psi+f_1\,\sinh\psi\right)+\lambda\,C_0\,f_3\right]}\right]  .
\end{align}
The four symmetric FEs will be used to determine the functions $\psi(z)$ and the $f_i(z)$ where $i=1,\,2,\,3$.

\item If $\sin\,\chi(z)\neq 0$, we get that: 
\small
\begin{subequations}
\begin{align}
& z\,\left(\ln\,T_0(z)\right)'\,\left[\cos\chi\,\left(C_0\,f_2\,\cosh\psi+f_1\,\sinh\psi\right)+\lambda\,C_0\,f_3\right]=2\,\lambda\,\left[\cos\chi\,\left(f_1\,\sinh\psi\right)-C_0\,z\,f_3'\right] , \label{3007a}
\\
& z\,\left(\ln\,T_0(z)\right)'\,\left[C_0\,f_2\,\sinh\psi+f_1\,\cosh\psi\right]=2\,\lambda\,\left[f_1\,\cosh\psi\right] .\label{3007b}
\end{align}
\end{subequations}
\normalsize
We substitute eqn. \eqref{3007a} into eqn. \eqref{3007b} and we obtain the relation:
\begin{align}\label{3008}
&\cos\chi=-\left[\frac{\sinh\psi}{f_1}\left(C_0\,z\,f_3'\right)+\frac{\cosh\psi}{f_2}\,\left(z\,f_3'+\lambda\,f_3\,\right)\right] .
\end{align}
Here again $z=z(t,r)$ is still a function of $t$ and $r$. It remains to substitute eqn. \eqref{3008} into the symmetric FEs as well as into eqn. \eqref{3006} to obtain the specific form of $T_0(z)$ applicable for this 2nd subsolution. 
\end{enumerate}

\subsection{The symmetric FEs}

The general symmetric FEs are:
\begin{subequations}
\begin{eqnarray}
0 &=& 2\,r^{-4\lambda}\,F''(T)\,\left[h_1\,T_0'(z)+m_1\,T_0(z)\right] + g_1\,r^{-2\lambda}\,F'(T), \label{3009a}
\\
\kappa\,P -\frac{F(T)}{2} &=& 2\,r^{-4\lambda}\,F''(T)\,h_2\,T_0'(z)+2\,g_2\,r^{-2\lambda}\,F'(T), \label{3009b}
\\
\kappa\,\rho + \frac{F(T)}{2} &=& 4\,r^{-4\lambda}\,F''(T)\,\left[h_3\,T_0'(z)+m_3\,T_0(z)\right]+2\,g_3\,r^{-2\lambda}\,F'(T), \label{3009c}
\\
0 &=& r^{-4\lambda}\,F''(T)\,\left[h_4\,T_0'(z)+m_4\,T_0(z)\right] + g_4\,r^{-2\lambda}\,F'(T), \label{3009d}
\end{eqnarray}
\end{subequations}
where the explicit forms of the functions $h_1,\,m_1,$ etc...  in the symmetric FE above in the case of the two subsolutions are given explicitly in Appendix \ref{appenb}.

By taking the eqns. \eqref{3009a} and \eqref{3009d}, we obtain the following expression:
\begin{align}\label{3010}
-\frac{F''(T)\,r^{-2\lambda}}{F'(T)}=\frac{g_1}{2\left[h_1\,T_0'(z)+m_1\,T_0(z)\right]}=\frac{g_4}{\left[h_4\,T_0'(z)+m_4\,T_0(z)\right]},
\end{align}
and hence formally:
\begin{align}\label{3011}
T_0(z)=T_0(0)\,\exp\left[\int_{0}^{z}\,dz\,\frac{2\,g_4\,m_1-g_1\,m_4}{g_1\,h_4-2\,g_4\,h_1}\right],
\end{align}
which yields the function $T_0(z)$ as a function of $\psi(z)$, $\chi(z)$ and $f_i(z)$. From the FEs expressed by eqns. \eqref{3009a} to \eqref{3009d}, for $\lambda \neq 0$, there are restrictions on the form of the function $F(T)$. Eqns. \eqref{3009b} and \eqref{3009c} become :
\begin{subequations}
\begin{eqnarray}
\kappa\,P &=& \left[\frac{2\,\alpha(z)}{\beta(z)}\,h_2\,T_0'(z)+\left(2\,g_2+\frac{\alpha(z)}{2}\right)\right]\,r^{-2\lambda}\,F'(T), \label{3014a}
\\
\kappa\,\rho &=& \left[\frac{4\,\alpha(z)}{\beta(z)}\,\left(h_3\,T_0'(z)+m_3\,T_0(z)\right)+\left(2\,g_3-\frac{\alpha(z)}{2}\right)\right]\,r^{-2\lambda}\,F'(T),  \label{3014b}
\end{eqnarray}
\end{subequations}
where $\frac{\alpha(z)}{\beta(z)}=\frac{T\,F''(T)}{T_0(z)\,F'(T)}$ and the functions $\alpha(z)$ and $\beta(z)$ are defined by eqns. \eqref{3009b} and \eqref{3009c}.

Hence the eqns. \eqref{3009b} and \eqref{3009c} are replaced by eqns. \eqref{3014a} and \eqref{3014b} giving $P$ and $\rho$ with $F(T)$. A possible solution of the FEs is that $F(T)$ is the sum of terms of the form:
\begin{equation}\label{3015}
F(T)=\sum\, F_{\gamma}(T)= \sum_{\gamma \geq 0}\, F_{0\;\gamma}\,T^{\gamma},
\end{equation}
with possible contraints on $\alpha(z)$ and $\beta(z)$ leading to additional ODEs. $P$ and $\rho$ are determined by eqns. \eqref{3014a} and \eqref{3014b}.

By adding the eqns. \eqref{3014a} and \eqref{3014b} together, we obtain the following eqn. in term of $r^{-2\lambda}\,F'(T)$:
\begin{eqnarray}
\kappa\,\left(P + \rho \right) &=& 2\,\left[-g_4\,\left[\frac{\left(h_2 +2\,h_3\right)\,T_0'(z)+ \left(2\,m_3\right)\,T_0(z)}{h_4\,T_0'(z)+m_4\,T_0(z)}\right] +\left(g_2+g_3\right)\right]\,r^{-2\lambda}\,F'(T),
\nonumber\\
 &\equiv& 2\,H(z)\,r^{-2\lambda}\,F'(T).   \label{3016}
\end{eqnarray}
We will study the different $H(z)$ giving the eqn. \eqref{3016} for the general symmetric FEs as well as for the two subsolutions.

For the situation without utilizing the antisymmetric FE solutions, we obtain:
\small
\begin{align}\label{3017}
H(z) =& -2\,C_0\,z^2\,\Bigg[\left(\ln\,f_3\right)'\,\left[\ln\left(\frac{f_1\,f_2}{z\,f_3'}\right)\right]'+\left[\ln\left(z^{\lambda}\right)\right]'\,\left[\ln\, f_2\right]'\Bigg]\,\Bigg[2\lambda\,T_0(z) \left[\frac{z\,f_3'+\lambda\,f_3}{f_2^2}+\frac{1}{f_2}\,\cosh\psi\,\cos\chi\right]
\nonumber\\
&+\left[\left(\frac{C_0\,\sinh\psi}{f_1}-\frac{\cosh\psi}{f_2}\right)\,\cos\chi-\left(\frac{C_0^2}{f_1^2}+\frac{1}{f_2^2}\right)\,z\,f_3'-\frac{\lambda\,f_3}{f_2^2}\right]\,z\,T_0'(z)\Bigg]
\nonumber\\
&\times\,\Bigg[-z\,T_0'(z)\,\left[\left(C_0\,f_2\,\cosh\psi-f_1\,\sinh\psi\right)\,\cos\chi+\lambda\,C_0\,f_3+2\,C_0\,z\,f_3'\right]+\lambda\,T_0(z)\,
\nonumber\\
&\times \left[2\,C_0\,z\,f_3'-f_1\,\sinh\psi\,\cos\chi\right]\Bigg]^{-1}+\Bigg[z^2\,\left(\ln\,f_3\right)'\,\left(\frac{C_0^2}{f_1^2}+\frac{1}{f_2^2}\right)\left[\ln \left(\frac{f_1\,f_2}{z\,f_3'}\right)\right]'+\frac{\lambda\,z}{f_2^2}\,\left[\ln \left(f_1\,f_2\,z^{\lambda}\right)\right]'\Bigg]
\end{align}
\normalsize

When the two subsolutions of the antisymmetric FE are utilized these expressions become much simpler. In particular, eqn. \eqref{3017} gives $H(z) \neq 0$ in general (for $\lambda \neq 0$). We note that when $\lambda = 0$ (special isometry case), we find that $H(z) = 0$, so that $\rho + P = 0$, so that from conservation laws $\rho = \Lambda_0 =$ constant and $P = - \Lambda_0$ \cite{TdS}.

For the $H(z) \neq 0$ case, one must take into account the conservation laws which can be summarized by the following relation:
\begin{eqnarray}\label{3018}
\left(\rho+P\right) &=& -\frac{f_1}{\left(\lambda\,f_1+z\,f_1'\right)}\,\left(r\,\partial_r\,P(t,r)\right) = \frac{f_2\,f_3}{C_0\,z\,\left(2\,f_2\,f_3'+f_2'\,f_3\right)}\,\left(\partial_t\,\rho(t,r)\right) .
\end{eqnarray}
We obtain from eqn. \eqref{3018} a relation of the form:
\begin{eqnarray}\label{3019}
\left(r\,\partial_r\,P(t,r)\right) \equiv -\Sigma(z)\,\left(\partial_t\,\rho(t,r)\right).
\end{eqnarray}
We must solve eqn. \eqref{3019} for $\Sigma(z)$ to determine the kind of fluid corresponding to the $H(z)$ obtained. To make further progress, we could assume an equation of state of the form $P=P(\rho)$ (or more generally for $\rho\equiv \sum_{n= 0}\,\rho_i(z)\,r^{-2n\,\lambda}$ and $P\equiv \sum_{n= 0}\,P_i(z)\,r^{-2n\,\lambda}$, eqns. of state for each component $\rho_i$, $P_i$) leading to further ODEs. Alternatively, we could investigate a particular $F(T)$.

\subsection{A particular quadratic $F(T)$ theory}

\noindent As an example, let us work with the following quadratic function of $T$ \cite{saridakis}:
\begin{equation}\label{3020}
F(T) = -\Lambda+T+\gamma\,T^2,
\end{equation}
where $\gamma$ is a constant (we could also consider a cubic function or an exponential function $F(T) = \frac{1}{2\gamma}\,\left[\exp\left(2\gamma\,T\right)-1\right]$). The symmetric FEs for the 1st subsolution is
\small
\begin{subequations}
\begin{align}
0 =& 4\,r^{-4\lambda}\,\gamma\,\left[h_1\,T_0'(z)+m_1\,T_0(z)\right] + g_1\,r^{-2\lambda}\,\Bigg(1+2\gamma\,T_0(z)\,r^{-2\lambda}\Bigg), \label{3020a}
\\
\kappa\,P =& \frac{1}{2}\Bigg(-\Lambda+T_0(z)\,r^{-2\lambda}+\gamma\,T_0^2(z)\,r^{-4\lambda}\Bigg)+4\,r^{-4\lambda}\,\gamma\,h_2\,T_0'(z)+2\,g_2\,r^{-2\lambda}\,\Bigg(1+2\gamma\,T_0(z)\,r^{-2\lambda}\Bigg), \label{3020b}
\\
\kappa\,\rho =& -\frac{1}{2}\Bigg(-\Lambda+T_0(z)\,r^{-2\lambda}+\gamma\,T_0^2(z)\,r^{-4\lambda}\Bigg) +8\,r^{-4\lambda}\,\gamma\,\left[h_3\,T_0'(z)+m_3\,T_0(z)\right]
\nonumber\\
&+2\,g_3\,r^{-2\lambda}\,\Bigg(1+2\gamma\,T_0(z)\,r^{-2\lambda}\Bigg), \label{3020c}
\\
0 =& 2\,r^{-4\lambda}\,\gamma\,\left[h_4\,T_0'(z)+m_4\,T_0(z)\right] + g_4\,r^{-2\lambda}\,\Bigg(1+2\gamma\,T_0(z)\,r^{-2\lambda}\Bigg) . \label{3020d}
\end{align}
\end{subequations}
\normalsize

The eqns. \eqref{3020a} and \eqref{3020d} give the following relation:
\begin{align}\label{3020ad}
\frac{r^{2\lambda}}{2\gamma}\Bigg(1+2\gamma\,T_0(z)\,r^{-2\lambda}\Bigg)=\frac{2\left[h_1\,T_0'(z)+m_1\,T_0(z)\right]}{g_1}=\frac{\left[h_4\,T_0'(z)+m_4\,T_0(z)\right]}{g_4} .
\end{align}
With the eq \eqref{3020ad}, the eqns. \eqref{3020b} and \eqref{3020c} become explicitly:
\small
\begin{subequations}
\begin{align}
\kappa\,P =& \frac{1}{2}\Bigg(-\Lambda+T_0(z)\,r^{-2\lambda}\Bigg)+4\,\gamma\,r^{-4\lambda}\,\Bigg[\left(h_2+\frac{g_2}{g_4}\,h_4\right)\,T_0'(z)+\left(\frac{g_2}{g_4}\,m_4\right)\,T_0(z)+\frac{T_0^2(z)}{8}\Bigg], \label{3020adb}
\\
\kappa\,\rho =& -\frac{1}{2}\Bigg(-\Lambda+T_0(z)\,r^{-2\lambda}\Bigg)+4\,\gamma\,r^{-4\lambda}\,\Bigg[\left(2 h_3+\frac{g_3}{g_4}\,h_4\right)\,T_0'(z)+\left(2 m_3+\frac{g_3}{g_4}\,m_4\right)\,T_0(z)-\frac{T_0^2(z)}{8}\Bigg]. \label{3020adc}
\end{align}
\end{subequations}
\normalsize
By adding the eq \eqref{3020adb} and \eqref{3020adc} into eq \eqref{3017}, we obtain the following eqn.:
\small
\begin{align}\label{3021}
\kappa\,\left(P + \rho \right) =& 4\,\gamma\,r^{-4\lambda}\,\Bigg[\left(\frac{\left(g_2+g_3\right)}{g_4}\,h_4+h_2+2 h_3\right)\,T_0'(z)+\left(\frac{\left(g_2+g_3\right)}{g_4}\,m_4+2 m_3\right)\,T_0(z)\Bigg],
\nonumber\\
=& 4\,\gamma\,r^{-4\lambda}\,\tilde{H}(z),
\end{align}
\normalsize
where $\tilde{H}(z)$ is the original $H(z)$ of eqn. \eqref{3016} multiplied by some terms inside $F'(T)$ (derivative of eqn. \eqref{3020} terms). The eqn. \eqref{3021} decribes the eqns. \eqref{3014a} and \eqref{3014b} solution for the quadratic theory described by eqn. \eqref{3020}.

\subsection{Simple Power-law solutions}

By using the power-law ansatz $f_1=a_0\,z^a$, $f_2=b_0\,z^b$, $f_3=c_0\,z^c$ and $\psi=e_0\,z^e$ with $P=P(z,r)$ and $\rho=\rho(z,r)$, the conservation laws become:
\begin{subequations}
\begin{eqnarray}
-\left(\lambda+a\right)\left(\rho(z,r)+P(z,r)\right) &=& \left(z\,\partial_z\,P(z,r)+r\,\partial_r\,P(z,r)\right)  , \label{3019a}
\\
\left(2c+b\right)\left(z\,\partial_z\,P(z,r)+r\,\partial_r\,P(z,r)\right) &=& \left(\lambda+a\right)\,z\,\partial_z\,\rho(z,r), \label{3019b}
\end{eqnarray}
\end{subequations}
where $\partial_r P(t,r)=\frac{z}{r}\,\partial_z\,P(z,r)+\partial_r\,P(z,r)$ and $\partial_t \rho(t,r)=-C_0\,z\,\partial_z\,\rho(z,r)$. The following cases arise:
\begin{enumerate}
\item $b=-2c$ : Leading to solutions with a static fluid $P=P(r)$ and $\rho=\rho(r)$.

\item $a= -\lambda$: We obtain that $P=P(t)$ and there is no constraint on $\rho$. 
\end{enumerate}

Additional solutions are possible with further assumptions on the form of $F(T)$ or the perfect fluid. For example, assuming a perfect fluid $P(z,r)=\alpha\,\rho(z,r)$ with a linear eqn. of state, we obtain:
\begin{subequations}
\begin{eqnarray}
-\left(\lambda+a\right)\left(1+\alpha\right)\rho(z,r) &=& \alpha\left(z\,\partial_z\,\rho(z,r)+r\,\partial_r\,\rho(z,r)\right)  , \label{3019c}
\\
\alpha \left(2c+b\right)\left(z\,\partial_z\,\rho(z,r)+r\,\partial_r\,\rho(z,r)\right) &=& \left(\lambda+a\right)\,z\,\partial_z\,\rho(z,r), \label{3019d}
\end{eqnarray}
\end{subequations}
where $-1 < \alpha \leq 1$. We find the solution is:
\begin{eqnarray}
\rho(z,r)&=& \rho(0,r)\,z^{-\left(2c+b\right)\,\left(1+\alpha\right)}, \label{3018n}
\end{eqnarray}
where $\rho(0,r)=g(r)$ and we find that:
\begin{eqnarray}
\rho(0,r)=g(r)=g(0)\,r^{\frac{\left(1+\alpha\right)}{\alpha}\left[\alpha\left(2c+b\right)-\left(\lambda+a\right)\right]}  . \label{3018q}
\end{eqnarray}
Further conditions can then be obtained from the FEs after solving the conservation laws. Typical solutions lead to power-law expressions for $F(T)$ of the form $F(T) =-\Lambda +\gamma\,T^{\beta}$.

\vspace*{1.0cm}

We note that all teleparallel solutions in this section are new.

\vspace*{1.0cm}

\section{Discussion and Conclusions}

In teleparallel $F(T)$  gravity, the geometrical quantities of interest (is the torsion computed from) the frame (or co-frame $\bh^a$) and a zero curvature, metric compatible, spin connection $\bomega^a_{~b}$. When $F(T)$ teleparallel gravity is defined in a gauge invariant manner, the ensuing field equations are fully Lorentz covariant.  Here we are interested in
teleparallel spherically symmetric solutions to the $F(T)$ teleparallel gravity FEs which are of particular physical importance.

To construct such gravitational models in teleparallel gravity, we first needed to determine the general forms for the the frame (or co-frame) and the spin connection that respect the assumed spherically symmetric (affine) symmetries. Since there is a non-trivial linear isotropy group, the determination of the group of affine symmetries necessarily requires the solution to a system of inhomogeneous DEs \eqref{Affine}. Following the approach outlined in \cite{MCH}, which relies on the existence and determination of a class of invariantly defined frames, it then follows from eqns. \eqref{Liederivative:frame} and \eqref{Liederivative:Con} that the Lie derivative of the torsion tensor and its covariant derivatives with respect to the affine frame symmetries ${\bf X_i}$ are identically zero.

We have presented the (diagonal) frame (or co-frame $\bh^a$) \eqref{VB:SS} and the metric compatible spin connection ($\bomega^a_{~b})$ \eqref{Con:SS} pair that describes the most general spherically symmetric teleparallel geometry, and that also satisfies  the flatness condition  \eqref{SS:TPcon}. This is a general geometric result and applies to all theories formulated with respect to a teleparallel geometry, and not just $F(T)$ teleparallel gravity. As an application, we then employ this general frame/spin connection pair to study the $F(T)$ spherically symmetric teleparallel gravity field equations.  In the general case we presented the solutions of the antisymmetric FEs (which split into two cases, which we consequently considered separately), and derived and subsequently analysed the resulting symmetric FEs.

In order to further study the applications of spherically symmetric teleparallel models in $F(T)$ teleparallel gravity, we studied $3$ subcases in which there is an additional affine symmetry (i.e., models with a 4-dimenional Lie group of affine symmetries) so that the FEs reduce to a system of ordinary differential equations (ODEs) suitable for further investigation.

First we studied static spherically symmetric spacetimes with an additional affine static spherically symmetric generator. We solved the antisymmetric FEs \eqref{2004} and derived the full set of  symmetric FEs \eqref{1003a} -- \eqref{1003c} and analysed their properties. In general, we assumed that the source is a perfect fluid and utilized a particular coordinate choice ($A_3=r$), and we also specifically investigated the vacuum FEs and obtained a number of new results. In particular, we studied power law solutions in detail and derived a number of new solutions (which, in some cases, reduce to the known power-law solution of Golovnev-Guzman \cite{golov1}).

We then considered an additional affine frame symmetry in order to expand the affine frame symmetry group of the spherically symmetric case to that of a spatially homogeneous Kantowski-Sachs geometry. A number of new results were obtained. Finally, we studied the special case of spherical symmetry with the 3 affine spherical symmetry vectors and an additional fourth affine symmetry vector (\ref{extra}) which commutes with them, and its similarity generalization. Again we studied the antisymmetric and symmetric FEs in this new class of teleparallel similarity spacetimes. We focused on simple power law models and obtained some new solutions.

\section*{Acknowledgments}
AAC and RvdH are supported by the Natural Sciences and Engineering Research Council of Canada. AL is supported by an AARMS fellowship. RvdH is supported by the Dr. W.F. James Chair of Studies in the Pure and Applied Sciences at St.F.X. DDM is supported by the Norwegian Financial Mechanism 2014-2021 (project registration number 2019/34/H/ST1/00636).




\begin{thebibliography}{999}



\bibitem{Bahamonde:2021gfp} Bahamonde, S., Dialektopoulos, K., Escamilla-Rivera, C., Farrugia, G., Gakis, V., Hendry, M., Hohmann, M., Said, J., Mifsud, J. \& Di Valentino, E., {Teleparallel Gravity: From Theory to Cosmology}, {\em Reports On Progress In Physics} \textbf{86}, 026901 (2023) [ArXiv:2106.13793 [gr-qc]].

\bibitem{cai2016f} Cai, Y., Capozziello, S., De Laurentis, M. \& Saridakis, E., {f(T) teleparallel gravity and cosmology.} {\em Reports On Progress In Physics} \textbf{79}, 106901 (2016) [arXiv:1511.07586 [gr-qc]]

\bibitem{Aldrovandi_Pereira2013} Aldrovandi, R. \& Pereira, J., {Teleparallel Gravity},  Fundamental Theories of Physics, Vol. 173 (Springer, Dordrecht, 2013)

\bibitem{Ferraro:2006jd} Ferraro, R. \& Fiorini, F., {Modified teleparallel gravity: Inflation without inflaton.} {\em Physical Review D} \textbf{75}, 084031 (2007) [arXiv:gr-qc/0610067]

\bibitem{Ferraro:2008ey} Ferraro, R. \& Fiorini, F., {On Born-Infeld Gravity in Weitzenbock spacetime.} {\em Physical Review D} \textbf{78}, 124019 (2008) [arXiv:0812.1981 [gr-qc]]

\bibitem{Linder:2010py} Linder, E., {Einstein's Other Gravity and the Acceleration of the Universe.} {\em Physical Review D} \textbf{81}, 127301 (2010), [Erratum: Physical Review D \textbf{82}, 109902 (2010)], [arXiv:1005.3039 [astro-ph]]

\bibitem{Krssak_Pereira2015} Krššák, M. \& Pereira, J., {Spin Connection and Renormalization of Teleparallel Action.} {\em The European Physical Journal C} \textbf{75}, 519 (2015) [arXiv:1504.07683 [gr-qc]]

\bibitem{Krssak:2018ywd} Krššák, M., van den Hoogen, R., Pereira, J., Boehmer, C. \& Coley, A., {Teleparallel Theories of Gravity: Illuminating a Fully Invariant Approach.} {\em Classical And Quantum Gravity} \textbf{36}, 183001 (2019) [arXiv:1810.12932 [gr-qc]]

\bibitem{Lucas_Obukhov_Pereira2009} Lucas, T., Obukhov, Y. \& Pereira, J., {Regularizing role of teleparallelism.} {\em Physical Review D} \textbf{80}, 064043 (2009) [arXiv:0909.2418 [gr-qc]]

\bibitem{hohmann2019modified} Hohmann, M., Järv, L., Krššák, M. \& Pfeifer, C., {Modified teleparallel theories of gravity in symmetric spacetimes.} {\em Physical Review D} \textbf{100}, 084002 (2019) [arXiv:1901.05472 [gr-qc]]

\bibitem{Coley:2019zld} Coley, A., van den Hoogen, R. \& McNutt, D.D., {Symmetry and Equivalence in Teleparallel Gravity.} {\em Journal of Mathematical Physics} \textbf{61}, 072503 (2020) [arXiv:1911.03893 [gr-qc]]

\bibitem{Hohmann:2021ast} Hohmann, M., {General covariant symmetric teleparallel cosmology.} {\em Physical Review D} \textbf{104}, 124077 (2021) [arXiv:2109.01525 [gr-qc]]

\bibitem{pfeifer2022quick} Pfeifer, C., {A quick guide to spacetime symmetry and symmetric solutions in teleparallel gravity.}, e-Boletim da Fisica, \textbf{10} 2 (2021) [arXiv:2201.04691 [gr-qc]]

\bibitem{BohmerJensko} Bohmer, C.G. \& Jensko, E., Modified gravity: a unified approach to metric-affine models, {\em Journal of Mathematical Physics} \textbf{64}, 082505 (2023) [arXiv:2301.11051 [gr-qc]].

\bibitem{bohmer1} Bohmer, C.G., Jensko, E. \& Lazkoz, R., Cosmological dynamical systems in modified gravity, {\em European Physical Journal C} \textbf{82} 6, 500 (2022), [arXiv:2201.09588 [gr-qc]].

\bibitem{MCH} McNutt, D. D., Coley, A.A. \& van den Hoogen, R.J., {A frame based approach to computing symmetries with non-trivial isotropy groups.} {\em Journal of Mathematical Physics} \textbf{64}, 032503 (2023) [arXiv:2302.11493 [gr-qc]]

\bibitem{Ruggiero} Ruggiero, M.L. \& Radicella, N., Weak-Field Spherically Symmetric Solutions in $f(T)$ gravity, {\em Physical Review D} \textbf{91} (2015) 104014, arXiv:1501.02198 [gr-qc]; 

\bibitem{Ruggiero2} Iorio, L., Radicella, N. \& Ruggiero, M.L., Constraining $f(T)$ gravity in the Solar System, {\em Journal of Cosmology and Astroparticle Physics} 08 (2015), 021, [arXiv:1505.06996 [gr-qc]].

\bibitem{coley03} Coley, A.A., “Dynamical systems and cosmology” (Kluwer Academic, Dordrecht: ISBN 1-4020-1403-1, 2003).

\bibitem{BahamondeBohmer} Bahamonde, S., Bohmer, C.G., Carloni, S., Copeland, E.J., Fang, W. \& Tamanini, N., Dynamical systems applied to cosmology: dark energy and modified gravity, {\em Physics Reports} 775-777 (2018), 1-122, [arXiv:1712.03107 [gr-qc]].

\bibitem{Kofinas} Kofinas, G., Leon, G. \& Saridakis, E.N., Dynamical behavior in $f(T,T_G)$ cosmology, {\em Classical and Quantum Gravity} \textbf{31} (2014) 175011, [arXiv:1404.7100 [gr-qc]].

\bibitem{golov1} Golovnev, A. \& Guzman, M.-J., Approaches to spherically symmetric solutions in $f(T)$-gravity, {\em Universe} \textbf{7} (5), 121 (2021), [ArXiv:2103.16970 [gr-qc]].

\bibitem{awad1} Awad, A., Golovnev, A., Guzman, M.-J. \& El Hanafy, W., Revisiting diagonal tetrads: New Black Hole solutions in $f(T)$-gravity, {\em European Physical Journal C} (2022) \textbf{82} 972, [ArXiv:2207.00059 [gr-qc]].

\bibitem{bahagolov1} Bahamonde, S., Golovnev, A., Guzm\'an, M.-J., Said, J.L. \& Pfeifer, C., Black Holes in $f(T,B)$ Gravity: Exact and Perturbed Solutions, {\em Journal of Cosmology and Astroparticle Physics} \textbf{01} (2022) 037, [ArXiv:2110.04087 [gr-qc]].

\bibitem{baha6} Bahamonde, S., Faraji, S. , Hackmann, E. \& Pfeifer, C., Thick accretion disk configurations in the Born-Infeld teleparallel gravity, {\em Physical Review D} \textbf{106} (2022), 084046, [ArXiv:2209.00020 [gr-qc]].

\bibitem{baha4} Bahamonde, S., L. Ducobu, L. \& Pfeifer, C., Scalarized Black Holes in Teleparallel Gravity, {\em Journal of Cosmology and Astroparticle Physics} 04 (2022), 018, [ArXiv:2201.11445 [gr-qc]].

\bibitem{baha1} Bahamonde, S. \& Camci, U., Exact Spherically Symmetric Solutions in Modified Teleparallel gravity, {\em Symmetry} 2019, \textbf{11} (12), 1462, [ArXiv:1911.03965 [gr-qc]].

\bibitem{chinea1988symmetries} Chinea, F.J., {Symmetries in tetrad theories.} {\em Classical And Quantum Gravity} \textbf{5}, 135 (1988)

\bibitem{estabrook1996moving} Estabrook, F. and  Wahlquist, H., {Moving frame formulations of 4-geometries having isometries.} {\em Classical and Quantum Gravity} \textbf{13}, 1333 (1996)

\bibitem{papadopoulos2012locally} Papadopoulos, G. \& Grammenos, T., {Locally homogeneous spaces, induced Killing vector fields and applications to Bianchi prototypes.} {\em Journal of Mathematical Physics} \textbf{53}, 072502 (2012) [arXiv:1106.3897 [math]]

\bibitem{olver1995equivalence} Olver, P., {Equivalence, invariants and symmetry.} (Cambridge University Press, 1995)

\bibitem{aaman1998riemann} Aman, J., Fonseca-Neto, J., MacCallum, M. \& Rebouças, M., {Riemann-Cartan spacetimes of Gödel type.} {\em Classical and Quantum Gravity} \textbf{15}, 1089 (1998) [arXiv:gr-qc/9711064]

\bibitem{fonseca1996algebraic} Fonseca-Neto, J., Reboucas, M. \& MacCallum, M., {Algebraic computing in torsion theories of gravitation.} {\em Mathematics And Computers In Simulation} \textbf{42}, 739 (1996)

\bibitem{Krssak_Saridakis2015} Krššák, M. \& Saridakis, E., {The covariant formulation of $f(T)$ gravity.} {\em Classical and Quantum Gravity} \textbf{33}, 115009 (2016) [arXiv:1510.08432 [gr-qc]]

\bibitem{Coley:2022aty} Coley, A.A. \& van den Hoogen, R.J.,  {Teleparallel Geometry with a Single Affine Symmetry.}, {\em Journal of Mathematical Physics} \textbf{64}, 022503 (2023) [arXiv:2205.07071 [gr-qc]]

\bibitem{vandenHoogen:2023pjs} van den Hoogen, R.J. \& Coley A.A.,  {Bianchi type cosmological models in $f(T)$ tele-parallel gravity.}, {\em Journal of Cosmology and Astroparticle Physics} \textbf{10}, 042 (2023) [arXiv:2307.11475 [gr-qc]]

\bibitem{Coley:2023ibm}  Coley, A.A. \& van den Hoogen R.J., {Spatially Homogeneous Teleparallel Gravity: Bianchi I.}, {\em Journal of Mathematical Physics} \textbf{64}, 102506 (2023), [arXiv:2305.12168 [gr-qc]]

\bibitem{golovadd1} Golovnev, A., Koivisto, T., Sandstad, M., On the covariance of teleparallel gravity theories, {\em Classical and Quantum Gravity} \textbf{34}, 145013 (2017), [arXiv:1701.06271 [gr-qc]]

\bibitem{sharif2009teleparallel} Sharif, M. \& Majeed, B. Teleparallel Killing Vectors of Spherically Symmetric Spacetimes. {\em Communications In Theoretical Physics}. \textbf{52}, 435 (2009), [arXiv:0905.3212 [gr-qc]]

\bibitem{pfeifer2021static} Pfeifer, C. \& Schuster, S., Static spherically symmetric black holes in weak $f(T)$-gravity. {\em Universe}. \textbf{7}, 153 (2021), [arXiv:2104.00116 [gr-qc]]

\bibitem{hohmann2021complete} Hohmann, M., Complete classification of cosmological teleparallel geometries. {\em International Journal Of Geometric Methods In Mathematical Physics}. \textbf{18}, 2140005 (2021), [arXiv:2008.12186 [gr-qc]]

\bibitem{hohmann2021teleparallel} Hohmann, M. \& Pfeifer, C. Teleparallel axions and cosmology. {\em European Physical Journal C}. \textbf{81}, 1-14 (2021), [arXiv:2012.14423 [gr-qc]]

\bibitem{hohmann2021general} Hohmann, M., General cosmological perturbations in teleparallel gravity. {\em European Physical Journal Plus}. \textbf{136}, 1-24 (2021), [arXiv:2011.02491  [gr-qc]]

\bibitem{paliathanasis2022f} Paliathanasis, A. \& Leon, G.,  $f(T,B)$ gravity in a Friedmann-Lemaitre-Robertson-Walker universe with nonzero spatial curvature, {\em Mathematical Methods in the Applied Sciences} 2022, 1-18 (2022), [arXiv:2201.12189 [gr-qc]]

\bibitem{McNutt:2022} McNutt, D.D., Coley, A.A. \& van den Hoogen, R.J., A frame based approach to computing symmetries with non-trivial isotropy groups, {\em Journal of Mathematical Physics} \textbf{64} 032503 (2023) [arXiv:2302.11493 [gr-qc]].

\bibitem{bahamonde2021exploring} Bahamonde, S., Valcarcel, J., Järv, L. \& Pfeifer, C., Exploring axial symmetry in modified teleparallel gravity. {\em Physical Review D}. \textbf{103}, 044058 (2021), [arXiv:2012.09193 [gr-qc]]

\bibitem{landryvandenhoogen1} Landry, A. \& van den Hoogen, R.J., Teleparallel Minkowski Spacetime with Perturbative Approach for Teleparallel Gravity on a Proper Frame, {\em Universe}, \textbf{9} (5), 232 (2023), [ArXiv:2303.16089 [gr-qc]]

\bibitem{Rodrigues2015} Rodrigues, M. E., Kpadonou, A.V., Rahaman, F., Oliveira, P.J. \& Houndjo, M.J.S., Bianchi type-I, type-III and Kantowski-Sachs solutions in $f(T)$ gravity, {\em Astrophysics and Space Science}, \textbf{357} (2015), 129, [arXiv:1408.2689 [gr-qc]].

\bibitem{Amir2015} Amir, M.J. \& Yussouf, M., Kantowski-Sachs Universe Models in $f(T)$ Theory of Gravity, {\em International Journal of Theoretical Physics} \textbf{54} (2015): 2798-2812, [arXiv:1502.00777 [gr-qc]].

\bibitem{TdS} Coley, A.A., Landry, A., van den Hoogen, R.J. \& McNutt, D.D., Generalized Teleparallel de Sitter geometries, {\em European Physical Journal C} \textbf{83}, 977 (2023), [ArXiv:2307.12930 [gr-qc]].

\bibitem{saridakis} Saridakis, E.N., Solving both $H_0$ and $\sigma_8$ tensions in $f(T)$ gravity, In \textit{The Sixteenth Marcel Grossmann Meeting}, World Scientific Publishing Company: Hackensack, NJ, USA, 2023, 1783–1791, 2023, [ArXiv:2301.06881 [gr-qc]]

\end{thebibliography}



\appendix

\section{The Field Equations}\label{appena}

Let us present an alternative form of the FEs for $F(T)$ teleparallel gravity. The FE can be decomposed into antisymmetric and symmetric parts:
\begin{subequations}
\begin{eqnarray}
0 &=& F''(T)S_{[ab]}^{\phantom{[ab]}\nu} \partial_{\nu} T  ,\label{temp2}
 \\
\kappa\Theta_{(ab)} &=&\frac{1}{2}g_{ab}\left[F(T)-TF'(T)\right] + F''(T)S_{(ab)}^{\phantom{(ab)}\nu} \partial_{\nu} T+F'(T)G_{ab}  ,\label{temp1}
\end{eqnarray}
\end{subequations}
where $G_{ab}$ is the usual Einstein tensor calculated from the metric and where it has been assumed that $\Theta_{[ab]}=0$.

In the spherically symmetric case studied here with coframe given by eqn. (\ref{VB:SS}) and spin connection given by eqns. (\ref{Con:SS}) and (\ref{SS:TPcon}) the torsion scalar has the form
\small
\begin{align}
T=&-4\left(\frac{\partial_t \chi}{A_1}\right)\left(\frac{\sin(\chi)\sinh(\psi)}{A_3}\right)
    +4\left(\frac{\partial_r \chi}{A_2}\right)\left(\frac{\sin(\chi)\cosh(\psi)}{A_3}\right)
    +4\left(\frac{\partial_t \psi}{A_1}\right)\Bigg(\frac{\cos(\chi)\cosh(\psi)}{A_3}
     \nonumber\\
  &    
    +\frac{\partial_r A_3}{A_2A_3}\Bigg)-4\left(\frac{\partial_r \psi}{A_2}\right)\left(\frac{\cos(\chi)\sinh(\psi)}{A_3}+\frac{\partial_t A_3}{A_1A_3}\right)
    -4\left(\frac{\partial_r A_1}{A_2 A_1}+\frac{\partial_r A_3}{A_2 A_3}\right)\left(\frac{\cos(\chi)\cosh(\psi)}{A_3}\right)
    \nonumber\\ 
  &
    +4\left(\frac{\partial_t A_2}{A_1 A_2}+\frac{\partial_t A_3}{A_1 A_3}\right)\left(\frac{\cos(\chi)\sinh(\psi)}{A_3}\right) 
  -2\left(\frac{\partial_r A_3}{A_2 A_3}\right)^2
  +2\left(\frac{\partial_t A_3}{A_1 A_3}\right)^2
  -4\left(\frac{\partial_r A_1}{A_2 A_1}\right)\left(\frac{\partial_r A_3}{A_2 A_3}\right)
    \nonumber\\ 
  &
  +4\left(\frac{\partial_t A_2}{A_1 A_2}\right)\left(\frac{\partial_t A_3}{A_1 A_3}\right)
  -\frac{2}{A_3^2}.
\end{align}
\normalsize

Assuming a co-moving fluid having energy density $\rho$ and pressure $P$, the antisymmetric part of the FEs yield the following two eqns.
\small
\begin{subequations}
\begin{eqnarray}
0 &=& F''(T)\left[\left(\frac{\partial_r T}{A_2}\right) \left(\frac{\cos(\chi)\,\sinh(\psi)}{A_3} +\frac{\partial_t A_3}{A_1 A_3}\right)-\left(\frac{\partial_t T}{A_1}\right) \left(\frac{\cos(\chi)\,\cosh(\psi)}{A_3}+\frac{\partial_r A_3}{A_2A_3}\right)\right],\quad
\\
0 &=& \frac{F''\left(T\right)\,\sin(\chi)}{A_3}\,\left[\left(\frac{\partial_r\,T}{A_2}\right) \cosh(\psi)-\left(\frac{\partial_t\,T}{A_1}\right)\sinh(\psi)\right].
\end{eqnarray}
\end{subequations}
\normalsize


The symmetric part of the FEs can be expressed in the following way
\small
\begin{subequations}
\begin{align}\label{Symmetric_FE}
&\kappa \rho=-\frac{1}{2}\left(F(T)-TF'(T)\vphantom{\frac{1}{A_3^2}}\right)
-2F''(T)\left(\frac{\partial_r T}{A_2}\right)\left(\frac{\cos(\chi)\cosh(\psi)}{A_3}+\frac{\partial_r A_3}{A_2A_3}\right) +F'(T)\biggl[-2\frac{\partial_r^2 A_3}{A_2^2A_3}
&\nonumber \\
&\qquad+2\left(\frac{\partial_r A_2}{A_2^2}\right)\left(\frac{\partial_r A_3}{A_2 A_3}\right)-\left(\frac{\partial_r A_3}{A_2A_3}\right)^2+\frac{1}{A_3^2}+2\left(\frac{\partial_t A_2}{A_1 A_2}\right)\left(\frac{\partial_t A_3}{A_1 A_3}\right)+\left(\frac{\partial_t A_3}{A_1 A_3}\right)^2\biggr] , \\
& 0=-F''(T)\left[\left(\frac{\partial_r T}{A_2}\right)\left(\frac{\cos(\chi)\sinh(\psi)}{A_3}+\frac{\partial_t A_3}{A_1A_3}\right)+\left(\frac{\partial_t T}{A_2}\right)\left(\frac{\cos(\chi)\cosh(\psi)}{A_3}+\frac{\partial_r A_3}{A_2A_3}\right)\right]
\nonumber \\
&\quad\ +F'(T)\biggl[-2\frac{\partial_t\partial_r A_3}{A_1A_2A_3}
+2\left(\frac{\partial_r A_1}{A_1A_2}\right)\left(\frac{\partial_t A_3}{A_1 A_3}\right)
+2\left(\frac{\partial_r A_3}{A_2A_3}\right)\left(\frac{\partial_t A_2}{A_1A_2}\right)\biggr], 
\\
&\kappa P=\frac{1}{2}\left(F(T)-TF'(T)\vphantom{\frac{1}{A_3^2}}\right)
-2F''(T)\left(\frac{\partial_t T}{A_1}\right)\left(\frac{\cos(\chi)\sinh(\psi)}{A_3}+\frac{\partial_t A_3}{A_1A_3}\right)+F'(T)
&\nonumber \\
&
\qquad\times\biggl[
2\left(\frac{\partial_r A_1}{A_2A_1}\right)\left(\frac{\partial_r A_3}{A_2 A_3}\right)
+\left(\frac{\partial_r A_3}{A_2A_3}\right)^2
-\frac{1}{A_3^2}+2\left(\frac{\partial_t A_1}{A_2A_1}\right)\left(\frac{\partial_t A_3}{A_1 A_3}\right)
-\left(\frac{\partial_t A_3}{A_1A_3}\right)^2
-2\left(\frac{\partial_t^2 A_3}{A_1^2A_3}\right)
\biggr],\\
&\frac{\kappa}{2} (\rho+3P)=\frac{1}{2}\left(F(T)-TF'(T)\vphantom{\frac{1}{A_3^2}}\right)+F''(T)\biggl[
\left(\frac{\partial_r T}{A_2}\right)\left(\frac{\partial_r A_1}{A_1A_2}-\frac{\partial_t \psi}{A_1}\right)+ \left(\frac{\partial_t T}{A_1}\right)
\nonumber \\
&\qquad\qquad
\times \Bigg(-2\frac{\cos(\chi)\sinh(\psi)}{A_3}-2\frac{\partial_t A_3}{A_1A_3}
 -\frac{\partial_t A_2}{A_1A_2}+\frac{\partial_r \psi}{A_2}\Bigg)\biggr]+F'(T)\biggl[\frac{\partial_r^2 A_1}{A_2^2A_1}
-\left(\frac{\partial_r A_1}{A_2A_1}\right)\left(\frac{\partial_r A_2}{A_2A_2}\right)
&\nonumber \\
&\qquad\qquad
+2\left(\frac{\partial_r A_1}{A_2A_1}\right)\left(\frac{\partial_r A_3}{A_2 A_3}\right)
-\frac{\partial_t^2 A_2}{A_1^2A_2}
-2\frac{\partial_t^2 A_3}{A_1^2A_3}
+\left(\frac{\partial_t A_1}{A_1^2}\right)\left(\frac{\partial_t A_2}{A_1A_2}\right)
+2\left(\frac{\partial_t A_1}{A_1^2}\right)\left(\frac{\partial_t A_3}{A_1A_3}\right)
\biggr].
\end{align}
\end{subequations}
\normalsize
The eqns. above can be supplemented with the energy-momentum conservation eqns.
\begin{subequations}
\begin{eqnarray}
0 &=&(\rho+P)\left(\frac{\partial_t{A_2}}{A_2}+2\frac{\partial_t A_3}{A_3}\right)+\partial_t \rho ,
\\
0 &=&(\rho+P)\left(\frac{\partial_rA_1}{A_1}\right)+\partial_r P .
\end{eqnarray}
\end{subequations}


\section{Symmetric FE components for the $\lambda \neq 0$ solutions}\label{appenb}

We present the functions defined in the symmetric FE in the 2 subsolutions for the general $\lambda \neq 0$ case.


\subsection{1st Subsolution}

\noindent For the case where $\sin \chi(z)=0$ and $\cos \chi=\delta$, the eqns. \eqref{3009a} -- \eqref{3009d} components are:
\small
\begin{subequations}
\begin{align}
g_1 =& \frac{1}{f_1^3\,f_2^3\,f_3^2} \Bigg[-\frac{z^2\,f_1\,f_2\,f_3\,f_3''}{2}\left(C_0^2\,f_2^2+f_1^2\right)+\frac{z^2\,f_1\,f_2\,f_3^2}{2}\left(C_0^2\,f_2\,f_2''-f_1\,f_1''\right)+\frac{z^2\,f_1\,f_2\,f_3'^2}{2}\left(f_1^2-C_0^2\,f_2^2\right)
\nonumber\\
&+f_3\,f_3'\,\Bigg(\frac{z^2\,f_2\,f_1'}{2}\left(C_0^2\,f_2^2+f_1^2\right)+f_1\left[\frac{z^2\,f_2'}{2}\,\left(C_0^2\,f_2^2+f_1^2\right)+z\,f_2\left(\left(\lambda-\frac{1}{2}\right)\,f_1^2-\frac{C_0^2\,f_2^2}{2}\right)\right]\Bigg)
\nonumber\\
& -\frac{f_3^2\,f_1'}{2}\left[z^2\,f_2'\,\left(C_0^2\,f_2^2-f_1^2\right)+z\,f_1^2\,f_2\right]+f_1\,\left[z\,f_3^2\,f_2'\left(\frac{C_0^2\,f_2^2}{2}+\lambda\,f_1^2\right)+f_1^2\,f_2\left(\lambda^2\,f_3^2-\frac{f_2^2}{2}\right)\right]\Bigg] , \label{3031a}
\end{align}

\newpage

\begin{align}
g_2 =&\frac{1}{f_1^3\,f_2^2\,f_3^2}\Bigg[-C_0^2\,z^2\,f_1\,f_2^2\,f_3\,f_3''+z^2\,f_1\,f_3'^2\,\left(f_1^2-C_0^2\,f_2^2\right)+f_3'\,\Bigg(2\,z^2\,f_3\,f_1'\left(\frac{C_0^2\,f_2^2}{2}+f_1^2\right)
\nonumber\\
&+4\,f_1\,\Bigg[-\frac{C_0^2\,z^2\,f_2\,f_3\,f_2'}{4}+z\,\left[\frac{\delta\,f_1\,f_2}{4}\left(C_0\,f_2\,\sinh\psi+f_1\,\cosh\psi\right)+f_3\,\left(-\frac{C_0\,f_2^2}{4}+\lambda\,f_1^2\right)\right]\Bigg]\Bigg)
\nonumber\\
&+3\,f_1^2\,f_3\,\Bigg(\frac{2\,z\,f_1'}{3}\left[\frac{\delta\,f_2\,\cosh\psi}{2}+\lambda\,f_3\right]+\frac{z\,f_2\,\psi'}{3}\left[\delta\,\left(C_0\,f_2\,\cosh\psi+f_1\,\sinh\psi\right)+\lambda\,C_0\,f_3\right]
\nonumber\\
&+\frac{\delta\,C_0\,z\,f_2\,f_2'\,\sinh\psi}{3}+\lambda\,f_1\,\left[\frac{2\,\delta\,f_2\,\cosh\psi}{3}+\lambda\,f_3\right]\Bigg)\Bigg] , \label{3031b}
\\
g_3 =& \frac{1}{f_1^2\,f_2^3\,f_3^2}\Bigg[-z^2\,f_1^2\,f_2\,f_3\,f_3''+z^2\,f_2\,f_3'^2\,\left(C_0^2\,f_2^2-f_1^2\right)+f_3'\,\Bigg(z^2\,f_3\,f_2'\,\left(2\,C_0^2\,f_2^2+f_1^2\right)
\nonumber\\
&-4\,f_1\,f_2\,\Bigg[\frac{z^2\,f_3\,f_1'}{4}+z\,\Bigg(\frac{\delta\,f_2}{4}\,\left(C_0\,f_2\,\sinh\psi+f_1\,\cosh\psi\right)+f_1\,f_3\,\left(\lambda+\frac{1}{4}\right)\Bigg)\Bigg]\Bigg)
\nonumber\\
&-2\,f_1\,f_3\Bigg(-\frac{z\,f_2'}{2}\left(-\delta\,C_0\,f_2^2\,\sinh\psi+\lambda\,f_1\,f_3\right)+f_2\,\Bigg[\frac{z\,f_1'}{2}\left(\delta\,f_2\,\cosh\psi+\lambda\,f_3\right)
\nonumber\\
&+\frac{z\,f_2\,\psi'}{2}\left(\delta\,\left(C_0\,f_2\,\cosh\psi+f_1\,\sinh\psi\right)+\lambda\,C_0\,f_3\right)+\lambda\,f_1\,\left(\delta\,f_2\,\cosh\psi+\lambda\,f_3\right)\Bigg]\Bigg)\Bigg] ,  \label{3031c}
\\
g_4 =& \frac{C_0}{f_1^2\,f_2^2\,f_3}\Bigg[-z^2\,f_1\,f_2\,f_3''+f_3'\,\left[z^2\,\left(f_1\,f_2\right)'-z\,f_1\,f_2\right]+\lambda\,z\,f_1\,f_2'\,f_3\Bigg] , \label{3031d}
\\
h_1=& \frac{1}{f_1^2\,f_2^2\,f_3} \Bigg[-\frac{z^2\,f_1\,f_3\,f_1'}{4}+\frac{C_0^2\,z^2\,f_2\,f_3\,f_2'}{4}-\frac{z\,f_1}{2}\,\left[\frac{\delta\,f_2}{2}\,\left(-C_0\,f_2\,\sinh\psi+f_1\,\cosh\psi\right)+\lambda\,f_1\,f_3\right]
\nonumber\\
&-\frac{z^2}{4}\,f_3'\,\left(C_0\,^2\,f_2^2+f_1^2\right)\Bigg] ,\label{3031e}
\\
h_2=& \frac{C_0}{f_1^2\,f_3}\,\Bigg[\delta\,z\,f_1\,\sinh\psi-C_0\,z^2\,f_3'\Bigg], \label{3031f}
\\
h_3=& -\frac{z}{2\,f_2^2\,f_3}\Bigg[z\,f_3'+\delta\,f_2\,\cosh\psi+\lambda\,f_3\Bigg], \label{3031g}
\\
h_4=& \frac{1}{f_1\,f_2\,f_3}\Bigg[-C_0\,z^2\,f_3'-\frac{z}{2}\,\left[\delta\,\left(C_0\,f_2\,\cosh\psi-f_1\,\sinh\psi\right)+\lambda\,C_0\,f_3\right]\Bigg], \label{3031h}
\\
m_1=& \frac{\lambda}{f_1\,f_2^2\,f_3}\Bigg[\frac{z\,f_3\,f_1'+C_0\,z\,f_2\,f_3\,\psi'}{2}+f_1\,\left[\frac{\delta\,f_2\,\cosh\psi}{2}+\lambda\,f_3\right]+\frac{z\,f_3'\,f_1}{2}\Bigg],\label{3031i}
\\
m_2=&0 , \label{3031j}
\\
m_3=& \frac{\lambda}{f_2^2\,f_3}\Bigg[z\,f_3'+\delta\,f_2\,\cosh\psi+\lambda\,f_3\Bigg],\label{3031k}
\\
m_4=& \frac{1}{f_1\,f_2\,f_3}\Bigg[C_0\,\lambda\,z\,f_3'-\delta\,\lambda\,f_1\,\sinh\psi\Bigg] . \label{3031l}
\end{align}
\end{subequations}
\normalsize

\newpage

\noindent The $T_0(z)$ expression is the following:
\small
\begin{align}\label{3032}
& T_0(z) =\Bigg[ \frac{1}{f_1^2\,f_2^2\,f_3^2}\Bigg[-8\,f_1\,f_2\,\delta\,\Bigg(\left[f_1\,\left(\lambda\,f_3+\frac{z\,f_3'}{2}\right)+\frac{z\,f_3\left(\psi'\,C_0\,f_2+f_1'\right)}{2}\right]\cosh\psi
\nonumber\\
&+\frac{z\,\sinh\psi}{2}\left[f_1\,f_3\,\psi'+C_0\,\left(f_2\,f_3\right)'\right]\Bigg)+f_1^2\,\left[-6\lambda^2\,f_3^2-8\lambda\,z\,f_3\,f_3'-2\,z^2\,f_3'^2-2\,f_2^2\right]
\nonumber\\
&-4\,f_1\,f_3\left[\lambda\,C_0\,z\,f_2\,f_3\,\psi'+f_1'\left(\lambda\,z\,f_3+z^2\,f_3'\right)\right]+2\,C_0^2\,z^2\,f_2\,f_3'\left(f_2\,f_3'+2\,f_3\,f_2'\right)\Bigg]\Bigg].
\end{align}
\normalsize


\subsection{2nd Subsolution}

\noindent For the case described by eq \eqref{3008}, eqns. \eqref{3009a} -- \eqref{3009d} components are:
\small
\begin{subequations}
\begin{align}
g_1 =& \frac{1}{f_1^3\,f_2^3\,f_3^2} \Bigg[-\frac{z^2\,f_1\,f_2\,f_3\,f_3''}{2}\left(C_0^2\,f_2^2+f_1^2\right)+\frac{z^2\,f_1\,f_2\,f_3^2}{2}\left(C_0^2\,f_2\,f_2''-f_1\,f_1''\right)+\frac{z^2\,f_1\,f_2\,f_3'^2}{2}\left(f_1^2-C_0^2\,f_2^2\right)
\nonumber\\
&+f_3\,f_3'\,\Bigg(\frac{z^2\,f_2\,f_1'}{2}\left(C_0^2\,f_2^2+f_1^2\right)+f_1\left[\frac{z^2\,f_2'}{2}\,\left(C_0^2\,f_2^2+f_1^2\right)+z\,f_2\left(\left(\lambda-\frac{1}{2}\right)\,f_1^2-\frac{C_0^2\,f_2^2}{2}\right)\right]\Bigg)
\nonumber\\
& -\frac{f_3^2\,f_1'}{2}\left[z^2\,f_2'\,\left(C_0^2\,f_2^2-f_1^2\right)+z\,f_1^2\,f_2\right]+f_1\,\left[z\,f_3^2\,f_2'\left(\frac{C_0^2\,f_2^2}{2}+\lambda\,f_1^2\right)+f_1^2\,f_2\left(\lambda^2\,f_3^2-\frac{f_2^2}{2}\right)\right]\Bigg] , \label{3041a}
\\
g_2 =& \frac{1}{f_1^3\,f_2^2\,f_3^2}\Bigg[-C_0^2\,z^2\,f_1\,f_2^2\,f_3\,f_3''+z^2\,f_1\,f_3'^2\,\left(f_1^2-C_0^2\,f_2^2\right)+f_3'\,\Bigg(2\,z^2\,f_3\,f_1'\left(\frac{C_0^2\,f_2^2}{2}+f_1^2\right)
\nonumber\\
&+4\,f_1\,\Bigg[-\frac{C_0^2\,z^2\,f_2\,f_3\,f_2'}{4}+z\,\Bigg[-\left[\frac{\sinh\psi}{f_1}\left(C_0\,z\,f_3'\right)+\frac{\cosh\psi}{f_2}\,\left(z\,f_3'+\lambda\,f_3\,\right)\right]\frac{f_1\,f_2}{4}
\nonumber\\
&\left(C_0\,f_2\,\sinh\psi+f_1\,\cosh\psi\right)+f_3\,\left(-\frac{C_0\,f_2^2}{4}+\lambda\,f_1^2\right)\Bigg]\Bigg]\Bigg)
\nonumber\\
&+3\,f_1^2\,f_3\,\Bigg(\frac{2\,z\,f_1'}{3}\left[-\left[\frac{\sinh\psi}{f_1}\left(C_0\,z\,f_3'\right)+\frac{\cosh\psi}{f_2}\,\left(z\,f_3'+\lambda\,f_3\,\right)\right]\frac{f_2\,\cosh\psi}{2}+\lambda\,f_3\right]
\nonumber\\
&+\frac{z\,f_2\,\psi'}{3}\left[-\left[\frac{\sinh\psi}{f_1}\left(C_0\,z\,f_3'\right)+\frac{\cosh\psi}{f_2}\,\left(z\,f_3'+\lambda\,f_3\,\right)\right]\left(C_0\,f_2\,\cosh\psi+f_1\,\sinh\psi\right)+\lambda\,C_0\,f_3\right]
\nonumber\\
&-\left[\frac{\sinh\psi}{f_1}\left(C_0\,z\,f_3'\right)+\frac{\cosh\psi}{f_2}\,\left(z\,f_3'+\lambda\,f_3\,\right)\right]\frac{C_0\,z\,f_2\,f_2'\,\sinh\psi}{3}
\nonumber\\
&-\frac{z\,f_2}{3}\sin\left(\arccos\left(-\left[\frac{\sinh\psi}{f_1}\left(C_0\,z\,f_3'\right)+\frac{\cosh\psi}{f_2}\,\left(z\,f_3'+\lambda\,f_3\,\right)\right]\right)\right)
\nonumber\\
&\times \left(\arccos\left(-\left[\frac{\sinh\psi}{f_1}\left(C_0\,z\,f_3'\right)+\frac{\cosh\psi}{f_2}\,\left(z\,f_3'+\lambda\,f_3\,\right)\right]\right)\right)'\left(C_0\,f_2\,\sinh\psi+f_1\,\cosh\psi\right)
\nonumber\\
&+\lambda\,f_1\,\left[-\left[\frac{\sinh\psi}{f_1}\left(C_0\,z\,f_3'\right)+\frac{\cosh\psi}{f_2}\,\left(z\,f_3'+\lambda\,f_3\,\right)\right]\frac{2\,f_2\,\cosh\psi}{3}+\lambda\,f_3\right]\Bigg)\Bigg] , \label{3041b}
\end{align}
\newpage

\begin{align}
g_3 =& \frac{1}{f_1^2\,f_2^3\,f_3^2}\Bigg[-z^2\,f_1^2\,f_2\,f_3\,f_3''+z^2\,f_2\,f_3'^2\,\left(C_0^2\,f_2^2-f_1^2\right)+f_3'\,\Bigg(z^2\,f_3\,f_2'\,\left(2\,C_0^2\,f_2^2+f_1^2\right)-4\,f_1\,f_2\,
\nonumber\\
&\times\Bigg[\frac{z^2\,f_3\,f_1'}{4}+z\,\Bigg(-\left[\frac{\sinh\psi}{f_1}\left(C_0\,z\,f_3'\right)+\frac{\cosh\psi}{f_2}\,\left(z\,f_3'+\lambda\,f_3\,\right)\right]\frac{f_2}{4}\,\left(C_0\,f_2\,\sinh\psi+f_1\,\cosh\psi\right)
\nonumber\\
&+f_1\,f_3\,\left(\lambda+\frac{1}{4}\right)\Bigg)\Bigg]\Bigg)-2\,f_1\,f_3\Bigg(-\frac{z\,f_2'}{2}\Bigg(C_0\,f_2^2\,\sinh\psi\,\Bigg[\frac{\sinh\psi}{f_1}\left(C_0\,z\,f_3'\right)
+\frac{\cosh\psi}{f_2}\,\left(z\,f_3'+\lambda\,f_3\,\right)\Bigg]
\nonumber\\
&+\lambda\,f_1\,f_3\Bigg)+f_2\,\Bigg[\frac{z\,f_1'}{2}\Bigg(-\Bigg[\frac{\sinh\psi}{f_1}\left(C_0\,z\,f_3'\right)+\frac{\cosh\psi}{f_2}\,\left(z\,f_3'+\lambda\,f_3\,\right)\Bigg]\,f_2\,\cosh\psi+\lambda\,f_3\Bigg)
\nonumber\\
&+\frac{z\,f_2\,\psi'}{2}\Bigg(-\left(C_0\,f_2\,\cosh\psi+f_1\,\sinh\psi\right)\,\left[\frac{\sinh\psi}{f_1}\left(C_0\,z\,f_3'\right)+\frac{\cosh\psi}{f_2}\,\left(z\,f_3'+\lambda\,f_3\,\right)\right]+\lambda\,C_0\,f_3\Bigg)
\nonumber\\
&-\frac{z\,f_2}{2}\sin\left(\arccos\left(-\left[\frac{\sinh\psi}{f_1}\left(C_0\,z\,f_3'\right)+\frac{\cosh\psi}{f_2}\,\left(z\,f_3'+\lambda\,f_3\,\right)\right]\right)\right)
\nonumber\\
&\times \left(\arccos\left(-\left[\frac{\sinh\psi}{f_1}\left(C_0\,z\,f_3'\right)+\frac{\cosh\psi}{f_2}\,\left(z\,f_3'+\lambda\,f_3\,\right)\right]\right)\right)'\left(C_0\,f_2\,\sinh\psi+f_1\,\cosh\psi\right)
\nonumber\\
&+\lambda\,f_1\,\left(-\left[\frac{\sinh\psi}{f_1}\left(C_0\,z\,f_3'\right)+\frac{\cosh\psi}{f_2}\,\left(z\,f_3'+\lambda\,f_3\,\right)\right]f_2\,\cosh\psi\,+\lambda\,f_3\right)\Bigg]\Bigg)\Bigg], \label{3041c}
\\
g_4 =& \frac{C_0}{f_1^2\,f_2^2\,f_3}\Bigg[-z^2\,f_1\,f_2\,f_3''+f_3'\,\left[z^2\,\left(f_1\,f_2\right)'-z\,f_1\,f_2\right]+\lambda\,z\,f_1\,f_2'\,f_3\Bigg],  \label{3041d}
\\
h_1=& \frac{1}{f_1^2\,f_2^2\,f_3} \Bigg[-\frac{f_3'\,z^2}{4}\,\left(C_0\,^2\,f_2^2+f_1^2\right)+\Bigg(-\frac{z^2\,f_1\,f_3\,f_1'}{4}+\frac{C_0^2\,z^2\,f_2\,f_3\,f_2'}{4}-\frac{z\,f_1}{2}\,
\nonumber\\
&\times\Bigg[-\left[\frac{\sinh\psi}{f_1}\left(C_0\,z\,f_3'\right)+\frac{\cosh\psi}{f_2}\,\left(z\,f_3'+\lambda\,f_3\,\right)\right]\frac{f_2}{2}
\,\left(-C_0\,f_2\,\sinh\psi+f_1\,\cosh\psi\right)+\lambda\,f_1\,f_3\Bigg]\Bigg)\Bigg], \label{3041e}
\\
h_2=&\frac{C_0}{f_1^2\,f_3} \Bigg[-\left[\frac{\sinh\psi}{f_1}\left(C_0\,z\,f_3'\right)+\frac{\cosh\psi}{f_2}\,\left(z\,f_3'+\lambda\,f_3\,\right)\right]z\,f_1\,\sinh\psi-C_0\,z^2\,f_3'\Bigg]  ,\label{3041f}
\\
h_3=&-\frac{z}{2f_2^2\,f_3}\Bigg[z\,f_3'-\left[\frac{\sinh\psi}{f_1}\left(C_0\,z\,f_3'\right)+\frac{\cosh\psi}{f_2}\,\left(z\,f_3'+\lambda\,f_3\,\right)\right]\, f_2\,\cosh\psi+\lambda\,f_3\Bigg] ,  \label{3041g}
\\
h_4=& \frac{1}{f_1\,f_2\,f_3}\Bigg[-z^2\,C_0\,f_3'-\frac{z}{2}\Bigg[-\left(C_0\,f_2\,\cosh\psi-f_1\,\sinh\psi\right)\,\left[\frac{\sinh\psi}{f_1}\left(C_0\,z\,f_3'\right)+\frac{\cosh\psi}{f_2}\,\left(z\,f_3'+\lambda\,f_3\,\right)\right]
\nonumber\\
&+\lambda\,C_0\,f_3\Bigg]\Bigg] , \label{3041h}
\\
m_1=& \frac{\lambda}{f_1\,f_2^2\,f_3} \Bigg[\frac{z\,f_3'\,f_1}{2}+\Bigg(\frac{z\,f_3\,f_1'+C_0\,z\,f_2\,f_3\,\psi'}{2}
\nonumber\\
&+f_1\,\left[-\left[\frac{\sinh\psi}{f_1}\left(C_0\,z\,f_3'\right)+\frac{\cosh\psi}{f_2}\,\left(z\,f_3'+\lambda\,f_3\,\right)\right]\frac{f_2\,\cosh\psi}{2}+\lambda\,f_3\right]\Bigg)\Bigg] , \label{3041i}
\\
m_2=&0 , \label{3041j}
\\
m_3=&\frac{\lambda}{f_2^2\,f_3}\Bigg[z\,f_3'-\left[\frac{\sinh\psi}{f_1}\left(C_0\,z\,f_3'\right)+\frac{\cosh\psi}{f_2}\,\left(z\,f_3'+\lambda\,f_3\,\right)\right]\, f_2\,\cosh\psi+\lambda\,f_3\Bigg] ,\label{3041k}
\\
m_4=& \frac{1}{f_1\,f_2\,f_3}\Bigg[C_0\,\lambda\,z\,f_3'+\lambda\,f_1\,\sinh\psi\,\left[\frac{\sinh\psi}{f_1}\left(C_0\,z\,f_3'\right)+\frac{\cosh\psi}{f_2}\,\left(z\,f_3'+\lambda\,f_3\,\right)\right]\Bigg] . \label{3041l}
\end{align}
\end{subequations}

\newpage

\noindent The $T_0(z)$ expression is the following:
\small
\begin{align}\label{3042}
&T_0(z)= \frac{1}{f_1^2\,f_2^2\,f_3^2}\Bigg[8\,f_1\,f_2\,\left[\frac{\sinh\psi}{f_1}\left(C_0\,z\,f_3'\right)+\frac{\cosh\psi}{f_2}\,\left(z\,f_3'+\lambda\,f_3\,\right)\right]\,
\nonumber\\
&\Bigg(\left[f_1\,\left(\lambda\,f_3+\frac{z\,f_3'}{2}\right)+\frac{z\,f_3\left(\psi'\,C_0\,f_2+f_1'\right)}{2}\right]\cosh\psi+\frac{z\,\sinh\psi}{2}\left[f_1\,f_3\,\psi'+C_0\,\left(f_2\,f_3\right)'\right]\Bigg)
\nonumber\\
&+4\,z\,\sin\left(\arccos\left(-\left[\frac{\sinh\psi}{f_1}\left(C_0\,z\,f_3'\right)+\frac{\cosh\psi}{f_2}\,\left(z\,f_3'+\lambda\,f_3\,\right)\right]\right)\right)\,\cosh\psi\,
\nonumber\\
&\left(\arccos\left(-\left[\frac{\sinh\psi}{f_1}\left(C_0\,z\,f_3'\right)+\frac{\cosh\psi}{f_2}\,\left(z\,f_3'+\lambda\,f_3\,\right)\right]\right)\right)'\,f_1^2\,f_2\,f_3
\nonumber\\
&+4\,z\,\sinh\psi\,\sin\left(\arccos\left(-\left[\frac{\sinh\psi}{f_1}\left(C_0\,z\,f_3'\right)+\frac{\cosh\psi}{f_2}\,\left(z\,f_3'+\lambda\,f_3\,\right)\right]\right)\right)\,
\nonumber\\
&\left(\arccos\left(-\left[\frac{\sinh\psi}{f_1}\left(C_0\,z\,f_3'\right)+\frac{\cosh\psi}{f_2}\,\left(z\,f_3'+\lambda\,f_3\,\right)\right]\right)\right)'\,C_0\,f_1\,f_2^2\,f_3
\nonumber\\
& +f_1^2\,\left[-6\lambda^2\,f_3^2-8\lambda\,z\,f_3\,f_3'-2\,z^2\,f_3'^2-2\,f_2^2\right]-4\,f_1\,f_3\left[\lambda\,C_0\,z\,f_2\,f_3\,\psi'+f_1'\left(\lambda\,z\,f_3+z^2\,f_3'\right)\right]
\nonumber\\
&+2\,C_0^2\,z^2\,f_2\,f_3'\left(f_2\,f_3'+2\,f_3\,f_2'\right)\Bigg].
\end{align}
\normalsize

\end{document}